\documentclass[fleqn,usenatbib]{mnras}

\usepackage{newtxtext,newtxmath}

\usepackage[T1]{fontenc}
\usepackage{ae,aecompl}

\usepackage{graphicx}	% Including figure files
\usepackage{amsmath}	% Advanced maths commands
\usepackage{bm}
\usepackage{tabularx}
\usepackage{xcolor}

\title[MASCARA-1b, KELT-16b, and Reanalyses with SPCA]{A Comprehensive Reanalysis of \textit{Spitzer}'s 4.5~$\mu$m Phase Curves, and the Phase Variations of the Ultra-hot Jupiters MASCARA-1b and KELT-16b}

\author[T.~J.~Bell et al.]{%
Taylor J.~Bell,$^{1,2}$\thanks{E-mail: taylor.bell@mail.mcgill.ca (TJB)}
Lisa Dang,$^{1,2}$
Nicolas B.~Cowan,$^{1,2,3}$
Jacob Bean,$^{4}$
\newauthor Jean-Michel D\'esert,$^{5}$
Jonathan J.~Fortney,$^{6}$
Dylan Keating,$^{1,2}$
Eliza Kempton,$^{7}$
\newauthor Laura Kreidberg,$^{8}$
Michael R.~Line,$^{9}$
Megan Mansfield,$^{10}$
Vivien Parmentier,$^{11}$
\newauthor Kevin B.~Stevenson,$^{12}$
Mark Swain,$^{13}$
and Robert T.~Zellem$^{13}$
\\
$^{1}$Department of Physics, McGill University, 3600 rue University, Montr\'eal, QC H3A 2T8, Canada\\
$^{2}$McGill Space Institute; Institute for Research on Exoplanets; Centre for Research in Astrophysics of Quebec\\
$^{3}$Department of Earth \& Planetary Sciences, McGill University, 3450 rue University, Montr\'eal, QC H3A 0E8, Canada\\
$^{4}$Department of Astronomy \& Astrophysics, University of Chicago, Chicago, IL 60637, USA\\
$^{5}$Anton Pannekoek Institute for Astronomy, University of Amsterdam, 1090 GE Amsterdam, The Netherlands\\
$^{6}$Other Worlds Laboratory, Department of Astronomy and Astrophysics, University of California, Santa Cruz, California 95064, USA\\
$^{7}$Department of Astronomy, University of Maryland, College Park, MD 20742, USA\\
$^{8}$Max-Planck-Institut f\"ur Astronomie, K\"onigstuhl 17, 69117 Heidelberg, Germany\\
$^{9}$School of Earth \& Space Exploration, Arizona State University, Tempe AZ 85287, USA\\
$^{10}$Department of Geophysical Sciences, University of Chicago, Chicago, IL 60637, USA\\
$^{11}$Atmospheric, Ocean, and Planetary Physics, Clarendon Laboratory, Department of Physics, University of Oxford, Oxford OX1 3PU, UK\\
$^{12}$Johns Hopkins University Applied Physics Laboratory, 11100 Johns Hopkins Rd, Laurel, MD 20723, USA\\
$^{13}$Jet Propulsion Laboratory, California Institute of Technology, 4800 Oak Grove Drive, Pasadena, CA 91109, USA\\
}

\date{Submitted 2020 October 1. Resubmitted 2021 February 26.}

\pubyear{2021}

\begin{document}
\label{firstpage}
\pagerange{\pageref{firstpage}--\pageref{lastpage}}
\maketitle

\begin{abstract}
We have developed an open-source pipeline for the analysis of \textit{Spitzer}/IRAC channel 1 and 2 time-series photometry, incorporating some of the most popular decorrelation methods. We applied this pipeline to new phase curve observations of ultra-hot Jupiters MASCARA-1b and KELT-16b, and we performed the first comprehensive reanalysis of 15 phase curves. We find that MASCARA-1b and KELT-16b have phase offsets of $6^{+11}_{-11}~^{\circ}$W and $38^{+16}_{-15}~^{\circ}$W, dayside temperatures of $2952^{+100}_{-97}$~K and $3070^{+160}_{-150}$~K, and nightside temperatures of $1300^{+340}_{-340}$~K and $1900^{+430}_{-440}$~K, respectively. We confirm a strong correlation between dayside and irradiation temperatures with a shallower dependency for nightside temperature. We also find evidence that the normalized phase curve amplitude (peak-to-trough divided by eclipse depth) is correlated with stellar effective temperature. In addition, while our different models often retrieve similar parameters, significant differences occasionally arise between them, as well as between our preferred model and the literature values. Nevertheless, our preferred models are consistent with published phase offsets to within $-8\pm21$ degrees ($-1.6\pm3.2$ sigma), and normalized phase curve amplitudes are on average reproduced to within $-0.01\pm0.24$ ($-0.1\pm1.6$ sigma). Finally, we find that BLISS performs best in most cases, but not all; we therefore recommend future analyses consider numerous detector models to ensure an optimal fit and to assess model dependencies.
\end{abstract}


%%%%%%%%%%%%%%%%%%%%%%%%%%%%%%%%%%%%%%%%%%%%%%%%%%

%%%%%%%%%%%%%%%%% BODY OF PAPER %%%%%%%%%%%%%%%%%%

\section{Introduction}

The thermal phase curve observations collected by \textit{Spitzer} have been one of its greatest scientific legacies. \textit{Spitzer} demonstrated that we can detect the variations in disk-integrated flux from an exoplanet as a function of orbital phase \citep[e.g.,][]{harrington2006,deming2020}, allowing us to probe atmospheric dynamics and heat transport \citep[e.g.,][]{parmentiercrossfield2018}. The success of phase curve observations from \textit{Spitzer} and \textit{Hubble} has ushered in the era of comparative atmospheric dynamics \citep[e.g.,][]{zhang2018,keating2019,beatty2019}, which \textit{JWST} and \textit{ARIEL} will carry on in the 2020s and beyond.

However, reaching the level of precision required to make phase curve observations with \textit{Spitzer} has been challenging, as strong intra-pixel sensitivity variations in \textit{Spitzer}'s Infrared Array Camera (IRAC) channels 1 and 2 can be an order of magnitude larger than the astrophysical signals \citep[e.g.,][]{charbonneau2005}. Many methods have been developed to model out these detector systematics, each with strengths and weaknesses, and most research groups have their own preferred method and code. Some of these codes are open source, but those who want to compare different decorrelation techniques are stuck learning (or building) new packages.

Here we present \texttt{SPCA}\footnote{Details about how to use and install \texttt{SPCA} can be found at \url{https://spca.readthedocs.io}}: the \textit{Spitzer} Phase Curve Analysis pipeline, developed by Lisa Dang and Taylor Bell. \texttt{SPCA} seeks to reduce the cost of entry for all while providing flexibility and effectiveness. \texttt{SPCA}'s routines have been developed for \textit{Spitzer}/IRAC channel 1 and channel 2 (3.6~$\mu$m and 4.5~$\mu$m, respectively) time-resolved photometry; these channels were used for the vast majority of \textit{Spitzer} phase curves and share similar detector noise characteristics. \texttt{SPCA} has implementations of 2D polynomial \citep{charbonneau2008}, Pixel Level Decorrelation \citep[PLD;][]{deming2015}, BiLinearly-Interpolated Sub-pixel Sensitivity mapping \citep[BLISS mapping;][]{stevenson2012a}, and Gaussian Process \citep[GP;][]{gibson2012a, evans2015} decorrelation methods, allowing the user to change between techniques by setting a single variable. The modular structure of the code also allows the user to integrate custom astrophysical models and decorrelation methods. Built with automation in mind, \texttt{SPCA} can reduce and decorrelate multiple data sets with a single command. Earlier versions of \texttt{SPCA} were described in \citet{dang2018} and \citet{bell2019}, but the pipeline has undergone significant development in the intervening years.

Our goal is to implement a collection of some of the most common decorrelation methods within a single framework so that it becomes feasible for anyone to perform uniform and repeatable reanalyses of phase curves with each of these decorrelation techniques. This allows for comparisons between detector model performances and results on phase curve observations with different observing techniques, exposure times, stellar fluxes, etc., while previous comparisons were restricted either to just the secondary eclipses of XO-3b \citep{ingalls2016} or individual phase curves \citep[e.g.,][]{wong2015, dang2018, bell2019, keating2020}. The automation within \texttt{SPCA} also makes it possible for us to test the reproducibility of literature phase curve values for most exoplanets, something that has only been done on an individual basis so far \citep[e.g.,][]{knutson2009hd189733b, knutson2012, mendonca2018a, morello2019, bell2019, may2020}.

In Section \ref{sec:data}, we introduce the data sets that we will analyze, and in Section \ref{sec:photometry} we present \texttt{SPCA}'s photometry techniques. In Section \ref{sec:analysis} we detail \texttt{SPCA}'s decorrelation methods and analysis techniques. In Section \ref{sec:eclipseAnalysis}, we validate our models against the collection of 10 \mbox{XO-3b} eclipses first published by \citet{wong2014} and later used in the IRAC Data Challenge 2015 and described in \citet{ingalls2016}. In Section \ref{sec:results} we present the results for our new phase curves of KELT-16b and MASCARA-1b \citep{talens2017,oberst2017}, as well as our reanalyses of most previously published phase curves, and in subsection \ref{sec:literatureComparisons} we compare our results to the literature values. Finally, Section \ref{sec:discussion} presents our discussion and conclusions.

\section{Observations}\label{sec:data}

As part of the final \textit{Spitzer} phase curve study that was conducted in Cycle 14 (PID 14059; PI Bean), we collected new \textit{Spitzer}/IRAC 4.5~$\mu$m phase curve for a total of 10 planets with a range of temperatures and orbital periods. \citet{mansfield2020} previously published the phase curve of KELT-9b from this program, and we present here the phase curves of ultra-hot Jupiters KELT-16b and MASCARA-1b. This pair of planets were selected to permit comparative studies of their atmospheric dynamics since they share similar radii, masses, and irradiation temperatures ($T_{\rm 0} = T_{\rm *,eff}\sqrt{R_{\rm *}/a}$, where $T_{\rm *,eff}$ is the stellar effective temperature, $R_{\rm *}$ is the stellar radius, and $a$ is the planet's orbital semi-major axis). Meanwhile, the two planets have orbital periods that differ by a factor of two and stellar effective temperatures differing by 1300~K. This pairing can, therefore, provide insight into the impacts of Coriolis forces and stellar spectra on the energy budgets of hot Jupiters.

We also present our reanalyses of nearly all previously published 4.5~$\mu$m phase curves: specifically those of
\mbox{CoRoT-2b} \citep[][PID 11073]{dang2018};
\mbox{HAT-P-7b} \citep[][PID 60021]{wong2016};
% \mbox{HD 149026b} \citep[][PID 60021]{zhang2018};
\mbox{HD 189733b} \citep[][PID 60021]{knutson2012};
\mbox{HD 209458b} \citep[][PID 60021]{zellem2014};
\mbox{KELT-1b} \citep[][PID 11095]{beatty2019};
\mbox{KELT-9b} \citep[][PID 14059]{mansfield2020};
\mbox{Qatar-1b} \citep[][PID 13038]{keating2020};
\mbox{WASP-12b} (\citealt{cowan2012}, PID 70060; \citealt{bell2019}, PID 90186);
\mbox{WASP-14b} \citep[][PID 80073]{wong2015};
\mbox{WASP-18b} \citep[][PID 60185]{maxted2013};
\mbox{WASP-19b} \citep[][PID 80073]{wong2016};
\mbox{WASP-33b} \citep[][PID 80073]{zhang2018};
\mbox{WASP-43b} \citep[][PID 11001]{stevenson2017}; and
\mbox{WASP-103b} \citep[][PID 11099]{kreidberg2018b}. We exclude the phase curve of \mbox{HD 149026b} \citep[][PID 60021]{zhang2018} as our initial attempts to fit these observations showed that they were especially challenging to fit and would hinder our attempts at a uniform treatment of each phase curve. We also exclude the observations of \mbox{55 Cnc e} \citep[][PID 90208]{demory2016b} due to the very different nature of that system and the enormous size of that dataset. Finally, we do not consider any phase curves that were not already published when we started this work.

All data sets we consider, except that of WASP-103b, used the subarray mode which produces datacubes of 64 frames, each \mbox{32 $\times$ 32} pixels (\mbox{39 arcsec $\times$ 39 arcsec}) in size. Meanwhile, the data set for WASP-103b was taken in full-frame mode, which gives individual frames that are 256 $\times$ 256 pixels (312 arcsec $\times$ 312 arcsec) in size. All data sets we consider were continuous, full-orbit phase curves, and all data sets start and end with a secondary eclipse (with the exception of WASP-18b which started mid-transit and ended shortly after a second transit). Information about the exposure times and other observing parameters of each previously published data set can be found in their respective papers referenced above. For both \mbox{KELT-16b} and \mbox{MASCARA-1b} we used a 2~s exposure time which resulted in 835 datacubes (53\,440 frames) and 1664 datacubes (106\,496 frames), respectively.

\section{Photometry and Data Reduction}\label{sec:photometry}

\texttt{SPCA} starts by unzipping the zip files for each phase curve downloaded from the \textit{Spitzer} Heritage Archive\footnote{\url{https://sha.ipac.caltech.edu/applications/Spitzer/SHA/}}, and then loads all of the files for one phase curve into RAM. For the subarray data sets, we perform an initial $4\sigma$ clipping and masking of each pixel along the time axis for each datacube to remove any artifacts like cosmic ray hits. Any frames where a masked pixel lies within the $5\times5$ pixel grid centered on the target star are masked entirely. For the full-frame photometry data set (WASP-103b), we extract just the \mbox{32 $\times$ 32} pixel stamp used in subarray mode: indices (9:40, 217:248). While \texttt{SPCA} allows oversampling the frames using bi-linear interpolation as is sometimes used in the literature \citep[e.g.,][]{stevenson2017}, we do not use the functionality in this work.

For the subarray data, we identify any subframes in which the aperture flux deviates by more than 4$\sigma$ from the median of the datacube after having performed a median average along the entire time axis. We then tried our photometry routines with and without these consistently bad frames and ultimately choose the photometry with the lowest scatter after being smoothed with a high-pass filter to remove any astrophysical signals. Our high-pass filter had a width of $5\times64$ data points (5 data cubes) for sub-array data or 64 data points for full-frame data. These timescales were selected to be shorter than the ingress/egress timescale which was greater than $5\times64$ frames for all sub-array data and greater than 64 frames for WASP-103b.

In order to compute photon noise limits, we convert all our data sets to electron counts using ${\rm Image}\times{\rm gain}\times\tau_{\rm exp}/{\rm FLUXCONV}$. This is an approximation of the photon limit, the full calculation of which is laid out in Section 3.3 of \citet{ingalls2016}. We then $5\sigma$ clip and mask each pixel along the entire time axis to remove any remaining artifacts. Any frames where a masked pixel lies within the $5\times5$ pixel grid centered on the target star are masked entirely. Finally, we subtract the background computed for each frame using the median of the frame's pixels, excluding a box (indices (11:19, 11:19)) around the target star. \texttt{SPCA} then performs its various photometry techniques, described in detail below. We then bin all of the sub-array mode data sets by datacube (64 frames) to reduce the computational cost of fitting the data with our many different decorrelation models, but we also save the unbinned data which we later use to test our decorrelation models. For the WASP-103b observations taken in full-frame mode, we chose not to temporally bin the data since the integration time was already much longer than the sub-array mode (12~s compared to 0.1--2~s).

\subsection{Aperture Photometry}\label{sec:aperture}
\texttt{SPCA}'s aperture photometry routine uses a flux-weighted mean (FWM) centroiding algorithm on the central $5\times5$ pixels:
\begin{equation*}
    {\rm x_{\rm cent}}= \frac{\sum_{i=0}^5\sum_{j=0}^5 i~\mathbf{I}_{i,j} }{\sum_{i=0}^5\sum_{j=0}^5 \mathbf{I}_{i,j}},
\end{equation*}
where $x_{cent}$ is the x-centroid in the 2D image, $\mathbf{I}$, and $i$ and $j$ are the x and y indices of each pixel. The similar equation for the y-centroid simply multiplies $\mathbf{I}$ by $j$ instead of $i$. The PSF width along each axis is also approximated using
\begin{equation*}
    {\rm \sigma_x}= \frac{\sum_{i=0}^5\sum_{j=0}^5 i^2~\mathbf{I}_{i,j} }{\sum_{i=0}^5\sum_{j=0}^5 \mathbf{I}_{i,j}},
\end{equation*}
where the equation for the PSF-width along the y-axis replaces $i^2$ with $j^2$.

Centroid and PSF widths are then put through a cleaning algorithm where the data are first 10$\sigma$ clipped. Any clipped data are then replaced by the median of the two preceding and two following data points. Subsequently, a copy of the data is smoothed using a high-pass filter with a width of $5\times64$ data points for sub-array data or 64 data points for full-frame data, any 5$\sigma$ outliers are identified, and the original data point is replaced by the median of the two preceding and two following data points. This data cleaning algorithm was inspired by that of \citet{zellem2014}.

\texttt{SPCA} makes accessible any \texttt{astropy} aperture, but little support is provided for non-circular apertures. For each of our data sets, we considered circular apertures with radii from 2.0 to 6.0 pixels in steps of 0.2 pixels, each of which was attempted with two types of aperture edges (hard, where a pixel is only included if its centre lies with the aperture, or exact, where a pixel is weighted by the fraction of the pixel which lies within the aperture). \texttt{SPCA} allows the aperture to either remain at a fixed location on the detector or to follow the centroid position, but initial tests suggested that having the aperture track the centroid gave cleaner photometry. The fluxes from all of these apertures were then subjected to the same cleaning algorithm as the centroid positions. Finally, \texttt{SPCA} chooses an aperture photometry technique by smoothing a copy of the fluxes with a high-pass filter with a width of $5\times64$ data points for sub-array data or 64 data points for full-frame data to remove transit, eclipse, and phase variation signals, and then selects the photometry with the lowest scatter under the premise that the data with the lowest high frequency noise will be the easiest to model cleanly. While this method is not guaranteed to give the cleanest possible photometry, it is more computationally efficient than trying all of our numerous detector models on each of the different photometry outputs. Moreover, previous comparisons \citep{bell2019,keating2020} have found that \texttt{SPCA}'s photometry routine gives qualitatively similar photometry to that from the Photometry for Orbits, Eclipses, and Transits (\texttt{POET}) pipeline \citep{stevenson2012a,cubillos2013}. \texttt{SPCA}'s algorithm also offers a potential improvement over the \texttt{POET} pipeline as we do not choose the photometry that best fits an assumed astrophysical model which could potentially bias the resulting phase curve parameters.

\begin{figure}
    \centering
    \includegraphics[width=\linewidth]{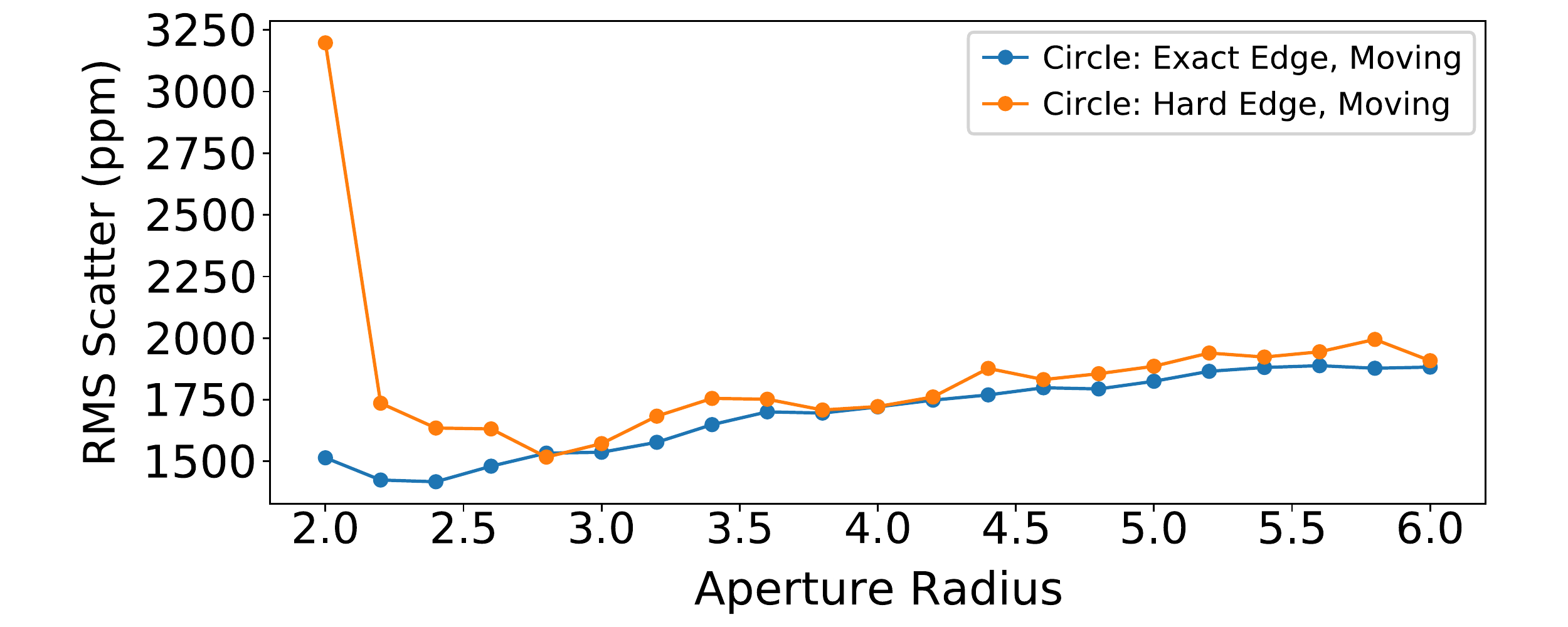}
    \includegraphics[width=\linewidth]{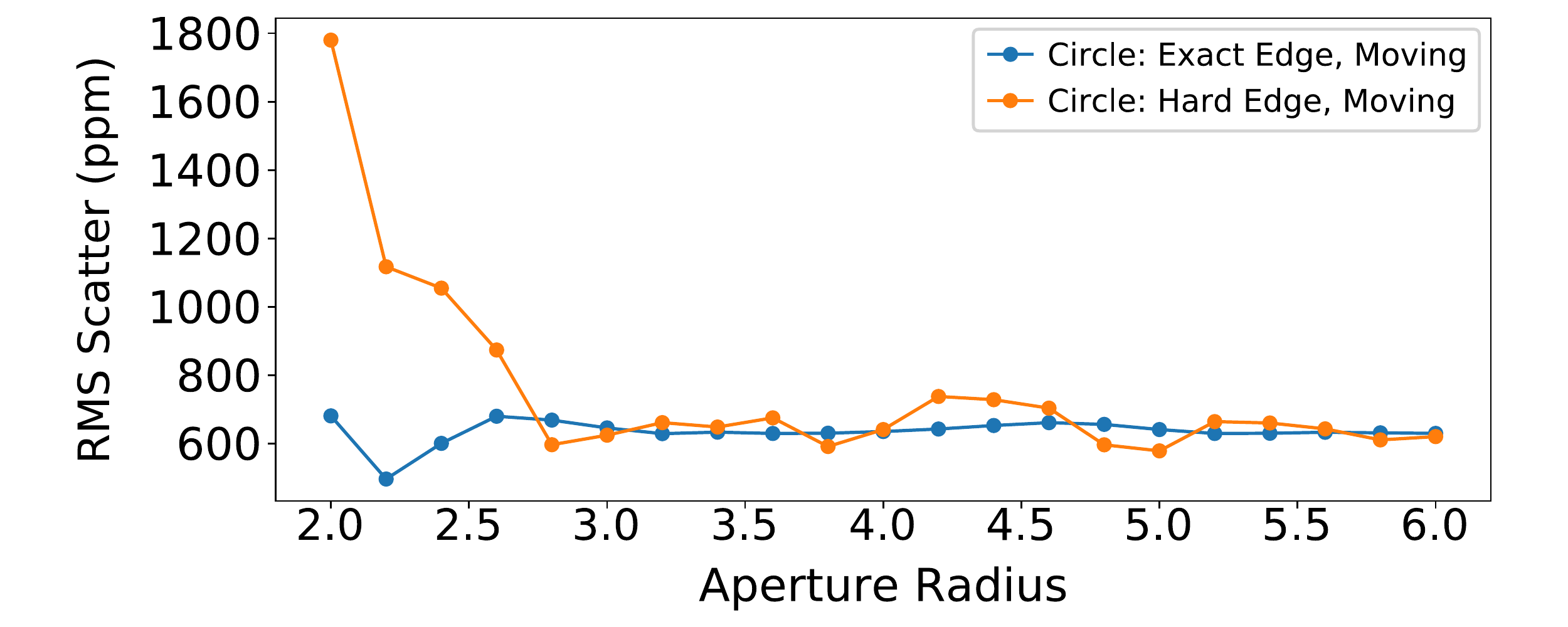}
    \caption{Comparison of the photometric root mean squared (RMS) after smoothing with a high-pass filter for the many apertures considered for \mbox{KELT-16b} (top) and \mbox{MASCARA-1b} (bottom). The aperture radius and edge combination that gives the lowest RMS after smoothing is considered to be our optimal aperture.}
    \label{fig:photometryComparison}
\end{figure}

\subsection{PSF Photometry}\label{sec:psf}
Our PSF photometry is initialized using the centroid and PSF-width algorithms described above, and then a 2D Gaussian is fitted to a $5\times5$ stamp centered at the pixel position (15,15) of each frame. The flux, centroid, and PSF width values are then cleaned using the same algorithm described above. As our PSF fitting fluxes are far noisier than the aperture fluxes, we only try using the centroids from this method to decorrelate the aperture photometry fluxes.

\subsection{PLD Photometry}\label{sec:pld_photo}
Our PLD photometry routine takes either a \mbox{$3\times3$} or \mbox{$5\times5$} stamp centered at the pixel position (15,15). Each pixel's lightcurve then undergoes the same cleaning routine described above. Additionally, we compute a total flux by summing the stamps and renormalize the sum of bad stamps using the same cleaning routine. When fitting observations with PLD, we ultimately use our aperture photometry as our flux measurement since it is much cleaner than the sum of the pixel stamps and then just use the individual pixel lightcurves as our covariates.

\section{Analyses}\label{sec:analysis}
\texttt{SPCA} models the photometry as a multiplicative combination of an astrophysical model and one or more detector models, each of which are described below. Except for the eclipse depth and phase curve coefficients, all astrophysical parameters are initialized to their best constrained values found on the NASA Exoplanet Archive\footnote{\url{https://exoplanetarchive.ipac.caltech.edu/}}. We set the initial eclipse depth to 3000~ppm which is typical of most of our phase curves. Finally, we set the initial phase curve semi-amplitude to 35\% of the eclipse depth and the phase offset to 0$^{\circ}$.

We chose to place a Gaussian prior on the linear ephemeris, $t_0$, the orbital period, $P$, the ratio of the semi-major axis to the stellar radius, $a/R_*$, and the orbital inclination, $i$, constraining them to the most precise values in the literature as these parameters are generally better constrained by the repeated transit observations used to discover these planets. We also constrain the orbital inclination, $i$, to be below 90$^{\circ}$. We place simple uniform priors constraining the planet-to-star radius ratio, $R_p/R_*$, the planet-to-star flux ratio, $F_p/F_*$, and the white noise amplitude normalized by the stellar flux, $\sigma_F$, to between 0 and 1 to ensure physicality.

After initializing our models, we begin with an initial stage of model optimization based on the method described by \citet{evans2015}. For all detector models except BLISS, we start by freezing the astrophysical parameters and perform an initial round of maximum likelihood estimation (MLE) on the detector models using \texttt{scipy.optimize.minimize}'s Nelder-Mead routine \citep{nelder1965} to ensure that our detector parameters begin in a reasonable location. We then run 10 rounds of optimization on all parameters, randomly drawing the starting position of all parameters within their uncertainty range or 10\% of the value where no uncertainty is known. We randomly draw starting phase curve semi-amplitudes between 0.2 and 0.5 and phase curve offsets between 10$^{\circ}$W and 30$^{\circ}$E. We then run 10 short MCMC chains containing 25\,500 samples using the \texttt{emcee.Ensemble\textunderscore Sampler} \citep{foremanMackey2013_emcee} initialized about the end points of the optimization runs to ensure that we are able to break free from any local minima; this proved to be very important and time-saving for our few model fits with the GP model. We then run 10 final rounds of optimization on all parameters, starting at the highest log-likelihood sample from each of the MCMC chains. The highest log-likelihood location found during this entire optimization routine is then used as the starting position for our MCMC marginalizations.

We start our MCMC with a dense, Gaussian ball about our maximum log-likelihood estimate, with a standard deviation of 0.01\% the parameter's value except for those parameters on which we have placed a Gaussian prior where we use the published uncertainty. We then run a 5000 step burn-in chain using \texttt{emcee.Ensemble\textunderscore Sampler} \citep{foremanMackey2013_emcee} with 150 walkers. Visual inspection of the tracks and distribution of MCMC walkers throughout this burn-in phase suggest that we had achieved convergence by the end of these chains. We then continue with a 1000 step production run with 150 walkers, providing us with a total of 150\,000 samples of the posterior. We use the maximum log-likelihood position from this chain as our fitted value, and use the 16th and 84th percentiles to compute our parameter uncertainties.

We name each of our model runs using a ``mode string'' to indicate the model choices that were made for that run. The mode string starts with a string describing the detector model used, followed by a description of the phase curve model, and potentially followed by ``\textunderscore PSFX'' when the PSF centroiding method used (when absent, FWM centroiding was used).

\subsection{Astrophysical Models}\label{sec:astro}

\texttt{SPCA}'s astrophysical model consists of a constant flux from the host star (except during transits), transit and eclipse signals modelled using \texttt{batman} \citep{kreidberg2015b}, and either a first order (single-peaked) or second order (double-peaked) sinusoidal phase variation. This can be written as
\begin{equation*}
	A(t) = F_*(t) + F_{\rm day}\Phi\big(\psi(t)\big),
\end{equation*}
where $F_*$ is the stellar flux, $F_{\rm day}$ is the planetary flux at a phase of 0.5, and $\Phi$ is our phase variation model which is a function of the orbital phase with respect to eclipse, $\psi(t)=2\pi(t-t_e)/P$, where $t_e$ is the time of eclipse and $P$ is the planet's orbital period. Our transit model assumes a reparameterized quadratic limb-darkening model (with parameters $q_1$ and $q_2$) to ensure efficient sampling and easy imposition of a physicality prior of $0 < \{~^{q_{1},}_{q_{2}}\} < 1$, following \citet{kipping2013}. We also fit for eccentricity using the parameters $e\cos{\omega}$ and $e\sin{\omega}$ to allow for efficient sampling and a simple prior of $-1 < \{~^{e\cos{\omega},}_{e\sin{\omega}}\} < 1$ \citep[e.g.,][]{butler2006}.

Our first order sinusoidal phase variation model is implemented as
\begin{equation*}
	\Phi_{\rm 1}(\psi) = 1 + C_1\bigg(\cos(\psi)-1\bigg) + D_1\sin(\psi),
\end{equation*}
and our second order phase variation model (permitting steeper day-night temperature transitions or ellipsoidal variations) is implemented as
\begin{equation*}
	\Phi_{\rm 2}(\psi) = \Phi_{\rm 1}(\psi) + C_2\bigg(\cos(2\psi)-1\bigg) + D_2\sin(2\psi),
\end{equation*}
where $C_1$, $C_2$, $D_1$, $D_2$ are fitted parameters. We add an appendix of ``\textunderscore v1'' to our mode string for first order phase variation models and ``\textunderscore v2'' for second order models. For our first order models, we compute the phase curve semi-amplitude using $\sqrt{C_1^2+D_1^2}$ and compute the phase offset in degrees east using $-{\arctan}2(D_1, C_1)$. For our second order models, we numerically compute the phase curve semi-amplitude and phase offset. When fitting, we require that the first order phase offset lie between -90$^{\circ}$ and 90$^{\circ}$. To ensure that our light curves remain physical, we also require that $\Phi\big(\psi(t)\big)$ be greater than zero for all phases; we do not require the physicality of an inferred temperature map.

\subsubsection*{Dilution Correction}\label{sec:dilution}
Three of the systems that we consider in this work (CoRoT-2b, WASP-12b, and WASP-103b) are host to a nearby star which acts to dilute the amplitude of the transit depth, eclipse depth, phase curve amplitude, and $\sigma_F$. CoRoT-2B, the stellar companion to the planet hosting star CoRoT-2A, is a K9 star with an effective temperature of 4000 K \citep{schroter2011} which is separated by 4.087$\arcsec$ at a position angle of 208.5$^{\circ}$ \citep{vizier2018}. \mbox{WASP-103B} is a K5V star with $T_{\rm eff}=4400$~K located 0.240$\arcsec$ away at a position angle of 208.5$^{\circ}$ \citep{cartier2017}. Finally, WASP-12A has two nearby M-dwarfs, WASP-12B,C, that are 1.06$\arcsec$ away at a position angle of 249.05$^{\circ}$ \citep{bergfors2011,crossfield2012,bechter2014} which have an effective temperature of $3660$~K \citep{stevenson2014b}.

We correct for the dilution from these nearby companions following a procedure similar to that described by \citet{stevenson2014b} and \citet{bell2019}. We start by making 10$\times$ oversampled simulated observations of the companion stars using the \texttt{STINYTIM}\footnote{\url{http://irsa.ipac.caltech.edu/data/SPITZER/docs/dataanalysistools/tools/contributed/general/stinytim/}} point response function modelling software for \textit{Spitzer}. We place the companion stars at the center of the subarray (24,232), and use the companion stars' blackbody temperatures described above. We use apertures that match the radius, $\beta$, of the selected aperture photometry for the three phase curves, and we place the apertures at the location where the host star would be to compute the fraction of the star's flux that falls within our aperture, $g(\beta)$. To compute the companion-to-host stellar flux ratio, $\alpha_{\rm comp}(\lambda)$, we integrate matching PHOENIX stellar models over the IRAC channel 2 bandpass using a uniform weighting. For CoRoT-2B, we assume $R_{\rm *} = 0.65 R_{\odot}$, $\log g = 4.28$, and $[Fe/H] = -0.17$ which are the parameters of of the star HD 113538 of the same spectral type \citep{moutou2011}. For the M-dwarf companions WASP-12B,C, we assume both stars have radii of $R_{\rm *} = 0.65 R_{\odot}$ \citep{stevenson2014a} and median M-dwarf values of $\log g = 5.0$ and $[Fe/H] = 0$ \citep{rajpurohit2018}.
For WASP-103B, we assume the star has the same parameters as 61 Cygni A which has the same spectral type: $R_{\rm *} = 0.665 R_{\odot}$, $\log g = 4.40$, and $[Fe/H] = -0.20$ \citep{kervella2008}. The dilution correction parameters are summarized in Table \ref{tab:dilution}. Our computed dilution parameters are generally consistent with those used by \citet{stevenson2014b}, \citet{bell2019}, and \citet{garhart2020} for WASP-12b and \citet{garhart2020} for WASP-103b, with only minor differences likely caused by different photometric aperture sizes.

We then correct the planet's radius using
$$
	\bigg(\frac{R_{\rm p}}{R_{\rm *}}(\lambda)\bigg)_{\rm corr} = \sqrt{C_{\rm corr}(\beta, \lambda)} \, \bigg(\frac{R_{\rm p}}{R_{\rm *}}(\lambda)\bigg)_{\rm meas},
$$
and the dayside flux was corrected using
$$
	\bigg(\frac{F_{\rm p}}{F_{\rm *}}(\lambda)\bigg)_{\rm corr} = C_{\rm corr}(\beta, \lambda) \, \bigg(\frac{F_{\rm p}}{F_{\rm *}}(\lambda)\bigg)_{\rm meas},
$$
with the white noise amplitude, $\sigma_{F}$, corrected similarly to the dayside flux. The correction factor is computed using
$$
	C_{\rm corr}(\beta, \lambda) = 1+g(\beta)\alpha_{\rm comp}(\lambda).
$$

\begin{table}
    \centering
    \begin{tabular}{l|c|c|c|c}
         System & $\beta^{\rm a}$ & $g^{\rm b}$ & ${\alpha_{\rm comp}}^{\rm c}$ & ${C_{\rm corr}}^{\rm d}$\\  \hline
         CoRoT-2        & 3.6 & 0.5921 & 0.3455 & 1.2046  \\ \hline
         WASP-12 (2010) & 2.2 & 0.7827 & 0.1161 & 1.09085 \\ \hline
         WASP-12 (2013) & 3.2 & 0.8593 & 0.1161 & 1.09976 \\ \hline
         WASP-103       & 2.6 & 0.8456 & 0.1460 & 1.1234  \\
    \end{tabular}
    \caption{Companion dilution correction parameters.\protect\\%
    $^{\rm a}$Aperture radius.\protect\\%
    $^{\rm b}$Fraction of companion's flux that falls within our aperture.\protect\\%
    $^{\rm c}$The companion-to-host stellar flux ratio.\protect\\%
    $^{\rm d}$The applied dilution correction factor.}
    \label{tab:dilution}
\end{table}

\subsection{Detector Models}\label{sec:detector}
\texttt{SPCA} currently has four of the most common decorrelation models used on \textit{Spitzer} IRAC 3.6~$\mu$m and 4.5~$\mu$m phase curves: 2D polynomials, BLISS mapping, a GP, and PLD. Each of these models and the decisions we made while implementing them are described in more detail below. While we have mostly followed the procedures laid out in the literature, we did make some judgement calls of our own where information was missing or unclear; as a result, the performance of our detector models may slightly differ from that of other pipelines.

It is possible to multiply our 2D polynomial, BLISS, and GP model with a simple linear model that depends on the PSF width. One can also add a linear slope in time to any detector model to capture long timescale stellar variability. There is also the possibility to add step functions at any of the Astronomical Observation Request (AOR) breaks where the telescope is re-pointed. Ultimately we decided not to consider the PSF width or any explicit function of time in this work to reduce the already very high computational cost of fitting all 17 phase curves with two different centroiding options each (FWM and PSF fitted), two different phase curve models, and 9--10 detector models, giving a total of 488 fits that take upwards of 20 wall-clock minutes each while using 12 parallel CPU threads. For a detailed look into the effects of PSF width and shape on \textit{Spitzer}/IRAC photometry and potential methods to decorrelate them, see \citet{challener2021} which was published after this work was submitted for review.

\subsubsection{2D Polynomials}\label{sec:poly}
The 2D polynomials are parametric models which use the centroid positions as its covariates and was first used for \textit{Spitzer}/IRAC data by \citet{charbonneau2008}. \texttt{SPCA} permits second- to fifth-order polynomial models, including all cross terms, and we try all four variants for all of our considered data sets. These models will be called ``Poly\#'' in the mode string where \# is the order of the polynomial model.

\subsubsection{Pixel Level Decorrelation}\label{sec:pld}
PLD is a parametric model which uses normalized lightcurves for each individual pixel as its covariates and is described in detail by \citet{deming2015}. While \texttt{SPCA} can use the sum of the PLD stamps as the raw flux which is decorrelated, we found that the flux from aperture photometry was less noisy to begin with and produced cleaner phase curves after decorrelation. \texttt{SPCA} has two different option pairs that can be selected which gives a total of four variants. One can choose either first order PLD where the individual pixel light curves are the covariates, or one can use second order PLD \citep{zhang2018} which also includes the square of each pixel light curve. Following \citet{zhang2018}, we do not include any cross-terms in our second order PLD model. \citet{deming2015} found that first order PLD performed best on eclipse observations when the centroid variation was less than 0.2 pixels. While second order PLD should help extend the applicability of PLD to larger centroid drifts, this has not been quantified. The other option is to use $3\times3$ or $5\times5$ pixel stamps to allow for a trade-off between capturing more stellar flux and capturing more background flux. For easier initialization of our detector models while fitting the data, we also put our pixel light curves (and their squared values where relevant) through a PCA algorithm and add a constant offset term. These models will be called ``PLD\#\textunderscore NxN'' in the mode string where \# is the order of the PLD model, and NxN is the size of the pixel stamps.

\subsubsection{BLISS Mapping}\label{sec:bliss}
BLISS mapping is a non-parametric model that uses the centroid positions as its covariates and is described in detail by \citet{stevenson2012a}. There is, however, a hyperparameter: the number of ``knots'' (x,y grid cells) used by the BLISS algorithm, which can be challenging to choose properly. With too few knots, the model becomes overly simple and results in discrepant retrieved astrophysical parameters, while too many knots can begin to over-fit the data, and in the limit you would have a knot for every single data point \citep{stevenson2012a}. We developed a routine similar to that described in \citet{stevenson2012a}, where we compare the performance of a nearest-neighbour interpolation (NNI) algorithm against the BLISS algorithm. We first fix the number of knots to an $8\times8$ square grid and perform our ten \texttt{scipy.optimize.minimize} fits to optimize the astrophysical parameters. We then consider several different knot spacings in the range of 0.01--0.06 pixels per knot (with the same scale in x and y), based on the findings of \citet{stevenson2012a} for the secondary eclipse observations of HD~149026b. We run a single optimization routine with each of these knot spacings, and then pick the least dense spacing where the fitted $\sigma_F$ with the BLISS algorithm is lower than that for NNI. However, for some data sets we find that BLISS outperforms NNI for all considered grid spacings, in which case we pick the most dense grid spacing that results in fewer than 50 utilized knots (limiting the model to be no more complex than our Poly5 models); in cases where there is a large spread in centroid position, the least dense BLISS model may still have more than 50 utilized knots, in which case we just select the least dense grid spacing. We then continue with the rest of our initial optimization routine (short MCMC runs and another round of MLE fits). These models will simply be called ``BLISS'' in the mode string.

\subsubsection{Gaussian Processes}\label{sec:gp}
The GP model we use is a non-parametric model that uses the centroid positions as its covariates and is based on \citet{gibson2012a} and \citet{evans2015}. We used the python package \texttt{george} \citep{foremanMackey2015_george} with the \texttt{BasicSolver} to implement the GP. We use a squared-exponential kernel with additive white noise in the form
\begin{equation*}
    \Sigma_{nm} = C^2\exp\Bigg(-\frac{(x_n-x_m)^2}{L_x^2} - \frac{(y_n-y_m)^2}{L_y^2} \Bigg)+\delta_{nm}\sigma_F^2,
\end{equation*}
where ($x_n$,$y_n$) is the centroid position of the $n$th datum (and similarly for the $m$th datum), $C$ is used to compute the covariance amplitude, $L_x$ and $L_y$ are the covariance lengthscales in the $x$ and $y$ directions, $\delta_{nm}$ is the Kronecker delta, and $\sigma_F^2$ is the aforementioned white noise amplitude normalized by the stellar flux. The choice of a squared-exponential kernel stems from the assumption that the detector sensitivity is a smooth function of the centroid position. This is similar in many ways to the Gaussian kernel regression methods used by \citet{ballard2010}, \citet{knutson2012}, and \citet{lewis2013}, but a GP is a more statistically robust, albeit computationally intensive, method. We chose not to include an additional Mat\'ern $\nu$ = 3/2 kernel as a function of time as was included by \citet{evans2015}.

We follow \citet{evans2015} by placing a uniform prior on the natural logarithm of the GP lengthscales to ensure that the GP does not over-fit the data and is fitting for intra-pixel sensitivity variations rather than larger lengthscales; we chose limits of -3 and 0 as \citet{evans2015} did not publish their limits. We also follow \citet{evans2015} in placing a Gamma prior on $C$ of the form $p(C) = {\rm Gam}(1, 100)$. During our initial 10 burn-ins, we randomly drew values of $\sqrt{C}$, $\ln L_x$, and $\ln L_y$ in the ranges (0.05,0.135), (-0.5,-1), and (-0.5,-1), respectively. As the GP model is exceptionally computationally expensive, we chose to reduce the number of burn-in steps in our MCMC runs to 1000; we confirmed that the MCMC had converged after this number of steps by visually examining the trace of the walkers afterwards. Even still, this required $\sim$100 CPU hours for each of the fits to MASCARA-1b's phase curve and $\sim$25 CPU hours for each the fits to KELT-16b's. As a result of this extremely high computational cost, we attempted to perform GP analyses on only HAT-P-7b from the previously published phase curves as our results for its phase curve appeared to be very strongly model dependent. These models will simply be called ``GP'' in the mode string.

\section{Validation Against XO-3\lowercase{b} Eclipses}\label{sec:eclipseAnalysis}
To test our photometry and decorrelation techniques, we first considered the 10 secondary eclipses of the eccentric planet XO-3b collected using IRAC channel 2 (PID 90032) which were first published by \citet{wong2014} and later extensively studied with many standard decorrelation techniques by \citet{ingalls2016}. We performed photometry on these data following the exact same methods as for the phase curve data, and treated each eclipse observation entirely independently. When fitting the data sets with our model suite, we chose to impose the following Gaussian priors in addition to all of the priors imposed on the phase curve data since these parameters were fairly poorly constrained by eclipse-only observations: $R_p = 0.08825\pm 0.00037$, $e\cos\omega = 0.2700 \pm 0.0024$, and $e\sin\omega = -0.0613\pm 0.0078$ \citep{wong2014}. We still fit for the phase variations to ensure that our models remain unbiased due to the downward curvature of the phase curve near eclipse, but these parameters are primarily constrained by our physicality priors. We also only considered a first order sinusoidal phase curve model since the phase variations were poorly constrained by the out-of-eclipse baseline.

\begin{figure*}
    \centering
    \includegraphics[width=\linewidth]{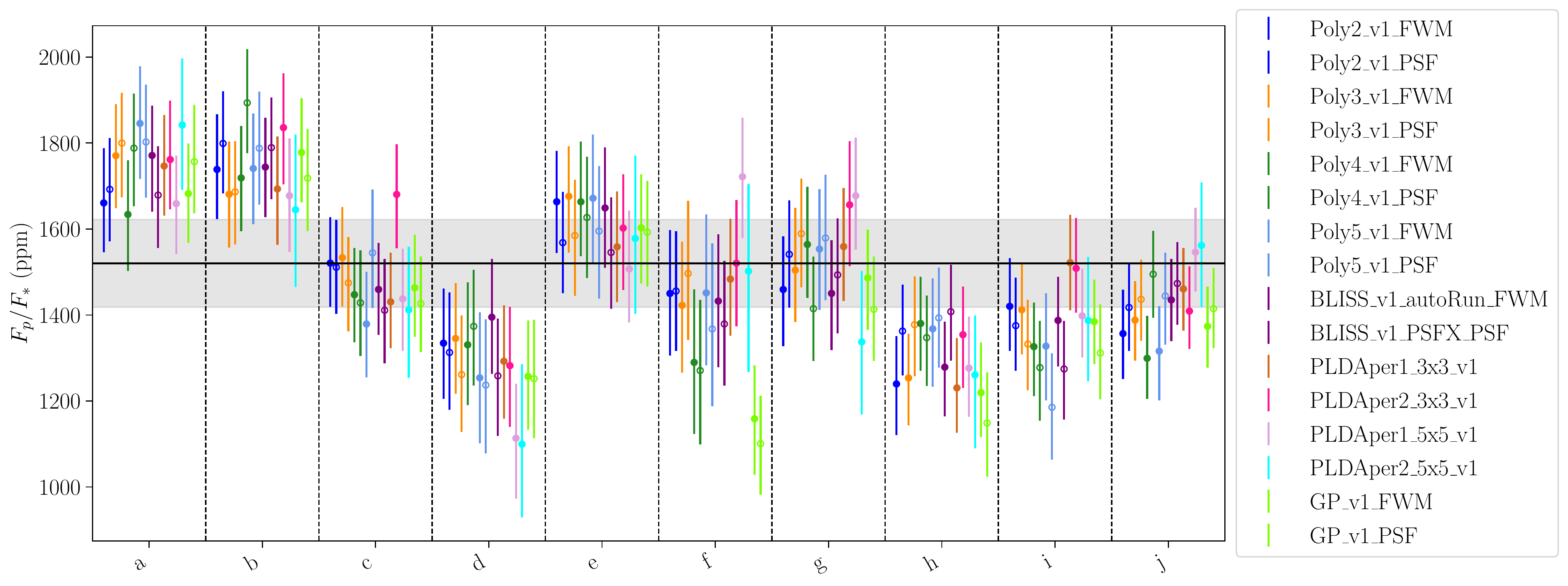}
    \caption{Eclipse depths for each of the 10 eclipse observations of XO-3b (named a--j), each independently analyzed with all of our detector models. The black line and shaded region show the median eclipse depth and median uncertainty on eclipse depth from \citet{ingalls2016}.}
    \label{fig:XO3b_fday}
\end{figure*}

\begin{figure*}
    \centering
    \includegraphics[width=0.7\linewidth]{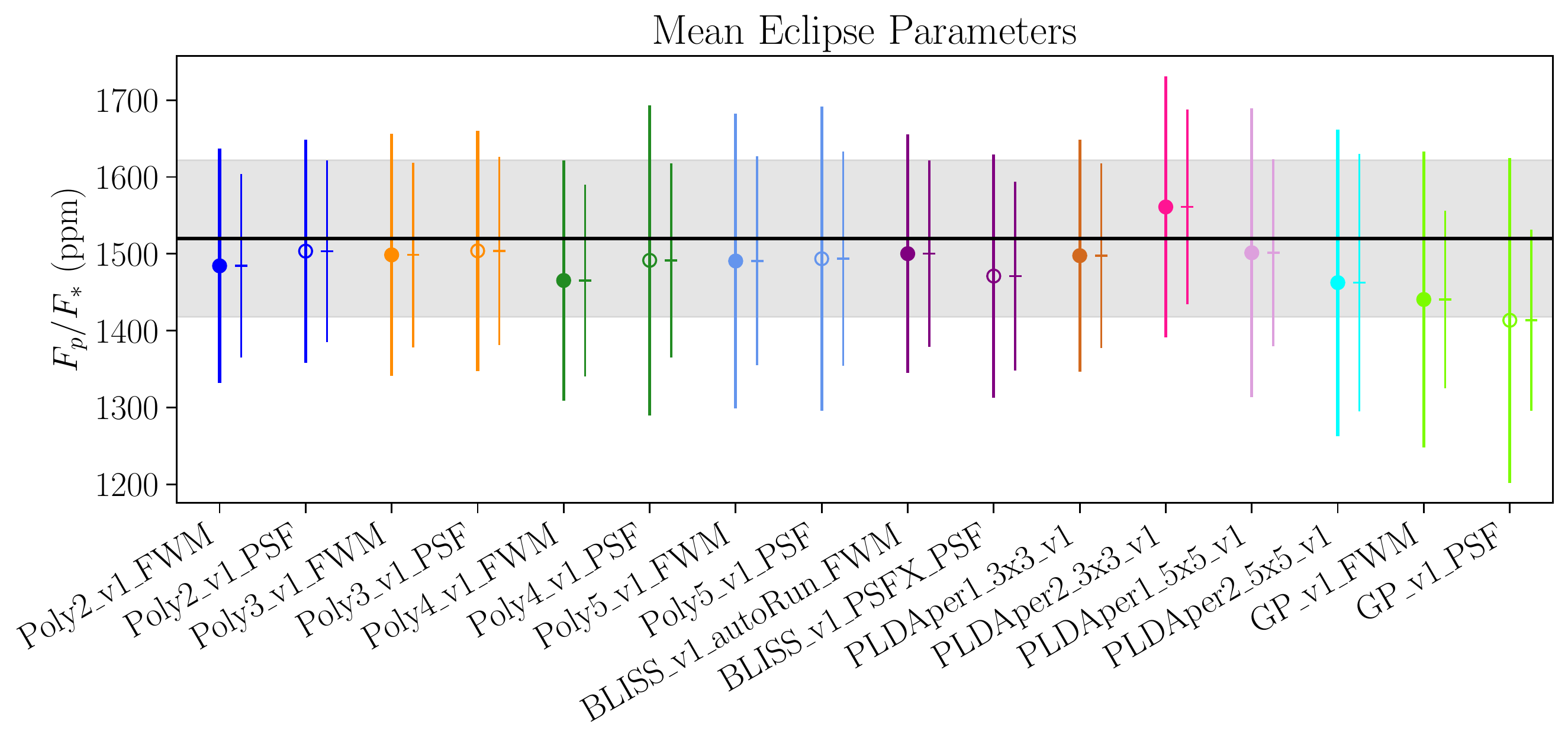}
    \caption{Mean and standard deviation of the 4.5~$\mu$m XO-3b eclipse depths fitted with each detector model are shown with solid points for FWM centroiding and hollow points for PSF centroiding. Meanwhile, there is another error bar with a horizontal line adjacent to each point which indicates the average uncertainty found from each independent eclipse fit. The black line and shaded region show the median eclipse depth and median uncertainty on eclipse depth from \citet{ingalls2016}. While there is a slight underestimation of the uncertainty on the eclipse depth with all models (the standard deviation in eclipse depths is larger that the reported uncertainty from individual eclipse observations), there is no clear bias in any of the models.}
    \label{fig:XO3b_fday_diff}
\end{figure*}

The retrieved eclipse depths for each of the 10 eclipses analyzed with all of our 16 detector models are plotted in Figure \ref{fig:XO3b_fday}. While there is slight variance between models and between the median model for each eclipse, no clear or consistent bias is evident. This is summarized in Figure \ref{fig:XO3b_fday_diff} which shows the mean and standard deviation of each models' fitted values for the 10 eclipses. Figure \ref{fig:XO3b_fday_diff} also shows that, while there is a slight tendency to underestimate our uncertainty on the eclipse depth, the correction factor is close to unity which is consistent with the findings of \citet{ingalls2016} and not 3 as had been suggested by \citet{hansen2014} for early \textit{Spitzer} eclipse observations. It is possible that this increased scatter between eclipse depths could be the result of astrophysical variations, but this would imply eclipse depth variability roughly twice as large as the maximum predicted level for hot Jupiters on circular orbits (\mbox{53\,ppm = 3.6\%} vs the $\leq$2\% predicted by \citet{komacek2020}; we note, however, that XO-3b is on a significantly eccentric orbit). Each of our models' average eclipse depth is consistent with the median eclipse depth from \citet{ingalls2016}, but our fitted uncertainties are all slightly larger than their median uncertainty. We also find no clear difference between decorrelating with PSF centroiding and FWM centroiding, but our aperture photometry was exclusively performed using FWM centroiding, so it is possible that aperture photometry performed using PSF centroiding would be better decorrelated with the PSF centroids.

We also computed the various statistics presented in \citet{ingalls2016} to quantitatively assess the performance of each of our models. Specifically, we computed the error-weighted average eclipse depth, $\overline{D}$, the average eclipse depth uncertainty, $\overline{\sigma}$, the standard deviation in the eclipse depths from the 10 observations, SD, and the weighted uncertainty in the mean eclipse depth given the uncertainty from our MCMC, $\sigma_{\rm orig}$. The expected level of scatter between eclipse depths assuming only photon noise is $\sigma_{\rm phot} \approx 64~{\rm ppm}$, so all decorrelation methods get within ${\sim}3\times$ the photon limit (179 ppm). We then computed the ``dispersion factor'', $f_{\rm dis}$, that multiplies our uncertainties to account for the observed eclipse depth scatter between different eclipse observations, the total uncertainty in the average eclipse depth after inflating our error bars, $\sigma_{\rm TOT}$, the ``repeatability'' of our fits, $R$, the ``reliability'' of our fits, $r$. Finally, we compute the ``accuracy'', $a$, of our fits with respect to the average eclipse depth from \citet{ingalls2016} which we consider to be the true eclipse depth. For the definitions of ``repeatability'', ``reliability'', and ``accuracy'' in the context of these model fits and their correlations, see Section 3.4 of \citet{ingalls2016}. Intriguingly, our lowest order polynomial models and our simplest PLD model rank the best in terms of repeatability, reliability, and accuracy, although there isn't a large spread in the performances of each of the 18 different detector models. It is unclear whether the performance of each of these models would extend equally well to longer duration phase curve observations which can either more densely sample the detector sensitivity if the telescope drifts slowly or is repointed, or can substantially drift across the detector resulting in larger pointing variations and a poorly sampled sensitivity map. However, \citet{ingalls2016} suggest that BLISS is likely to perform best under situations with larger pointing variations.

\begin{table*}
    \centering
    \begin{tabular}{l|ccccccccc}
        Mode                    &  $\overline{D}^{\rm a}$ & $\overline{\sigma}^{\rm b}$ & SD$^{\rm c}$ & ${\sigma_{\rm orig}}^{\rm d}$ & ${f_{\rm dis}}^{\rm e}$ & ${\sigma_{\rm TOT}}^{\rm f}$ & $R^{\rm g}$ & $r^{\rm h}$ & $a^{\rm i}$   \\
                                &  (ppm)          & (ppm)              &(ppm)& (ppm)               &               & (ppm)              &(ppm)&     & \\ \hline
    % -------------------------------------------------------------------------\\
    Poly2        &   1480 & 120 & 153 & 37 & 1.4 & 51 & 216 & 0.42 & 0.41 \\
    Poly2\textunderscore PSFX   &   1500 & 118 & 145 & 37 & 1.3 & 48 & 205 & 0.44 & 0.44 \\
    Poly3        &   1485 & 120 & 158 & 37 & 1.4 & 53 & 223 & 0.41 & 0.40 \\
    Poly3\textunderscore PSFX   &   1496 & 123 & 156 & 38 & 1.4 & 51 & 221 & 0.41 & 0.41 \\
    Poly4        &   1455 & 125 & 157 & 38 & 1.3 & 51 & 221 & 0.41 & 0.39 \\
    Poly4\textunderscore PSFX   &   1490 & 126 & 202 & 39 & 1.7 & 65 & 285 & 0.32 & 0.31 \\
    Poly5        &   1483 & 136 & 192 & 42 & 1.5 & 65 & 271 & 0.33 & 0.33 \\
    Poly5\textunderscore PSFX   &   1494 & 139 & 198 & 43 & 1.6 & 66 & 280 & 0.32 & 0.32 \\
    BLISS        &   1490 & 121 & 155 & 37 & 1.4 & 51 & 219 & 0.41 & 0.41 \\
    BLISS\textunderscore PSFX   &   1476 & 123 & 158 & 38 & 1.4 & 52 & 224 & 0.40 & 0.39 \\
    PLDAper1\textunderscore 3x3 &   1492 & 120 & 151 & 37 & 1.3 & 50 & 213 & 0.42 & 0.42 \\
    PLDAper2\textunderscore 3x3 &   1549 & 127 & 170 & 39 & 1.4 & 56 & 240 & 0.38 & 0.37 \\
    PLDAper1\textunderscore 5x5 &   1496 & 122 & 188 & 38 & 1.5 & 58 & 266 & 0.34 & 0.34 \\
    PLDAper2\textunderscore 5x5 &   1465 & 168 & 200 & 52 & 1.3 & 67 & 282 & 0.32 & 0.31 \\
    GP           &   1435 & 116 & 193 & 36 & 1.7 & 61 & 272 & 0.33 & 0.31 \\
    GP\textunderscore PSFX      &   1411 & 118 & 211 & 37 & 1.8 & 68 & 299 & 0.30 & 0.27 \\\hline
    % -------------------------------------------------------------------------
    Average                 &   1481 & 126 & 174 & 39 & 1.5 & 57 & 246 & 0.37 & 0.36 \\
    \end{tabular}
    \caption{XO-3b eclipse depth repeatability statistics following \citet{ingalls2016}. The expected level of scatter between eclipse depths assuming only photon noise is $\sigma_{\rm phot} \approx 64~{\rm ppm}$.\protect\\%
    $^{\rm a}$The error-weighted average eclipse depth.\protect\\%
    $^{\rm b}$The average eclipse depth uncertainty.\protect\\%
    $^{\rm c}$The standard deviation in the eclipse depths.\protect\\%
    $^{\rm d}$The weighted uncertainty in the mean eclipse depth given the uncertainty from our MCMC.\protect\\%
    $^{\rm e}$The ``dispersion factor'' that multiplies our uncertainties to account for the observed eclipse depth scatter between different eclipse observations.\protect\\%
    $^{\rm f}$The total uncertainty in the average eclipse depth after inflating our error bars by $f_{\rm dis}$.\protect\\%
    $^{\rm g}$The ``repeatability'' of our fits.\protect\\%
    $^{\rm h}$The ``reliability'' of our fits.\protect\\%
    $^{\rm i}$The ``accuracy'' of our fits with respect to the average eclipse depth from \citet{ingalls2016}.}
    \label{tab:xo3stats}
\end{table*}

Our model fits indicate that no one model consistently produces lower scatter in the residuals for the 64$\times$ binned data that we fitted. We also compare our fitted models to the unbinned data, adjusting only $\sigma_F$ to give a $\chi^2/N_{\rm data}$ of 1. These values suggest that the lower order polynomial models and BLISS models outperform the higher order polynomial models and PLD models. For the higher order polynomial models, this may be indicative of the impact of centroiding uncertainty. For the PLD models, this may be the result of noisy pixel lightcurves that are better behaved in binned data \citep[e.g.][]{deming2015,zhang2018}. Overall though, \texttt{SPCA}'s photometry and decorrelation techniques perform well on this validation test, and no one model clearly outperformed any others on the binned data; this is consistent with the findings of \citet{ingalls2016}, where BLISS, GP, and PLD models performed quite similarly (they did not consider polynomial models).

\section{Results}\label{sec:results}
For each data set, we start by selecting the model with the lowest Bayesian Information Criterion (BIC), which we defined as:
\begin{equation*}
	{\rm BIC} = -2\ln(L) + N_{\rm par}\ln(N_{\rm dat}),
\end{equation*}
where $\ln(L)$ is the model log-likelihood, $N_{\rm par}$ is the number of fitted parameters, and $N_{\rm dat}$ is the number of fitted data. For our BLISS models, we consider $N_{\rm par}$ to be the number of BLISS knots which had one or more data points since \citet{schwartz2017a} showed that you can achieve the same results as BLISS by treating each knot as a fittable parameter in your MCMC. As is shown in Figure \ref{fig:BIC}, for each phase curve there is always one model which vastly out-performs all other models; this model is typically BLISS. For that reason, we do not use averages or weighted averages from our different model fits.

\subsection{KELT-16\lowercase{b} and MASCARA-1\lowercase{b}}

\begin{table*}
    \centering
    \bgroup
    \def\arraystretch{1.4}
    \begin{tabular}{l|ccccccc}
         Planet & $t_0$ (BJD) & $P$ (days) & $e\cos{\omega}$ & $e\sin{\omega}$ & $a/R_{\rm*}$ & $i$ (degrees) \\ \hline
         KELT-16b   & $2457247.24795^{+0.00018}_{-0.00019}$ & $0.96899225^{+0.00000047}_{-0.00000046}$ & $0.0016^{+0.0016}_{-0.0015}$ & $0.013^{+0.015}_{-0.016}$ & $3.171^{+0.082}_{-0.093}$ & $83.5^{+1.8}_{-1.6}$ \\
         MASCARA-1b & $2457097.2782^{+0.0018}_{-0.0020}$    & $2.1487760^{+0.0000029}_{-0.0000027}$ & $0.00041^{+0.00052}_{-0.00055}$ & $-0.0070^{+0.0059}_{-0.0050}$ & $4.08^{+0.12}_{-0.15}$ & $85.7^{+1.9}_{-1.7}$\\
    \end{tabular}
    \egroup
    \caption{Updated orbital parameters from the new phase curves of KELT-16b and MASCARA-1b.}
    \label{tab:orbit}
\end{table*}

\begin{figure*}
    \centering
    \begin{minipage}{0.5\textwidth}
        \centering
        \Large
        ~~~~~KELT-16b
    \end{minipage}%
    \begin{minipage}{0.5\textwidth}
        \centering
        \Large
        ~~~~~~~~~~~~~MASCARA-1b
    \end{minipage}
    \includegraphics[width=0.482\textwidth]{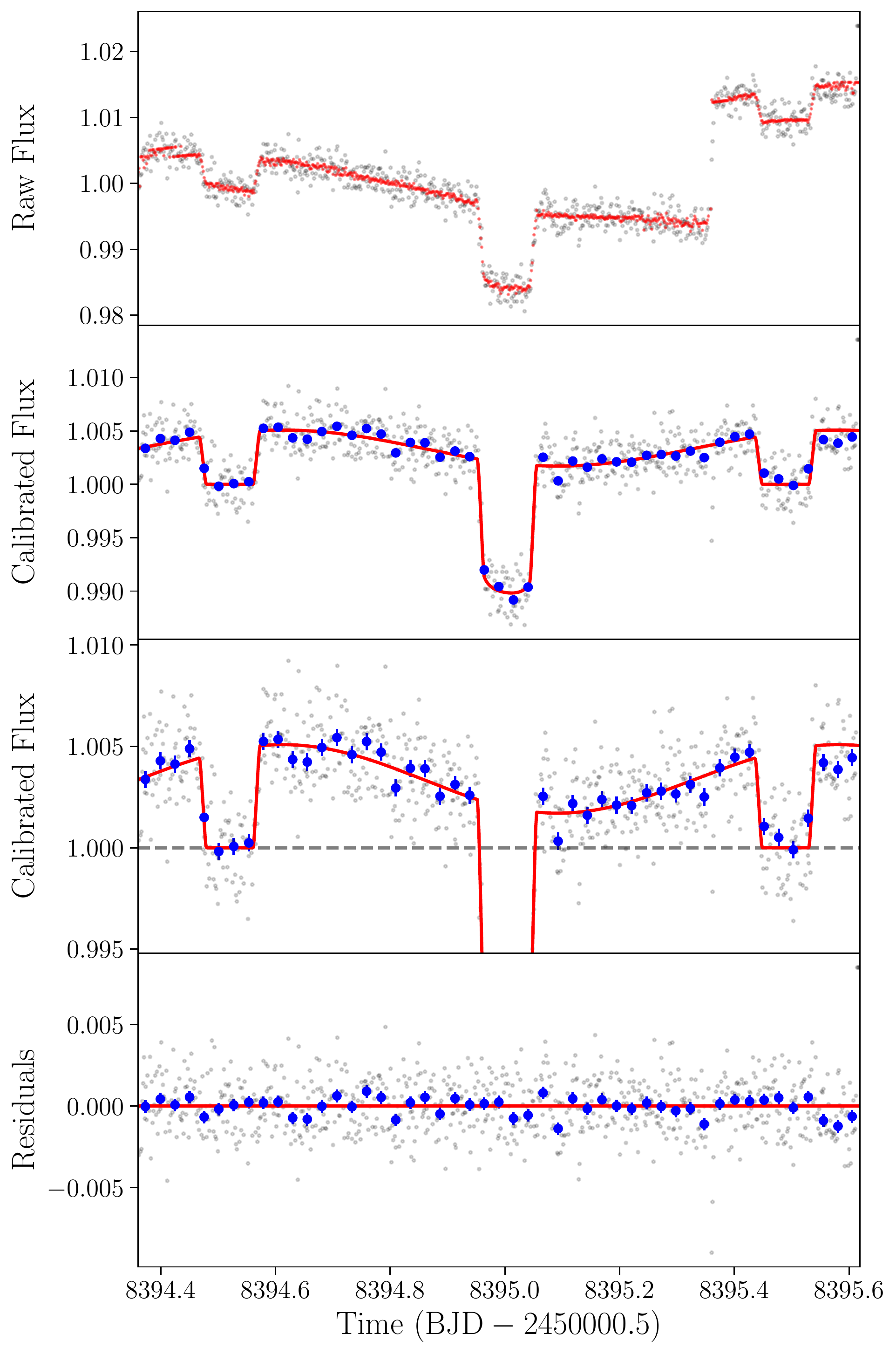}\hfill%
    \includegraphics[width=0.47\textwidth]{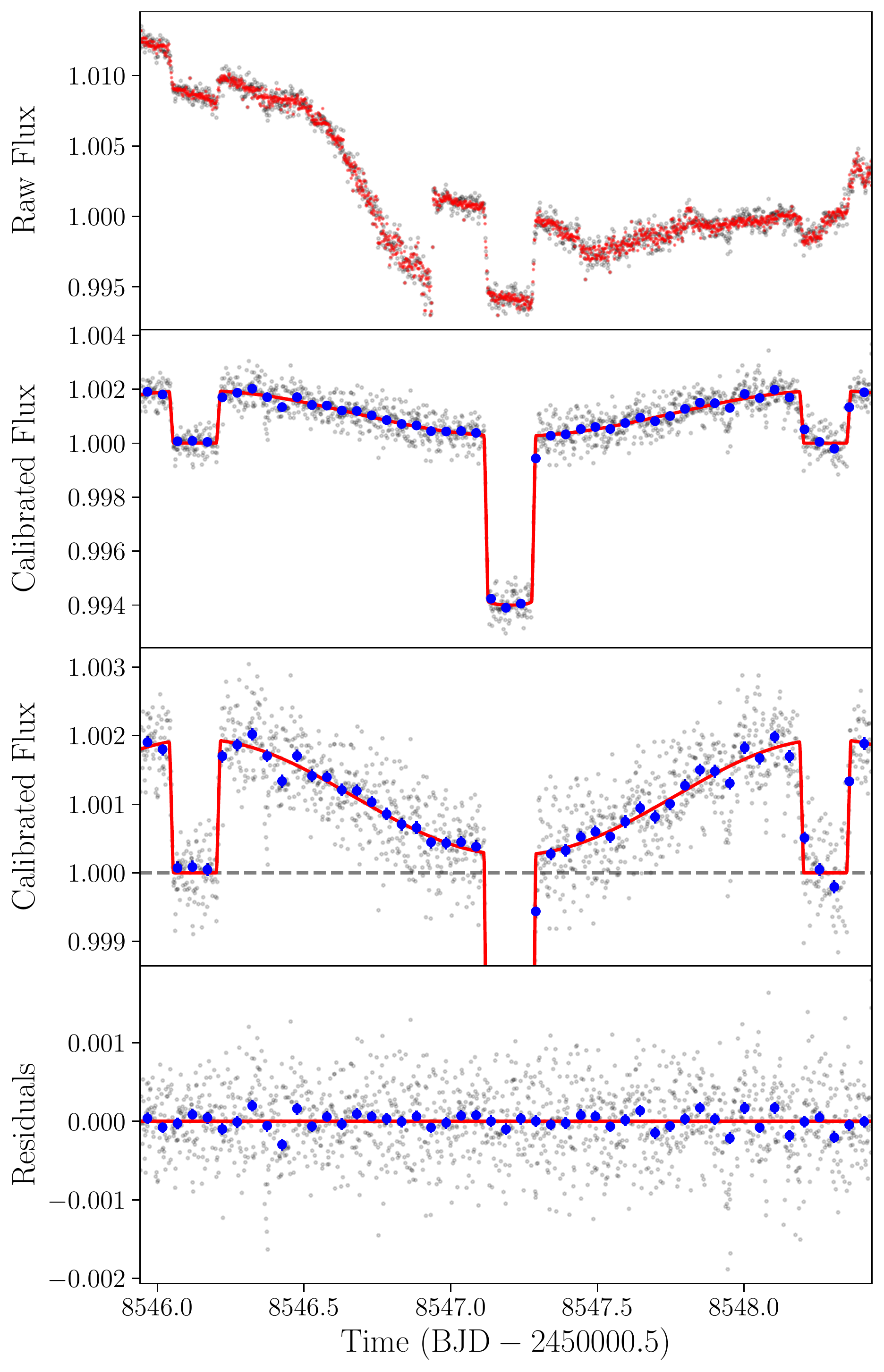}
    \includegraphics[width=0.47\textwidth]{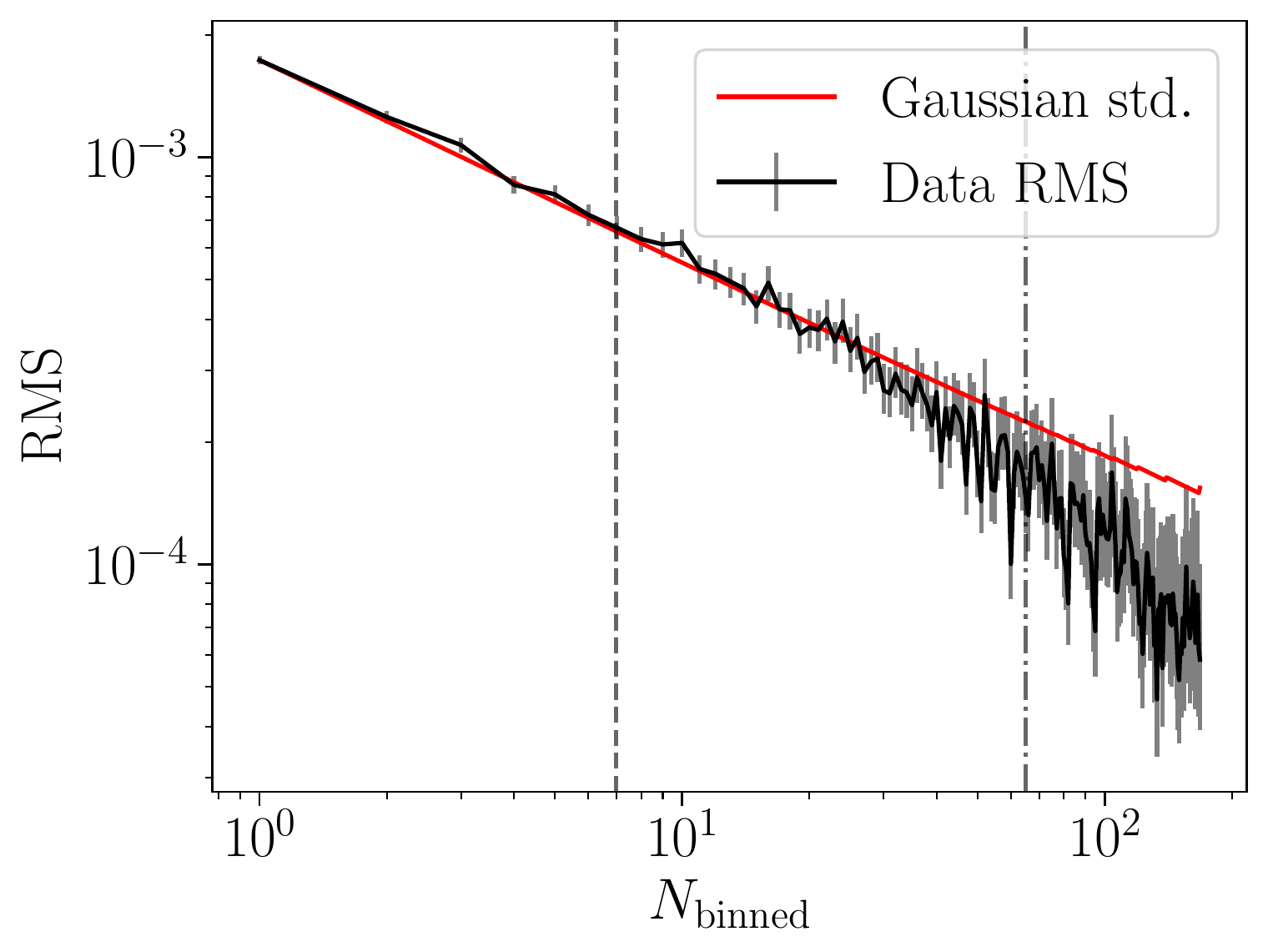}\hfill%
    \includegraphics[width=0.47\textwidth]{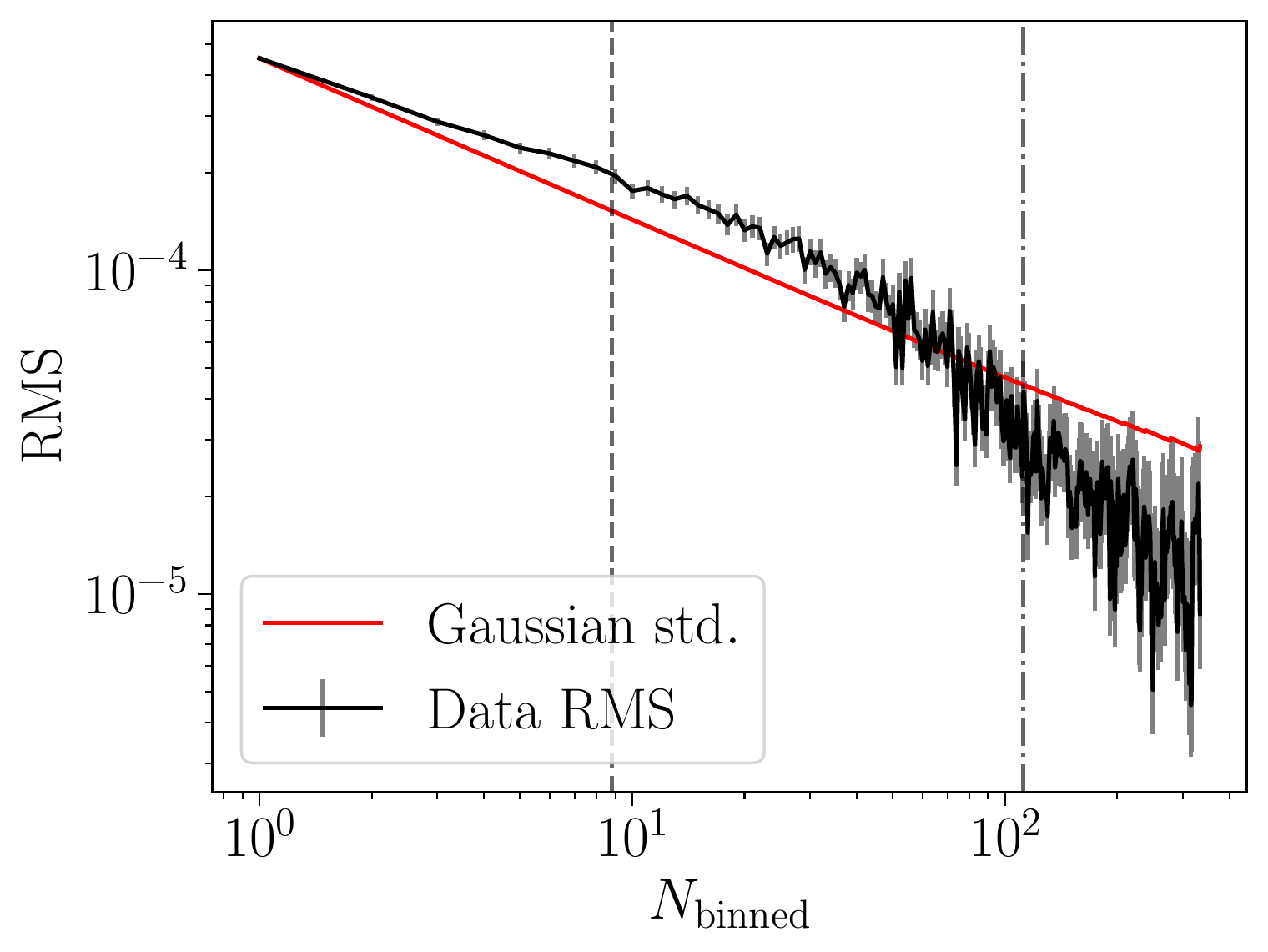}
    \caption{\textit{Top:} Preferred model fits (BLISS\textunderscore v1 using FWM centroiding) for KELT-16b and MASCARA-1b. The top four panels show raw photometry, photometry after correcting for detector systematics, a zoom-in on the calibrated data to show the phase variations, and the residuals from the fit. Vertical dashed lines indicate breaks between AORs, grey points are the 64$\times$ binned data which were fitted, blue points are further binned to 50 points per phase curve to show lower frequency noise levels, and the red lines indicate the best-fit model. \textit{Bottom:} Red noise tests for the above fits, showing the decrease in the RMS of the residuals as the number of datapoints binned together ($N_{\rm binned}$) increases, starting from our 64$\times$ binning. The red lines show the expected decrease in RMS assuming white noise. The timescale for transit/eclipse ingress and egress is indicated with a vertical, dashed line, while the full \mbox{$t_1$--$t_4$} transit duration is shown with a vertical, dash-dotted line.}
    \label{fig:bestfit_pcs}
\end{figure*}

For both of our newly observed and analyzed phase curves of KELT-16b and MASCARA-1b, the BLISS model with a first order sinusoidal phase curve and using FWM centroiding was the preferred mode; these fitted models are plotted in Figure \ref{fig:bestfit_pcs}. The fitted parameters for all considered models are available as \texttt{numpy} zip files in the Supplementary Data, and parameters of interest for the preferred phase curves are presented in Table \ref{tab:preferred}. We also find updated orbital parameters for KELT-16b and MASCARA-1b using the orbital parameters of \citet{talens2017} and \citet{oberst2017} as Gaussian priors, respectively; our updated parameters are summarized in Table \ref{tab:orbit}.

MASCARA-1b has strong systematics shortly after the first eclipse at a phase of $\sim$0.65 (BMJD = 58546.5) which do not show any clear correlation with sudden or unusual changes in centroid position or PSF width. This systematic noise is poorly handled by many of the detector models which results in strongly correlated residuals and wildly discrepant astrophysical parameters. Our BLISS and GP models, however, perform far better for this data set and are consistent with each other, with the BLISS model giving lower scatter in the model residuals. We also find that the results from our preferred BLISS model are robust to removing the affected data points from our fit. Meanwhile, the KELT-16b data are much simpler to fit and all models we consider are broadly consistent with each other, although the BLISS model gives a slightly more westward offset than the other models.

The pair of ultra-hot Jupiters MASCARA-1b and KELT-16b were chosen to allow for comparative studies since they share many physical characteristics in common. Both planets are highly irradiated (with irradiation temperatures of $\sim$3500~K), highly inflated ($R_{\rm p} \approx 1.4 R_{\rm jup}$), and have similar masses ($M_{\rm p} \approx 3 M_{\rm jup}$). The two main distinctions between the systems are the planets' orbital periods ($\sim$1 day for KELT-16b, and $\sim$2 days for MASCARA-1b) and their host stars' effective temperatures (6200~K for KELT-16 and 7500~K for MASCARA-1b) which balance each other out to give roughly the same incident flux. As a result, any significant differences in the normalized phase curve amplitude or offset would potentially be due to differences in Coriolis forces or stellar spectra. Assuming a wind speed of 5\,km/s for both planets and lengthscales equal to the planetary radii, we estimate mid-latitude Rossby numbers (Ro) of 0.91 and 0.47 for MASCARA-1b and KELT-16b, respectively. We also calculated the equatorial deformation radius ($L_{\rm D}$) following \citet{tan2020} who showed that the equatorial jet width scales as roughly 1.8 times $L_{\rm D}$; we find radii of 247\,000 and 62\,000 km (2.15 and 0.62 $R_{\rm p}$) for MASCARA-1b and KELT-16b, respectively, assuming a Brunt–V\"ais\"al\"a frequency of $N\approx\sqrt{g/H}$. Both Ro and $L_{\rm D}$ suggest that MASCARA-1b would possess a significantly larger jet than KELT-16b which would result in an increased phase offset and a warmer nightside temperature for MASCARA-1b.

However, we find no significant differences between the dayside temperatures of the two planets and similarly no differences between the nightside temperatures. There is a preference for a westward offset in the phase curve of KELT-16b ($38^{+16}_{-15}$ degrees W) which is also seen to a lesser extent for MASCARA-1b ($6^{+11}_{-11}$ degrees W), but the two values differ by only 1.7$\sigma$. We therefore find no significant evidence for the impact of different Coriolis forces and stellar spectra in the comparisons between the phase curve properties of these two particular planets. We also computed KELT-16b and MASCARA-1b's Bond albedos ($-0.16\pm0.38$ and $0.26\pm0.14$, respectively) and recirculation efficiencies ($0.309\pm0.074$ and $0.118\pm0.038$, respectively) after increasing our uncertainty on the effective temperatures following \citet{pass2019} and then inverting equations 4 and 5 from \citet{cowan2011b}. Ultimately, we find that both planets have poor heat recirculation and, while the Bond albedos of these planets are poorly constrained with these single wavelength observations, they are consistent with zero reflected light as would be expected for ultra-hot Jupiters.

\subsection{Uniform Reanalyses and Model Comparisons}
For each phase curve, we start by choosing the best phase curve model (first or second order sinusoid) for each of our 9--10 different detector models using the BIC; this reduces the number of models we are comparing by a factor of 2. We also found no clear differences between the results using PSF centroiding and FWM centroiding, so we decided to focus solely on our FWM results to reduce the number of models we are comparing by another factor of 2.

\begin{figure*}
    \centering
    \includegraphics[width=\linewidth]{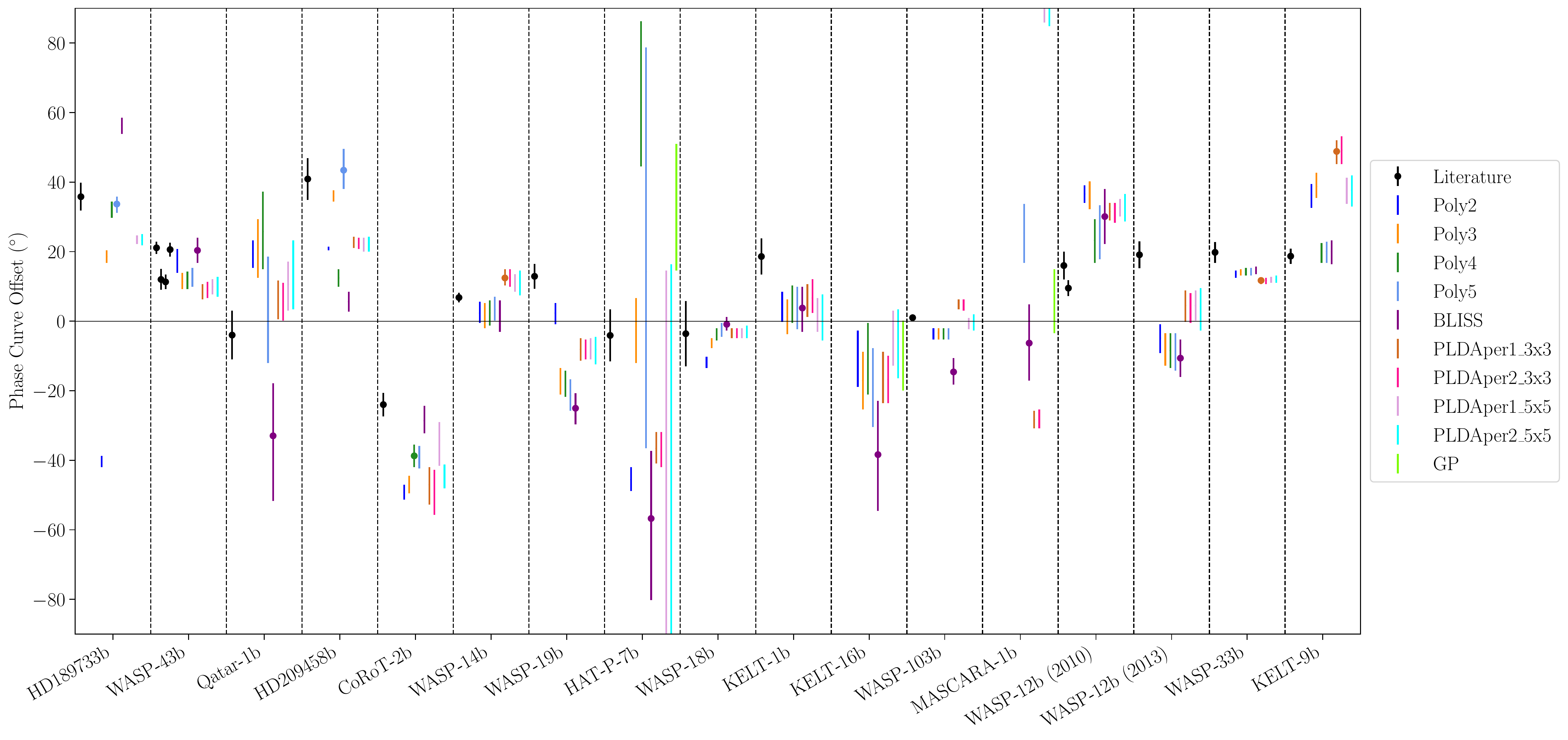}
    \caption{Phase curve offsets for all detector models using FWM centroiding and showing only the preferred astrophysical model. These offsets are compared to the previously published offsets for each phase curve indicated with black points. The first literature value for WASP-12b (2010) is from \citet{cowan2012}, and the second is from \citet{bell2019}. The literature values for WASP-43b are from \citet{stevenson2017}, \citet{mendonca2018a}, \citet{morello2019}, and \citet{may2020} from left to right. Similar figures for other astrophysical parameters can be found in the Appendix.}
    \label{fig:offset}
\end{figure*}

In Figure \ref{fig:offset} we highlight the different models' phase curve offsets for each planet compared to the literature values, while similar plots for phase curve semi-amplitude, eclipse depth, planet--star radius ratio, and nightside temperature are shown in Figures \ref{fig:amp}--\ref{fig:tnight} in the Appendix. Reassuringly, in most cases the retrieved parameters and uncertainties for each phase curve do not strongly depend on the detector model used, with most of the differences between model parameters being consistent at a ${\sim}1\sigma$ level.

Comparing individual model performances for different planets, we can see that HD 189733b, HD 209458b, HAT-P-7b, and MASCARA-1b show especially large dispersion between different models' phase offsets. In the case of HAT-P-7b, this is driven by our models preferring an unusually flat phase curve compared to the literature; as the phase curve amplitude approaches zero, the phase offset becomes undefined in a manner similar to the argument of periapse becoming undefined for a circular orbit. It is unclear why our HAT-P-7b models differ so greatly from the published fit \citep{wong2016} as the raw photometry appears fairly clean and there do not appear to be unusual correlations with PSF width or other covariates. For MASCARA-1b, the large scatter in retrieved phase offset is a result of the previously mentioned strong detector systematics at a phase of $\sim$0.65 which is only well fit by the BLISS and GP models. Finally, for HD 189733b and HD 209458b, the large scatter is the result of the other detector models poorly fitting the strong ``saw-tooth''-like systematic noise in these data sets. These ``saw-tooth'' systematics are sharply peaked, high frequency systematics present only in earlier \textit{Spitzer} observations before changes were made to the cycling of the spacecraft battery's heater to mitigate this effect\footnote{\url{http://ssc.spitzer.caltech.edu/warmmission/news/21oct2010memo.pdf}}.

\begin{figure*}
    \centering
    \includegraphics[width=\linewidth]{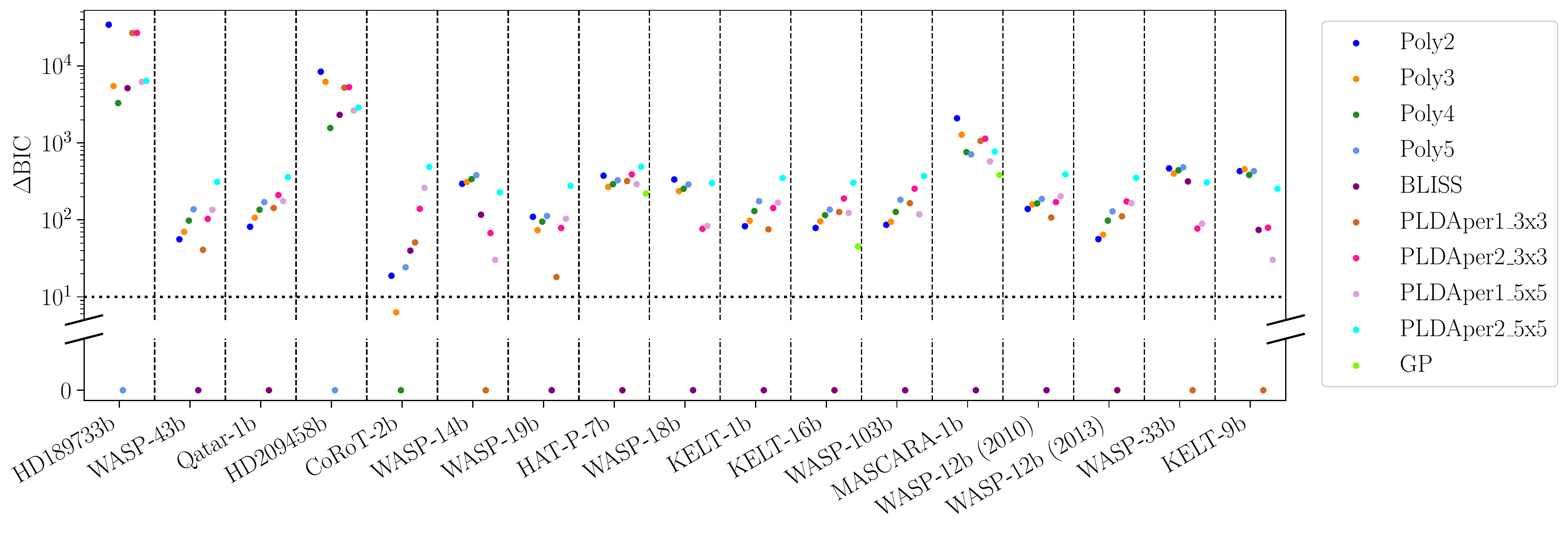}
    \caption{A comparison of the performance of each detector model for the full suite of models using the $\Delta$BIC with respect to the preferred model. A dotted horizontal line indicates the minimum $\Delta$BIC where there is no strong preference between models.}
    \label{fig:BIC}
\end{figure*}

While our BLISS model is typically preferred, for HD 189733b, HD 209458b, CoRoT-2b, WASP-14b, WASP-33b, and KELT-9b it is strongly disfavoured compared to the preferred models (Poly5, Poly5, Poly4, PLDAper1\textunderscore3x3, PLDAper1\textunderscore3x3, and PLDAper1\textunderscore3x3 respectively). While a better fit to these data sets with a BLISS model could likely be made using a more tailored approach---indeed, the phase curves of HD 189733b and HD 209458b were originally published using the Gaussian Kernel Regression technique which is similar in many respects to BLISS mapping---this lies beyond the scope of our current uniform reanalysis where we haven't tailored our algorithms to any data set in particular. The model fits to WASP-14b show an interesting feature where the Poly models and BLISS models all agree with each other, but all of the PLD models (where PLDAper1\textunderscore 3x3 is the preferred model) prefer larger phase curve semi-amplitudes, larger phase offsets, smaller eclipse depths, larger radii, and colder nightside temperatures. A similar effect is seen for some phase curves, but typically only for a single parameter (e.g. the phase offset for the 2013 observations of WASP-12b). Finally, while the BLISS models for WASP-33b are disfavoured, the retrieved phase curve parameters are consistent between the preferred PLDAper1\textunderscore 3x3 models and the BLISS models.

\begin{figure*}
    \centering
    \includegraphics[width=\linewidth]{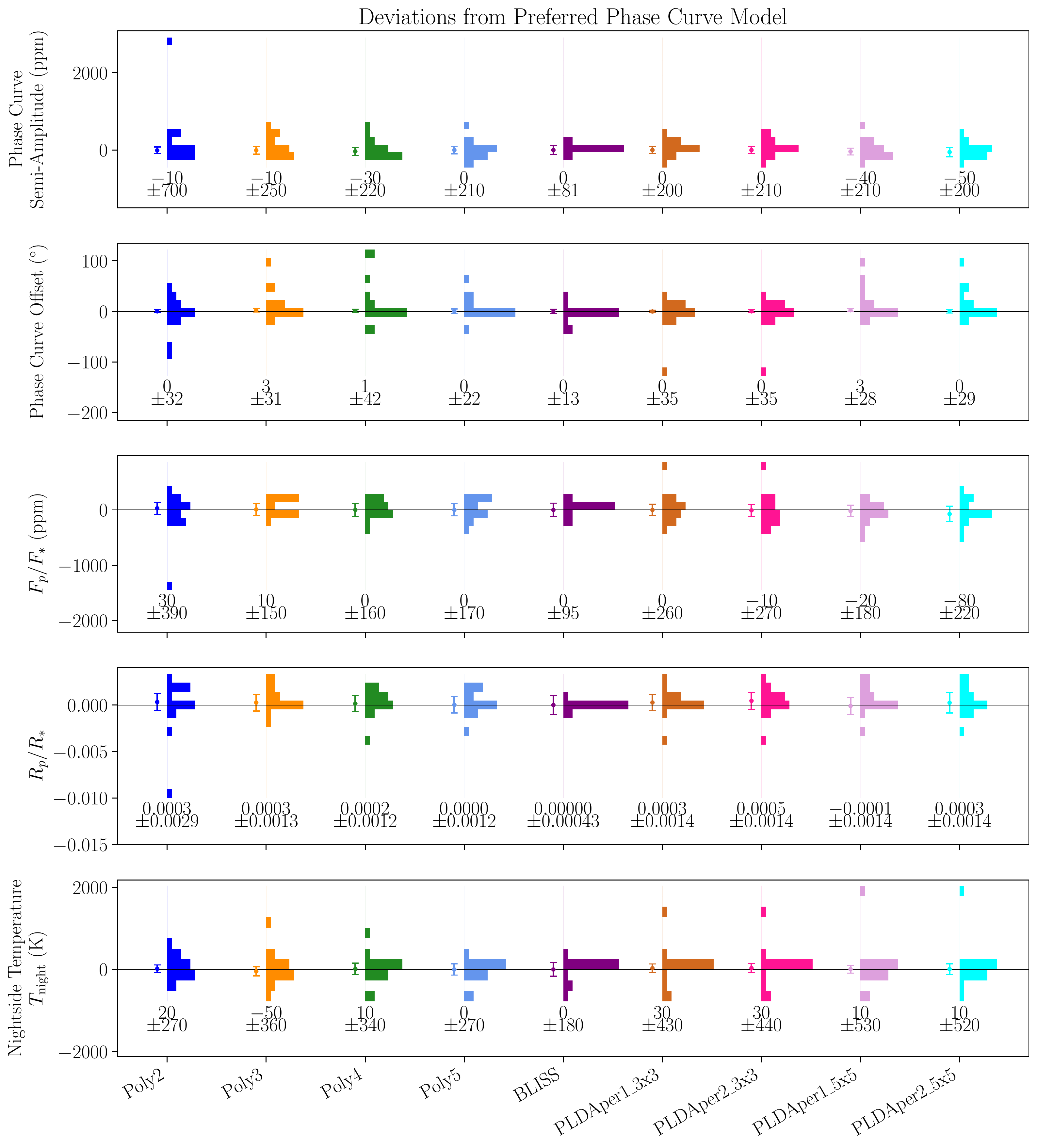}
    \caption{Histograms showing the bias and scatter of each model compared to the preferred model of each phase curve. Each histogram contains 17 values: one for each of the 17 phase curves. Beside each histogram is an error bar showing the average uncertainty for all fits with that model, and underneath each histogram is the observed median bias and scatter with respect to the preferred model. As can clearly be seen, some models occasionally produce wildly discrepant results. It is important to note, however, that this plot gives no indication as to how well each model fits the data sets. The strong performance by BLISS in these plots is mostly driven by the fact that the vast majority of data sets have BLISS as their preferred model.}
    \label{fig:diffs}
\end{figure*}

To further simplify comparisons between models, we decide to compare the fitted parameters from each model to the preferred model for that data set. In Figure \ref{fig:diffs}, we plot histograms of these differences to search for model biases and compare model performances; we look in particular at phase curve semi-amplitude, phase offset, eclipse depth, radius, and nightside temperature. We also make population plots using our preferred models, showing the dependencies of the dayside temperature, nightside temperature, and phase offset on the irradiation temperature of the systems (Figure \ref{fig:Tb_T0} and \ref{fig:offset_T0}).

Compared to the preferred model, our Poly2 model's offsets, phase curve semi-amplitudes, and eclipse depths are frequently discrepant, and our Poly2 model often leaves noisy residuals compared to the preferred model. Meanwhile, our Poly3--Poly5 models typically perform quite well compared to the preferred model. Our BLISS algorithm also performs very well and is the preferred model for most phase curves, although there are cases where BLISS significantly differs from the preferred model. Our PLD models have larger than typical scatter about the preferred model's phase curve semi-amplitude, phase offset, and eclipse depth, and they also result in noisier residuals than the preferred model. For the three phase curves that we fitted with the GP models (HAT-P-7b, MASCARA-1b, and KELT-16b), the GP model was largely consistent with the preferred BLISS model, although the GP model prefers a positive phase offset for HAT-P-7b.

Aside from HD 189733b, HD 209458b, HAT-P-7b, and MASCARA-1b's phase curves, we find that the scatter between different models' phase offsets is on average only $1.17\pm0.75$ times (or $0.9\pm3.8$ degrees) larger than the fitted uncertainty from the bestfit model for each lightcurve. For HD 189733b, HD 209458b, HAT-P-7b, and MASCARA-1b we find that our fitted uncertainty underpredicts the scatter between models by 21, 2.0, 1.5, and 5.9 times, respectively (or 51, 5.2, 12, and 53 degrees, respectively). Taking all phase curves into consideration, we find that the scatter is $1.5\pm4.8$ times larger or $2\pm17$ degrees larger than the fitted uncertainties. In summary, for the majority of phase curve observations there is no evidence for a need to inflate phase offset uncertainties, but in rare cases the scatter between different models' offsets suggests that uncertainties computed using only a single detector model could be underestimated by a factor of 3 or more. These comparisons are complicated, however, by the fact that in almost every case there is a single model which drastically outperforms all others (see Figure \ref{fig:BIC}). For this reason, we recommend that all future phase curve analyses explore a large range of detector models to simultaneously ensure that an optimal fit is found and to assess the dependence of phase offset on the decorrelation method used.

\subsection{Comparisons with Literature Values}\label{sec:literatureComparisons}
\begin{figure*}
    \centering
    \includegraphics[width=\linewidth]{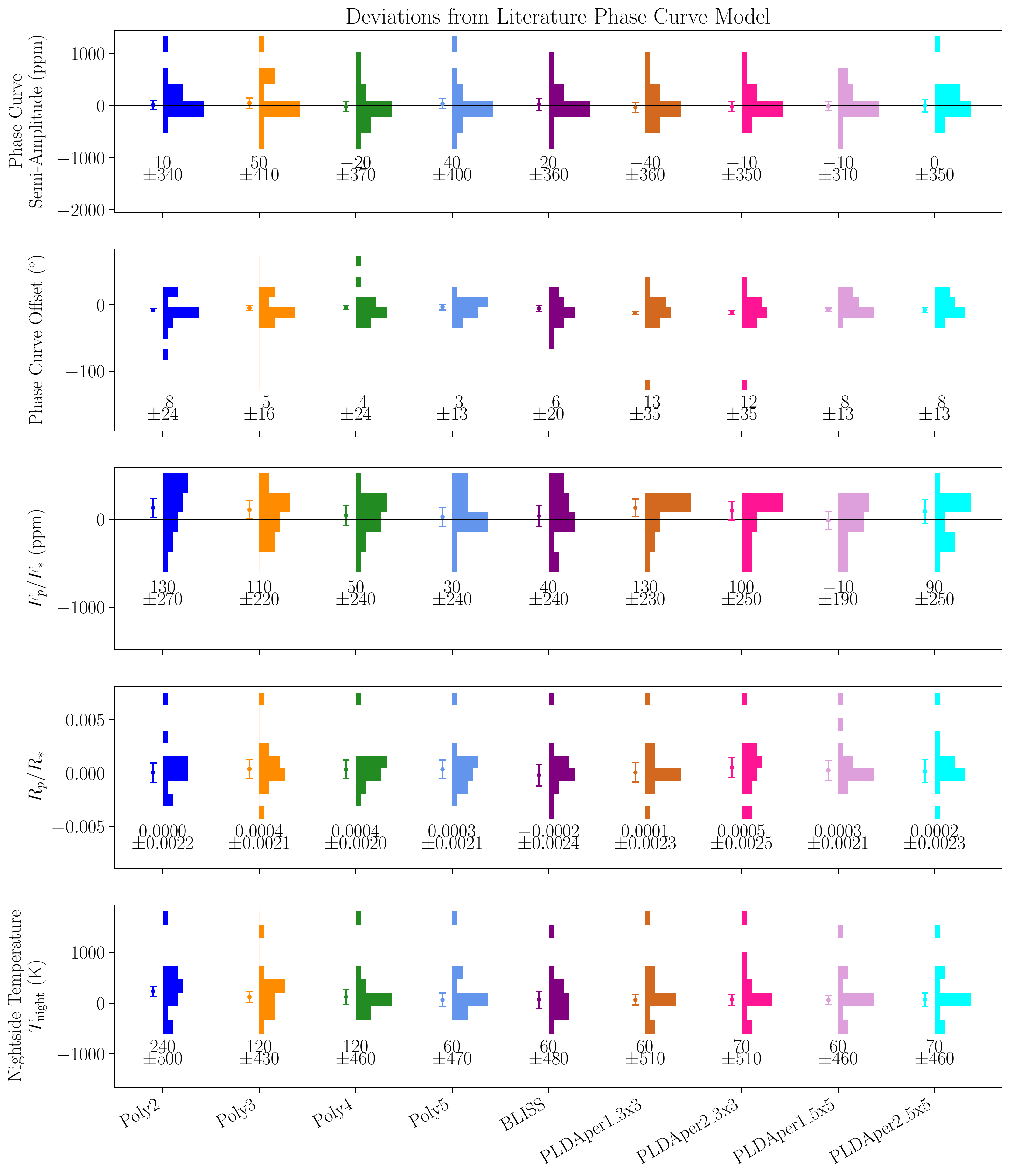}
    \caption{Histograms showing the bias and scatter of each model compared to the first published literature value for each phase curve. Each histogram contains 15 values: one for each of the 15 previously published phase curves.}
    \label{fig:diffs_literature}
\end{figure*}

Our preferred phase curve parameters from our two new and 15 reanalysed phase curves are presented in Table \ref{tab:preferred}, while we have compiled the literature values from the 15 previously published phase curves that we have reanalyzed in Table \ref{tab:literature}. Since there is no consistent parameterization for phase curves and different works define the terms ``dayside'' and ``nightside'' differently, we needed to convert or compute some values from most papers. We define dayside as the observer facing hemisphere at mid-eclipse, and nightside as the observer facing hemisphere during mid-transit. We chose to compute our tabulated values using the published values and to use a Monte-Carlo simulation to propagate uncertainties. As a result, we chose to only tabulate/compute symmetric uncertainties for the literature values. Overall, we do not find significant evidence for biases or severe underestimation of uncertainties for all phase curve parameters, with phase offsets on average reproduced to within $-8\pm21$ degrees ($-1.6\pm3.2$ sigma) and normalized phase curve amplitudes (peak-to-trough divided by eclipse depth) on average reproduced to within $-0.01\pm0.24$ ($-0.1\pm1.6$ sigma). We also compare each model's performance against the literature values in a manner similar to Figure \ref{fig:diffs} in Figure \ref{fig:diffs_literature}.

\begin{table*}
    \centering
    \bgroup
    \def\arraystretch{1.4}
    \begin{tabular}{l|c|c|c|c|c|c|c|c}
         & Detector & & $F_{\rm p}/F_{\rm *}$ & Semi-Amplitude & Max Flux & $T_{\rm 0}$ & $T_{\rm day}$ & $T_{\rm night}$  \\
        Planet & Model & $R_{\rm p}/R_{\rm *}$ & (ppm) & (ppm) & Offset ($^{\circ}$E) & (K) & (K) & (K) \\ \hline
        HD189733b & Poly5 & $0.15639^{+0.00021}_{-0.00021}$ & $1797^{+24}_{-26}$ & $654^{+36}_{-32}$ & $33.7^{+2.5}_{-2.2}$ & $1699^{+27}_{-27}$ & $1216.9^{+6.1}_{-6.4}$ & $929^{+26}_{-26}$ \\
		WASP-43b & BLISS & $0.15935^{+0.00095}_{-0.0011}$ & $3650^{+140}_{-140}$ & $1822^{+97}_{-110}$ & $20.4^{+3.6}_{-3.6}$ & $1994^{+90}_{-90}$ & $1476^{+47}_{-46}$ & $640^{+100}_{-110}$ \\
		Qatar-1b & BLISS & $0.1464^{+0.0020}_{-0.0019}$ & $3090^{+270}_{-260}$ & $1570^{+310}_{-250}$ & $-33^{+19}_{-15}$ & $2005^{+38}_{-38}$ & $1535^{+61}_{-61}$ & $900^{+180}_{-180}$ \\
		HD209458b & Poly5 & $0.12047^{+0.00039}_{-0.00042}$ & $1376^{+46}_{-40}$ & $489^{+72}_{-68}$ & $43.4^{+5.4}_{-6.1}$ & $2052^{+23}_{-23}$ & $1418^{+23}_{-19}$ & $1009^{+71}_{-81}$ \\
		CoRoT-2b & Poly4 & $0.1704^{+0.0014}_{-0.0017}$ & $4880^{+200}_{-190}$ & $2680^{+160}_{-130}$ & $-38.7^{+3.2}_{-3.2}$ & $2175^{+82}_{-82}$ & $1756^{+44}_{-43}$ & $873^{+51}_{-41}$ \\
		WASP-14b & PLDAper1\textunderscore3x3 & $0.09561^{+0.00049}_{-0.00052}$ & $2327^{+69}_{-69}$ & $843^{+39}_{-41}$ & $12.4^{+2.2}_{-2.5}$ & $2631^{+86}_{-86}$ & $2401^{+50}_{-49}$ & $1391^{+56}_{-61}$ \\
		WASP-19b & BLISS & $0.1384^{+0.0019}_{-0.0019}$ & $5400^{+240}_{-250}$ & $2170^{+220}_{-200}$ & $-25.0^{+4.7}_{-4.3}$ & $2993^{+52}_{-52}$ & $2291^{+67}_{-66}$ & $1380^{+120}_{-140}$ \\
		HAT-P-7b & BLISS & $0.0774^{+0.0011}_{-0.0011}$ & $2220^{+110}_{-110}$ & $480^{+160}_{-170}$ & $-57^{+23}_{-19}$ & $3145^{+57}_{-57}$ & $2930^{+100}_{-100}$ & $2520^{+240}_{-290}$ \\
		WASP-18b & BLISS & $0.09831^{+0.00051}_{-0.00054}$ & $3935^{+100}_{-97}$ & $1831^{+75}_{-89}$ & $-0.9^{+1.8}_{-2.2}$ & $3388^{+53}_{-53}$ & $3151^{+59}_{-58}$ & $960^{+140}_{-170}$ \\
		KELT-1b & BLISS & $0.0742^{+0.0014}_{-0.0014}$ & $2400^{+120}_{-120}$ & $1020^{+110}_{-110}$ & $3.8^{+6.8}_{-6.1}$ & $3435^{+77}_{-77}$ & $3240^{+140}_{-140}$ & $1350^{+230}_{-260}$ \\
		\textbf{KELT-16b} & BLISS & $0.1074^{+0.0019}_{-0.0022}$ & $4810^{+330}_{-310}$ & $1740^{+480}_{-460}$ & $-38^{+16}_{-15}$ & $3469^{+74}_{-74}$ & $3070^{+160}_{-150}$ & $1900^{+430}_{-440}$ \\
		WASP-103b & BLISS & $0.11551^{+0.00093}_{-0.00095}$ & $5240^{+150}_{-150}$ & $2500^{+120}_{-110}$ & $-14.6^{+3.6}_{-4.0}$ & $3540^{+100}_{-100}$ & $2971^{+88}_{-87}$ & $920^{+140}_{-160}$ \\
		\textbf{MASCARA-1b} & BLISS & $0.07881^{+0.00084}_{-0.00087}$ & $1947^{+82}_{-85}$ & $850^{+140}_{-130}$ & $-6^{+11}_{-11}$ & $3600^{+300}_{-300}$ & $2952^{+100}_{-97}$ & $1300^{+340}_{-340}$ \\
		WASP-12b (2010) & BLISS & $0.1047^{+0.0015}_{-0.0014}$ & $4230^{+230}_{-230}$ & $1790^{+270}_{-250}$ $^{\dagger}$ & $30.1^{+7.9}_{-7.9}$ $^{\dagger}$ & $3673^{+81}_{-81}$ & $2950^{+120}_{-120}$ & $1550^{+250}_{-270}$ \\
		WASP-12b (2013) & BLISS & $0.1047^{+0.0016}_{-0.0017}$ & $3940^{+210}_{-210}$ & $1920^{+190}_{-180}$ $^{\dagger}$ & $-10.6^{+5.4}_{-5.4}$ $^{\dagger}$ & $3674^{+82}_{-82}$ & $2920^{+120}_{-120}$ & $1110^{+250}_{-260}$ \\
		WASP-33b & PLDAper1\textunderscore3x3 & $0.11009^{+0.00045}_{-0.00046}$ & $4431^{+56}_{-57}$ & $1884^{+37}_{-39}$ & $11.71^{+1.1}_{-0.72}$ & $3932^{+53}_{-53}$ & $3232^{+49}_{-49}$ & $1559^{+39}_{-39}$ \\
		KELT-9b & PLDAper1\textunderscore3x3 & $0.08044^{+0.00057}_{-0.00056}$ & $2889^{+46}_{-43}$ & $703^{+48}_{-45}$ & $48.8^{+3.6}_{-3.2}$ & $5720^{+250}_{-250}$ & $4450^{+220}_{-210}$ & $3290^{+170}_{-170}$ \\
    \end{tabular}
    \egroup
    \caption{Preferred \texttt{SPCA} model parameters for each of our fitted phase curves. The planet names for our two new phase curves are bolded. Note that no fits were made the the $\delta$ Scuti pulsations of WASP-33.\newline
    $^{\dagger}$ WASP-12b's offsets and semi-amplitude are only from the first order sinusoid as there is a strong second order term which causes two peaks near quadrature.}\label{tab:preferred}
\end{table*}

\begin{table*}
    \centering
    \bgroup
    \def\arraystretch{1.4}
    \begin{tabular}{l|c|c|c|c|c|c|c}
         &  & & $F_{\rm p}/F_{\rm *}$ & Semi-Amplitude & Max Flux & $T_{\rm day}$ & $T_{\rm night}$ \\
        Planet & Reference & $R_{\rm p}/R_{\rm *}$ & (ppm) & (ppm) & Offset ($^{\circ}$E) & (K) & (K) \\ \hline
        HD189733b & \citet{knutson2012} & $0.15580\pm0.00019$ & $[1793\pm55]$ & $491\pm45$ & $[35.8\pm4.0]$ & $1192.0\pm9.0$ & $928\pm26$ \\
        WASP-43b & \citet{stevenson2017} & $0.15890\pm0.00050$ & $3830\pm80$ & $[1999\pm62]$ & $21.1\pm1.8$ & $1512\pm25$ & $<650$ @ $2\sigma$ \\
        WASP-43b & \citet{mendonca2018a} & -- & $[4060\pm100]$ & $[1630\pm120]$ & $[12.0\pm3.0]$ & $[1545\pm47]$ & $[914\pm75]$ \\
        WASP-43b & \citet{morello2019} & $[0.1572\pm0.0010]$ & $[3870\pm120]$ & $[1800\pm96]$ & $11.3\pm2.1$ & $[1522\pm+47]$ & $[730\pm97]$ \\	
        WASP-43b & \citet{may2020} & -- & $[3660\pm120]$ & $1613\pm83$ & $20.6\pm2.0$ & $[1478\pm45]$ & $[838\pm65]$ \\	
        Qatar-1b & \citet{keating2020} & $0.1450\pm0.0010$ & $3000\pm200$ & $920\pm110$ & $-4.0\pm7.0$ & $1557\pm35$ & $1167\pm71$ \\
        HD209458b & \citet{zellem2014} & $0.12130\pm0.00030$ & $1317\pm50$ & $[545\pm58]$ & $40.9\pm6.0$ & $1499\pm15$ & $972\pm44$ \\
        CoRoT-2b & \citet{dang2018} & $0.16970\pm0.00090$ & $4400\pm200$ & $[1700\pm200]$ & $-24.0\pm3.4$ & $1693\pm17$ & $[730\pm140]$ \\
        WASP-14b & \citet{wong2015} & $0.09421\pm0.00059$ & $2247\pm86$ & $786\pm23$ & $[6.8\pm1.4]$ & $2402\pm35$ & $1380\pm65$ \\
        WASP-19b & \citet{wong2016} & $0.1427\pm0.0021$ & $[5840\pm290]$ & $2370\pm220$ & $[12.9\pm3.6]$ & $2357\pm64$ & $[1180\pm160]$ \\
        HAT-P-7b & \citet{wong2016} & $0.07769\pm0.00078$ & $[1900\pm60]$ & $1040\pm175$ & $[-4.1\pm7.5]$ & $2682\pm49$ & $[1010\pm290]$ \\
        WASP-18b & \citet{maxted2013} & $0.09870\pm0.00072$ & $3790\pm210$ & $[1830\pm110]$ & $[-3.6\pm9.4]$ & $[3050\pm110]$ & $[980\pm230]$ \\
        KELT-1b & \citet{beatty2019} & $0.07710\pm0.00030$ & $2083\pm70$ & $979\pm54$ & $18.6\pm5.2$ & $2902\pm74$ & $1050\pm200$ \\
        WASP-103b & \citet{kreidberg2018b} & $0.1164\pm0.0011$ & $5690\pm140$ & $[2360\pm150]$ & $1.00\pm0.40$ & $3154\pm99$ & $1440\pm110$ \\
        WASP-12b (2010) & \citet{cowan2012} & $0.1054\pm0.0014$ & $3900\pm300$ & $[2000\pm150]$ $^{\dagger}$ & $16.0\pm4.0$ $^{\dagger}$ & $[2840\pm150]$ & $[960\pm250]$ \\
        WASP-12b (2010) & \citet{bell2019} & $0.10656\pm0.00085$ & $4360\pm140$ & $[2163\pm98]$ $^{\dagger}$ & $9.5\pm2.3$ $^{\dagger}$ & $2989\pm66$ & $790\pm150$ \\
        WASP-12b (2013) & \citet{bell2019} & $0.1049\pm0.0010$ & $3920\pm150$ & $[1640\pm150]$ $^{\dagger}$ & $19.1\pm3.9$ $^{\dagger}$ & $2854\pm74$ & $1340\pm180$ \\
        WASP-33b & \citet{zhang2018} & $0.1030\pm0.0011$ & $4250\pm160$ & $1792\pm94$ & $19.8\pm3.0$ & $3209\pm88$ & $1500\pm120$ \\
        KELT-9b & \citet{mansfield2020} & $0.08004\pm0.00041$ & $3131\pm62$ & $[953\pm37]$ & $18.7\pm2.2$ & $4566\pm138$ & $2556\pm99$ \\
    \end{tabular}
    \egroup
    \caption{Previously published model parameters for each of the phase curves we consider. Parameters reported using a different phase curve parameterization are converted and indicated with brackets. A dash indicates where the values cannot be computed from the published values. The offset and eclipse depth from \citet{mendonca2018a} were not originally published and come from \citet{may2020}. The day and nightside temperatures for \citet{mendonca2018a} and \citet{may2020} were calculated using the radius from \citet{stevenson2017} since they did not publish their radius.\newline
    $^{\dagger}$ WASP-12b's offsets and semi-amplitude are only from the first order sinusoid as there is a strong second order term which causes two peaks near quadrature.}
    \label{tab:literature}
\end{table*}

WASP-43b is the most heavily scrutinized phase curve, with four analyses of this data set already published \citep{stevenson2017, mendonca2018a, morello2019, may2020}. Our phase curve semi-amplitude, eclipse depth, and radius are consistent with all of these works. The more contentious issue is that of the phase curve's phase offset and nightside temperature. \citet{stevenson2017} initially reported only a $2\sigma$ upper limit on the nightside temperature of 650~K, while all subsequent reanalyses (including ours) favour a significantly detectable nightside temperature of $\sim$800 K. As for the planet's phase offset, \citet{stevenson2017} and \citet{may2020} favour a larger phase offset ($21\pm2~^{\circ}$E) than \citet{mendonca2018a} and \citet{morello2019} ($12\pm3~^{\circ}$E and $11\pm2~^{\circ}$E). \citet{may2020} claimed that the differences between the retrieved phase offsets is the result of temporal binning which was not used by \citet{stevenson2017} and \citet{may2020} but was used by \citet{mendonca2018a}, \citet{morello2019}, and this work. Fitting the temporally binned photometry for all 17 phase curves with each of our detector models already required more than 2\,000 CPU hours, and expanding this to unbinned photometry for all phase curve fits would require more than 125\,000 CPU hours (or 434 days using our 12$\times$ multi-threading computer) optimistically assuming all of detector models scaled linearly with the number of input data. However, we did try fitting just the WASP-43b unbinned phase curve with our preferred detector model (BLISS) and found that our phase offset and nightside temperature was unchanged. Including a linear slope in time also did not affect our phase offset or nightside temperature. Instead, we find that the phase offset inferred by our models depends on the choice of phase curve model, as our 4-parameter (v2) phase curve models are consistent with those of \citet{stevenson2017} and \citet{may2020}, while our 2-parameter phase curve models (v1) are consistent with \citet{mendonca2018a} and \citet{morello2019}. Ultimately, we cannot decide between these two discrepant offsets as the $\Delta$BIC between the two phase curve models for our preferred BLISS detector model is only 3.7 (insignificantly favouring the $20.4\pm3.6$ offset from the v2 model). For reference, \citet{stevenson2014c} found phase offsets ranging from roughly -6 to 17 degrees east in the \textit{Hubble}/WFC3 bandpass.

For HD 189733b, we retrieve a slightly larger phase curve semi-amplitude (2.9$\sigma$) than that reported by \citet{knutson2012}. For Qatar-1b, our models prefer a larger phase curve semi-amplitude (2.2$\sigma$) and larger uncertainty on the phase offset ($\pm$17$^{\circ}$ vs $\pm$7$^{\circ}$) than published by \citet{keating2020}, making it appear more consistent with WASP-43b. For HD 209458b, we find a significantly colder dayside temperature (3.2$\sigma$) than that published by \citep{zellem2014}. We retrieve a significantly westward phase offset for CoRoT-2b, consistent with the findings of \citet{dang2018}, but with a larger phase offset (3.1$\sigma$) than their reported value. We also find a significantly larger phase curve semi-amplitude (4.0$\sigma$) for CoRoT-2b than was reported by \citet{dang2018}. Our preferred model's values for WASP-14b were all consistent with their previously published values to within 2$\sigma$ \citep{wong2015}. For WASP-19b, we find the phase offset changes direction with respect to that published by \citet{wong2016} ($25.0^{+4.7}_{-4.3}$ degrees \textit{west} rather than $12.9\pm3.6$ degrees \textit{east}; 6.6$\sigma$). It is unclear why the phase offset is so different for this dataset as there were not particularly strong detector systematics or unusual variations in centroid position or PSF width. For HAT-P-7b, our models suggest a much shallower phase curve (1.9$\sigma$) than that reported by \citet{wong2016}. As a result of the smaller phase curve semi-amplitude, we also find a far larger uncertainty on the phase offset and larger scatter between our detector models. For WASP-18b, our preferred model's values are consistent with the literature values \citep{maxted2013} to within 1$\sigma$. For reference, the \textit{Hubble}/WFC3 phase offset reported by \citep{arcangeli2019} for WASP-18b was $4.5\pm0.5$ east, while we find an offset of $-0.9\pm2.2$ degrees east at 4.5~$\mu$m.

For KELT-1b, we find a hotter dayside temperature (2.2$\sigma$) than that reported by \citet{beatty2019}. Our models for the WASP-103b data suggest a marginally westward offset ($-14.6^{+3.6}_{-4.0}$ degrees) compared to the previously published eastward offset at 4.5~$\mu$m ($1.00\pm0.40$ degrees; differing by 2.2$\sigma$) and a colder nightside (2.2$\sigma$) than that reported by \citet{kreidberg2018b}. For reference, \citet{kreidberg2018b} found phase offsets of $-0.3\pm0.1$ degrees east in the \textit{Hubble}/WFC3 bandpass and $2.0\pm0.7$ degrees east in the 3.6~$\mu$m bandpass. For WASP-12b, we retrieve a moderately westward first order sinusoidal phase offset for the 2013 observations which is discrepant at 4.3$\sigma$ compared to the moderately eastward offset from \citet{bell2019}. Interestingly, this would be consistent with the observed change from an eastward phase offset in 2010 to westward phase offset in 2013 seen for the channel 1 observations of WASP-12b \citep{bell2019}. We also still find evidence for very strong second order phase variations in both 4.5~$\mu$m phase curves of WASP-12b, consistent with the findings of \citet{cowan2012} and \citet{bell2019}.

For WASP-33b, only our retrieved radius varied significantly (6.0$\sigma$) from the published value from \citet{zhang2018}. Leaving unmodelled the strong variability of the host star WASP-33A (seen clearly in our residuals in the Supplementary Information) could potentially have led to this difference. It is notable, however, that no other phase curve parameters were strongly affected. Finally, our models for KELT-9b prefer a lower semi-amplitude (4.2$\sigma$) and a larger phase offset (7.4$\sigma$) with a smaller eclipse depth (3.2$\sigma$) and hotter nightside temperature (3.7$\sigma$) than that reported by \citet{mansfield2020}. Given that modelling the stellar pulsations reported by \citet{wong2020} had only a negligible effect on the retrieved phase curve parameters for \citet{mansfield2020}, the differences for KELT-9b are unlikely to be the result of our choice to neglect them. An increased nightside temperature for KELT-9b would only further increase the evidence that the latent heat-like effects of H$_2$ dissociation/recombination operate in ultra-hot Jupiter atmospheres as was predicted by \citet{bell2018}.

\subsection{Population Level Trends}\label{sec:populationTrends}
We also used the Pearson's correlation coefficient ($r$) to re-evaluate population level trends in phase curve parameters using our reanalyses. We summarize here the most relevant pairs for which there is a p-value below 0.05. To fit trends, we use an orthogonal distance regression routine (\texttt{scipy.odr}) to find the best-fit linear trend given the uncertainties in both x and y directions while performing a Monte Carlo over the x and y values to determine the uncertainty in the fitted parameters.

First, we confirm a positive correlation between irradiation temperature and radius ($r=0.69$; $p=0.0020$) which is consistent with the well known phenomenon of hot Jupiter radius inflation \citep[e.g.,][]{guillot2002,laughlin2011}. We also find tentative evidence for a negative correlation between normalized phase curve amplitude (peak-to-trough divided by eclipse depth) and stellar effective temperature ($r=-0.51$; $p=0.034$), while the normalized phase curve amplitude does not appear to be correlated with irradiation temperature or dayside temperature. This could potentially be explained through the lower energy photons preferentially emitted by cooler stars being absorbed higher in the planetary atmosphere where radiative timescales are much more rapid.

We confirm that 4.5~$\mu$m dayside brightness temperature is strongly correlated with irradiation temperature ($r=0.96$; $p<10^{-9}$), and we find that the best-fit slope of $T_{\rm day,bright}$ vs $\,T_{\rm 0}$ is $0.818\pm0.011$ when neglecting the extreme outlier KELT-9b. Meanwhile, the equilibrium temperature (assuming zero albedo and uniform recirculation) follows $T_{\rm eq} \equiv 0.71\,T_{\rm 0}$. Previously, \citet{beatty2019} found a slope of $0.94\pm0.08$ for the 4.5~$\mu$m dayside brightness temperatures from 11 hot Jupiter phase curves, \citet{garhart2020} found a median slope of 0.79 for 36 hot Jupiters using the error-weighted average of the 3.6 and 4.5~$\mu$m brightness temperatures from eclipse observations, and \citet{baxter2020} found a slope of $0.84\pm0.04$ using 4.5~$\mu$m eclipse observations of 78 hot Jupiters. The steep slope at 4.5~$\mu$m dayside brightness temperature, combined with a shallower slope at 3.6~$\mu$m, is believed to be caused by changing temperature--pressure profiles \citep{garhart2020} resulting in a transition between seeing CO in absorption for colder planets and emission for hotter planets \citep{baxter2020}.

\begin{figure*}
    \centering
    \includegraphics[width=\linewidth]{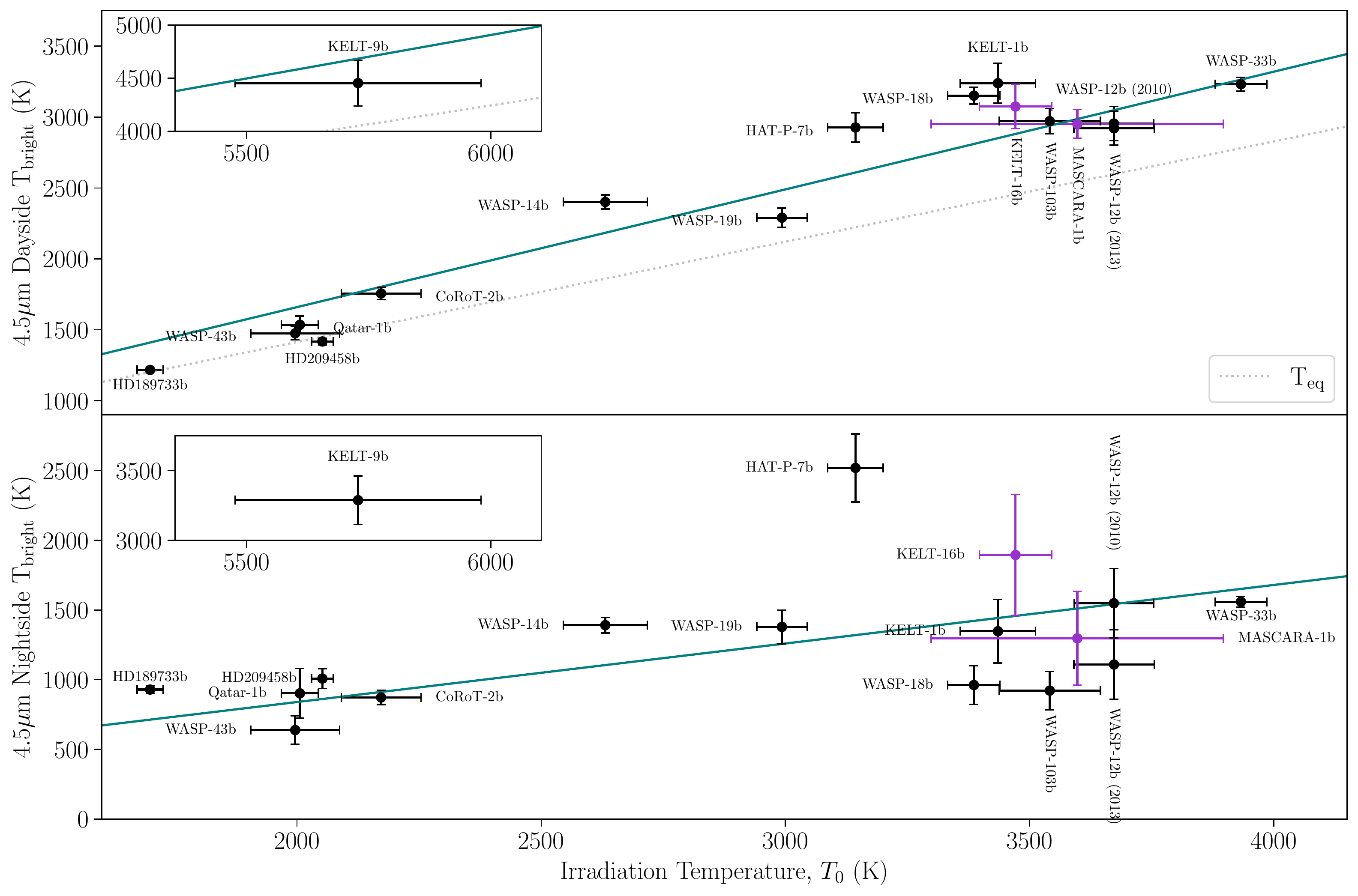}
    \caption{Day and nightside brightness temperatures as a function of irradiation temperature for all considered planets, using the preferred model selected by \texttt{SPCA}. Our new planets KELT-16b and MASCARA-1b are highlighted in purple. KELT-9b has been place in an inset with the same scale size as it lies far beyond the bounds of the plot. A dotted line in the top panel shows the relationship between irradiation temperature and equilibrium temperature (assuming zero Bond albedo and uniform recirculation) and is present in the KELT-9b inset figure as well. A teal line indicates the fitted slopes of $0.818\pm0.011$ for $T_{\rm day,bright}$ vs $T_{\rm 0}$ and $0.421\pm0.011$ for $T_{\rm night,bright}$ vs $T_{\rm 0}$.}
    \label{fig:Tb_T0}
\end{figure*}

We also confirm a significant, fairly shallow dependence of nightside brightness temperature on irradiation temperature ($r=0.73$; $p=0.00089$) which has a slope of $0.421\pm0.011$ when neglecting the extreme outlier KELT-9b; a nearly flat trend was previously reported by \citet{keating2019} and \citet{beatty2019}. \citet{keating2019} didn't compute a slope, but using the effective nightside temperatures published in their Table 1 we compute a slope of $0.44\pm0.01$. Meanwhile, \citet{beatty2019} applied different phase curve inversion methods and found a much shallower slope of $0.08\pm0.11$ for the 4.5~$\mu$m brightness temperatures. The interpretation from these two works was that this weak dependence of nightside temperatures on irradiation temperature is driven by a cloud layer that ubiquitously covers hot Jupiter nightsides; silicate clouds were a preferred species as they condense at the $\sim$1000~K temperatures observed on the nightsides of these planets. The extremely hot nightside temperature of KELT-9b has been attributed to the latent heat-like effects of H$_2$ dissociation/recombination \citet{mansfield2020} as was predicted by \citet{bell2018}.

\begin{figure*}
    \centering
    \includegraphics[width=\linewidth]{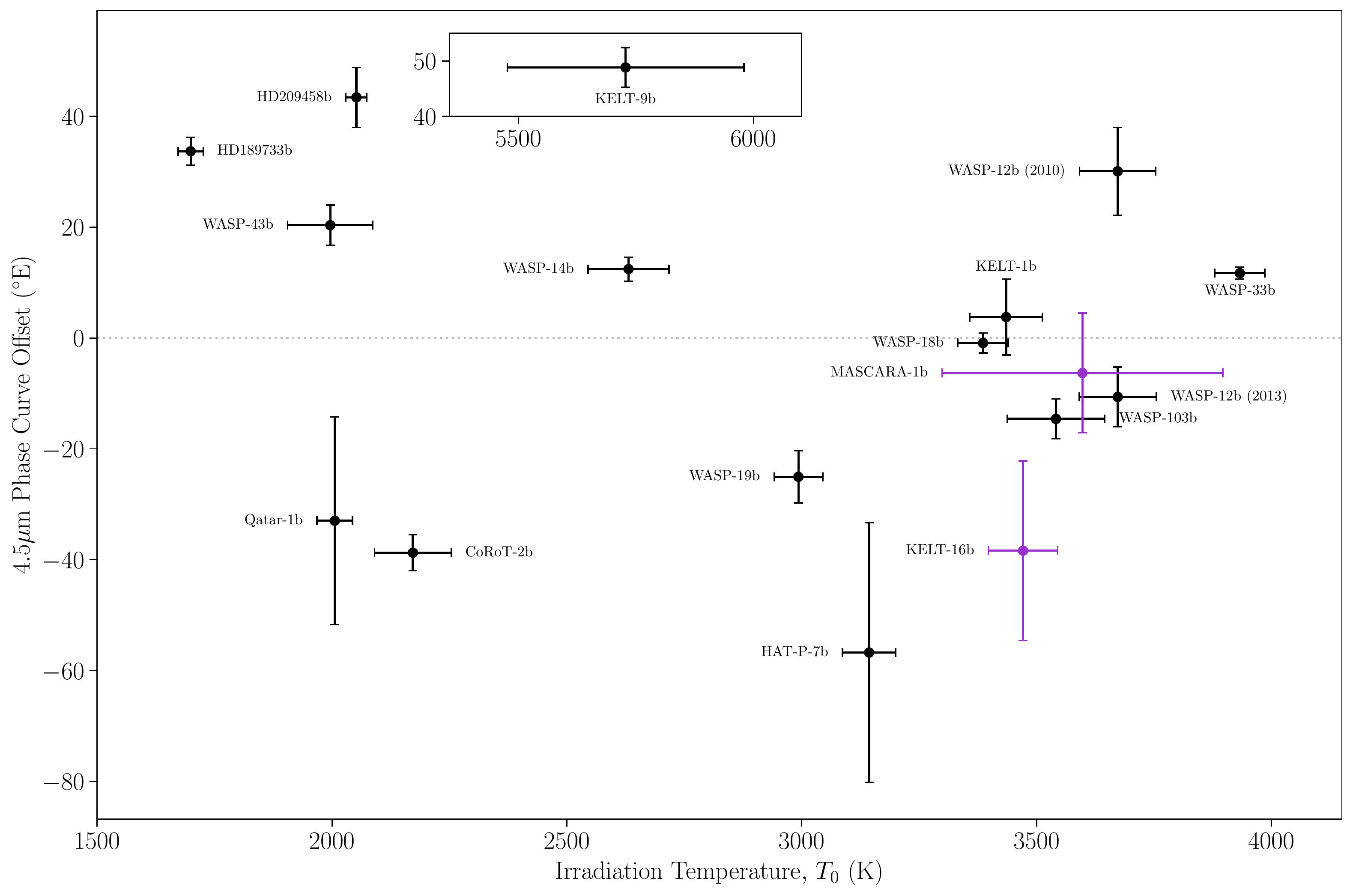}
    \caption{Phase curve offsets as a function of irradiation temperature for all considered planets, using the preferred model selected by \texttt{SPCA}. Our new planets KELT-16b and MASCARA-1b are highlighted in purple. KELT-9b has been place in an inset with the same scale size and vertical position as it lies far beyond the bounds of the plot.}
    \label{fig:offset_T0}
\end{figure*}

\begin{figure*}
    \centering
    \includegraphics[width=\linewidth]{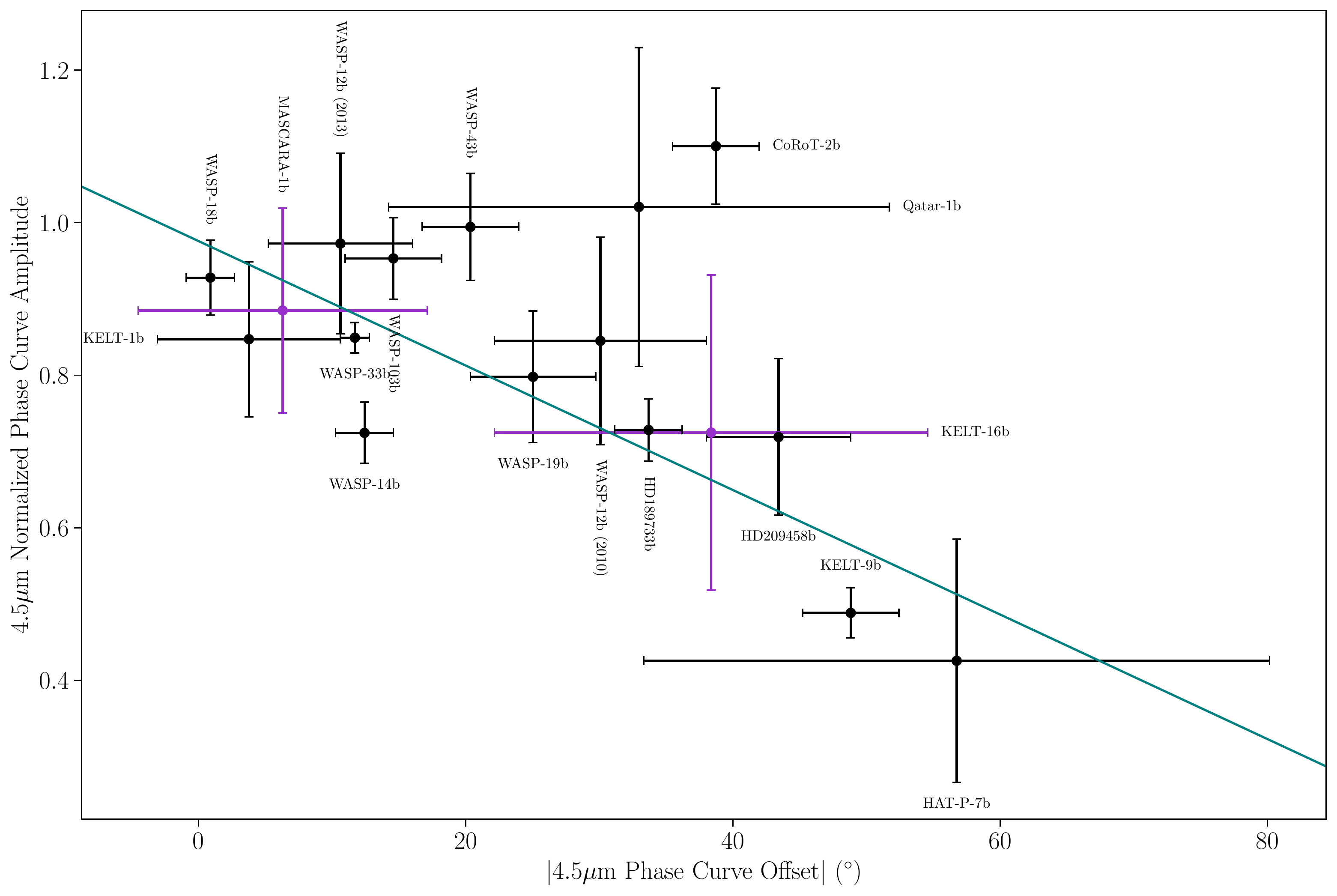}
    \caption{Normalized phase curve amplitudes (peak-to-trough divided by eclipse depth) as a function of the absolute value of phase curve offset for all considered planets, using the preferred model selected by \texttt{SPCA}. Our new planets KELT-16b and MASCARA-1b are highlighted in purple. A teal line indicates the fitted relationship with a slope of $-0.0082\pm0.0015$ and a y-intercept of $0.976\pm0.027$. While WASP-43b and CoRoT-2b have normalized phase curve amplitudes greater than unity, this is caused by the significant phase offset of the systems which cause the eclipse depth to be significantly lower than the phase curve maximum.}
    \label{fig:amp_offset}
\end{figure*}

Unlike \citet{zhang2018}, we find no correlation between phase offset and irradiation temperature, nor is any obvious trend visible by eye (Figure \ref{fig:offset_T0}). We do, however, find that the orbital period is correlated with the heat recirculation efficiency ($r=0.61$, $p=0.0087$). This positive correlation between heat recirculation efficiency and orbital period is consistent with that predicted by \citet{komacek2017}, although they also predicted a strong dependence on irradiation temperature for which we do not find evidence. We find no significant correlation between phase offset and normalized phase curve amplitude ($r=-0.20$; $p=0.45$), but we do find evidence for a correlation between the more physically meaningful absolute magnitude of the phase offset and normalized phase curve amplitude relationship ($r=-0.55$; $p=0.021$; Figure \ref{fig:amp_offset}). When we fit for a trend between the normalized phase curve amplitude and the absolute magnitude of the phase offset, we find a slope of $-0.0082\pm0.0015$ and a y-intercept of $0.976\pm0.027$.

We find that the Bond Albedo is not strongly correlated with the planetary mass, the logarithm of the planetary mass, or the logarithm of the surface gravity ($r=-0.39, -0.44, -0.38$; $p=0.12, 0.08, 0.13$, respectively). \citet{zhang2018} previously reported a negative correlation between Bond Albedo and planetary mass, and they suggested this could be the result of decreased lofting of cloud particles with increased surface gravity (although they also found no significant correlation with surface gravity). The dependence of cloud particle lofting on surface gravity has been predicted \citep[e.g.,][]{marley1999,heng2013} and has been observed for brown dwarfs where lower surface gravity objects exhibit increased cloudiness \citep{faherty2016}.

\begin{figure*}
    \centering
    \includegraphics[width=0.45\linewidth]{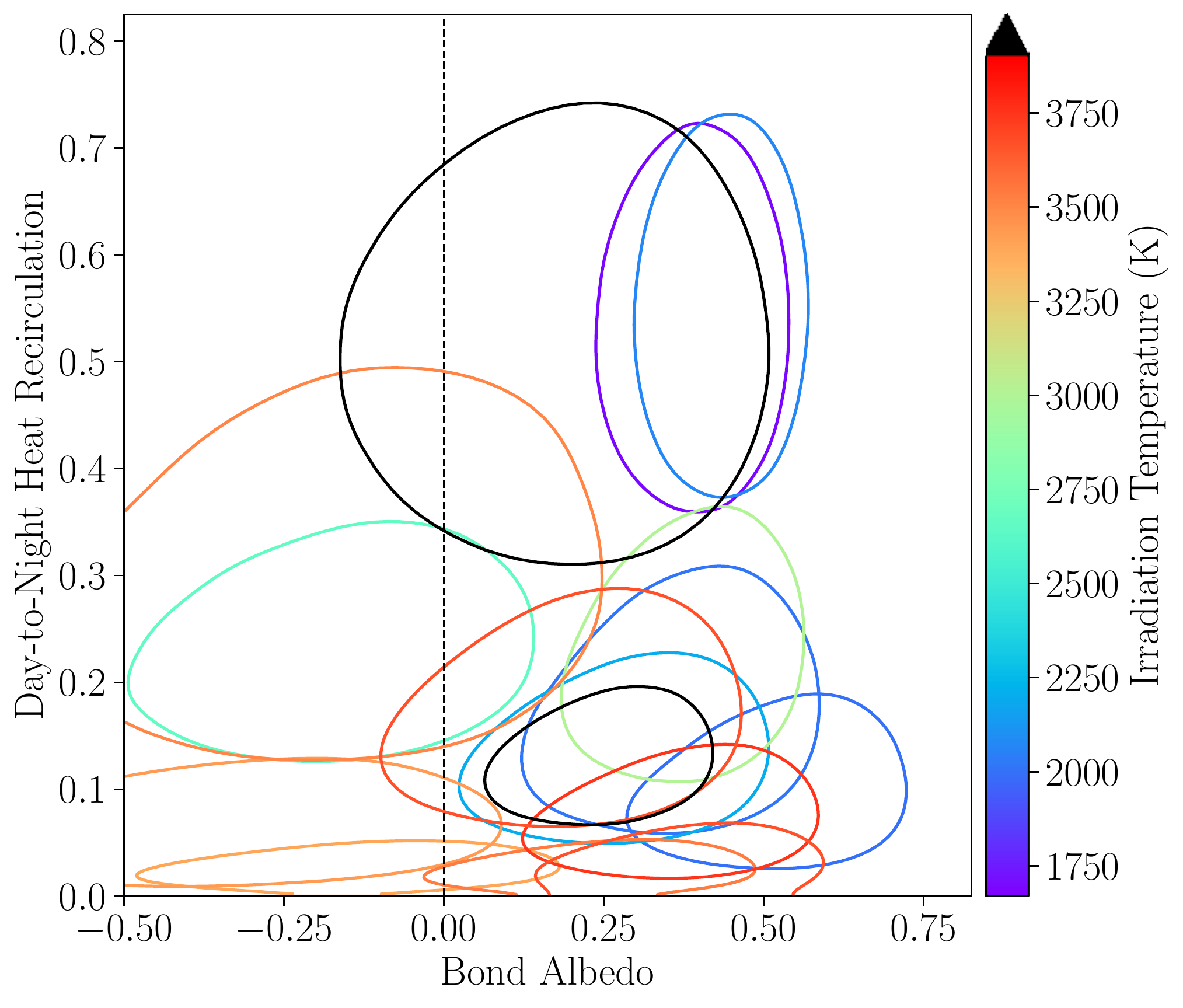}\hfill%
    \includegraphics[width=0.45\linewidth]{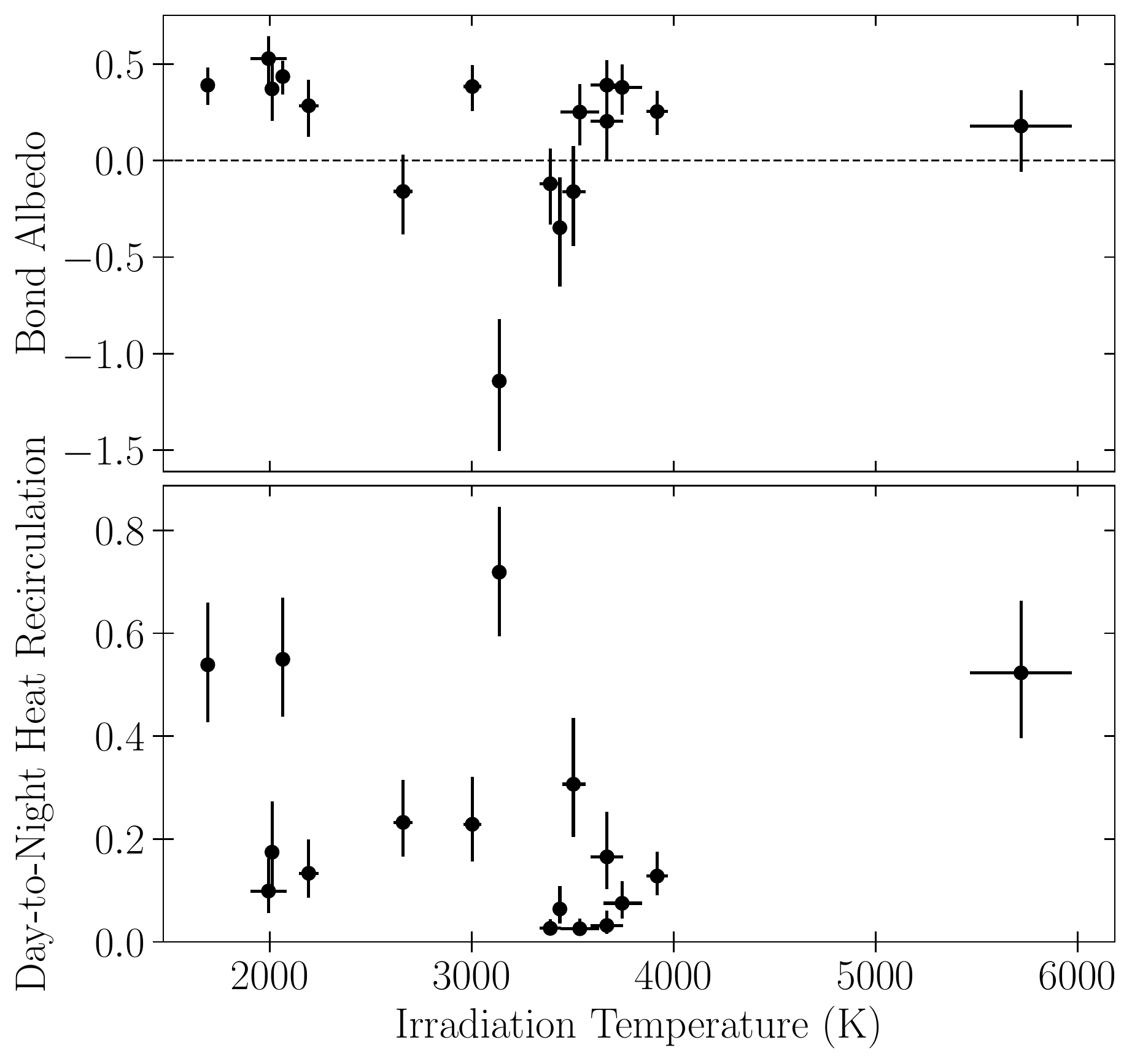}
    \caption{Left: An updated version of Figure 3 from \citet{schwartz2017b} showing the relationships between Bond albedo and day--night heat recirculation. Significant differences exist between this figure and that of \citet{schwartz2017b} as we consider only the 4.5~$\mu$m phase curves and have followed the procedure of \citet{pass2019} to account for underestimated uncertainties in inferring effective temperatures using only one or a few photometric bands. Right: the same parameters shown as 1D trends with irradiation temperature.}
    \label{fig:schwartz_cowan}
\end{figure*}

\section{Discussion and Conclusions}\label{sec:discussion}

We have developed an open-source, modular pipeline for the reduction and decorrelation of \textit{Spitzer}/IRAC channel 1 and 2 photometry, incorporating versions of some of the most popular decorrelation methods in the literature. We invite anyone interested in contributing their decorrelation method to visit our GitHub (\url{https://github.com/lisadang27/SPCA}). We first validated the implementation of our pipeline on the ten repeated eclipse observations of XO-3b, finding all our models perform equally well on these data with our fitted uncertainty on each eclipse depth only slightly underestimating the scatter between the ten eclipse observations. We then used this pipeline to perform the uniform reanalysis of 15 \textit{Spitzer} phase curve observations and analyse the new phase curves of ultra-hot Jupiters MASCARA-1b and KELT-16b. We use these analyses to test for the reproducibility of the literature values and perform a comparison of decorrelation models across 17 different phase curves; something previously only done for individual phase curves \citep[e.g.][]{wong2015,dang2018,bell2019,keating2020} or the 10 repeated eclipse observations of XO-3b \citep{ingalls2016}.

For our decorrelation model comparisons, we find that our BLISS model tends to perform the best as evaluated by the BIC, where we consider each of the occupied BLISS knots a fitted parameter. For most phase curves, our higher complexity 2D Polynomial models (Poly3--5), our PLD models, and our BLISS model all give consistent results. However, there are cases like HD 189733b, HD 209458b, HAT-P-7b, and MASCARA-1b where the retrieved results do strongly depend on the model used.

We find that our reanalysis of WASP-43b's channel 2 phase curve is consistent to within $\sim$2$\sigma$ of all of the values published by \citet{mendonca2018a}, \citet{morello2019}, and \citet{may2020}, but we do find a significantly hotter nightside than was published by \citet{stevenson2017}. Using WASP-43b as a test case, we found that our BLISS results were not affected by temporal binning; this is consistent with the findings of \citet{may2020} which showed that phase curve offsets and nightside temperatures are not affected by temporal binning when using their BLISS algorithm without an additional PSF-width model. We instead find that the retrieved offset for the WASP-43b phase curve changes significantly depending on the phase curve model used, with first order models reproducing the phase offsets of \citet{mendonca2018a} and \citet{morello2019} and second order models reproducing the phase offsets of \citet{stevenson2017} and \citet{may2020}; there is inadequate statistical evidence to differentiate these two models, but the second order model's offset of $20.4\pm3.6~^{\circ}$E is marginally preferred (\mbox{$\Delta$BIC $\sim$ 3.7}). We find that Qatar-1b, WASP-14b, WASP-18b, WASP-103b, and the 2010 observations of WASP-12b the only other phase curves for which we reproduce all literature values within $\sim$2$\sigma$, and we find that our retrieved phase offsets and nightside temperatures often differ from their published values, while eclipse depths and radii are typically consistent with the literature.

Our novel observations of MASCARA-1b and KELT-16b suggest these two ultra-hot Jupiters have quite similar phase curves, despite their orbital period, and thus likely their rotational periods, differing by a factor of two. KELT-16b's and MASCARA-1b's energy budgets are poorly constrained but consistent with zero Bond albedo and fairly inefficient recirculation. We also find that there is minimal diversity in the phase curves of similarly irradiated ultra-hot Jupiters WASP-18b, KELT-1b, KELT-16b, WASP-103b, and MASCARA-1b, with all planets having similar dayside temperatures, nightside temperatures, and phase offsets (Figures \ref{fig:Tb_T0} and \ref{fig:offset_T0}) despite masses ranging from to 1.5 to 27~$M_{\rm Jup}$ and periods ranging from 1 to 2 days. While these cooler ultra-hot Jupiters don't show strong evidence for the effects of H$_2$ dissociation/recombination, the hot nightsides and large phase offsets of WASP-33b and KELT-9b do imply heat transport far greater than would be predicted in the absence of H$_2$ dissociation/recombination.

Using our reanalyzed and new phase curve observations, we confirm significant trends in the $4.5~\mu$m brightness temperatures of the dayside and nightside hemispheres as a function of irradiation temperature. However, we do not find clear evidence for previously reported trends in phase offset with irradiation temperature. We also find evidence that normalized phase curve amplitude is correlated with stellar effective temperature and that day--night heat recirculation is correlated with orbital period. Finally, we find that normalized phase curve amplitude does not appear to be correlated with phase offset but does appear to be correlated with the absolute value of phase offset

Overall, while our different decorrelation models often retrieve similar phase curve parameters, significant differences can arise between different models as well as between our preferred model and the literature values. We find differences of up to $\sim$30$^{\circ}$ in the phase offset between our preferred model and the literature value, but ultimately, our preferred models are consistent with published phase offsets to within $-8\pm21$ degrees ($-1.6\pm3.2$ sigma) and normalized phase curve amplitudes are on average reproduced to within $-0.01\pm0.24$ ($-0.1\pm1.6$ sigma). Additional studies on the reproducibility of phase curve parameters (and especially offsets) with and without temporal binning need to be performed on a large number of phase curves to ensure that any conclusions hold for the entire collection of 4.5~$\mu$m \underline{and} 3.6~$\mu$m \textit{Spitzer} phase curves. Finally, we recommend that the principles of open-source and modular code be applied in the coming era of \textit{JWST}, reducing redundant labour and increasing reproducibility and uniformity.

\section*{Acknowledgements}

This work is based on observations made with the \textit{Spitzer} Space Telescope, which was operated by the Jet Propulsion Laboratory, California Institute of Technology under a contract with NASA. T.J.B.~acknowledges support from the McGill Space Institute Graduate Fellowship, the Natural Sciences and Engineering Research Council of Canada's Postgraduate Scholarships-Doctoral Fellowship, and from the Fonds de recherche du Qu\'ebec -- Nature et technologies through the Centre de recherche en astrophysique du Qu\'ebec. Support for this work was also provided to US-based investigators by NASA through an award issued by JPL/Caltech. J.M.D acknowledges support from the Amsterdam Academic Alliance (AAA) Program, and the European Research Council (ERC) European Union’s Horizon 2020 research and innovation program (grant agreement no. 679633; Exo-Atmos). This work is part of the research programme VIDI with project number 016.Vidi.189.174, which is (partly) financed by the Dutch Research Council (NWO). Finally, we have also made use of open-source software provided by the Python, Astropy, SciPy, and Matplotlib communities.

\section*{Data Availability}
The raw observations used in this work are freely accessible on the \textit{Spitzer} Heritage Archive. The fitted parameters for each considered model are available as numpy zip files in the Supplementary Data. The data presented in each figure will be shared on reasonable request to the corresponding author.

\bibliographystyle{mnras}
\bibliography{mega_science} % if your bibtex file is called example.bib

\appendix

\section{Priors}
The priors used throughout our fitting are described in Table \ref{tab:priors}.

\begin{table}
\caption{A summary of all the priors used in the model fitting. Uniform priors were used where there are inequalities below, Gaussian priors were used to constrain astrophysical parameters to the most precise published values from the NASA Exoplanet Archive (\url{https://exoplanetarchive.ipac.caltech.edu/}), and parameters were unconstrained where Free is written. The $p_{i,1}$ parameters are the first order PLD terms, and $p_{i,2}$ are the second order PLD terms}\label{tab:priors}
\begin{tabular}{c|c}
Parameter                    & Prior                              \\ \hline
$t_0$ (BMJD)                 & Gaussian                           \\ \hline
$R_p/R_*$                    & $0 < R_p/R_* < 1$                  \\ \hline
$a/R_*$                      & Gaussian                           \\ \hline
$i$ (degrees)                & Gaussian                           \\ \hline
$P$ (days)                   & Gaussian                           \\ \hline
$F_p/F_*$                    & $0 < F_p/F_* < 1$                  \\ \hline
$C_1$                        & Positive Phase Curve                \\ \hline
$D_1$                        & \begin{tabular}[c]{@{}c@{}}Positive Phase Curve;\\$|{\arctan}2(D_1, C_1)| < 90^{\circ}$\end{tabular}                \\ \hline
$C_2$                        & Positive Phase Curve (if present)   \\ \hline
$D_2$                        & Positive Phase Curve (if present)   \\ \hline
$\sigma_F$ (white noise)     & $0 < \sigma_F < 1$                 \\ \hline
Limb Darkening               & \begin{tabular}[c]{@{}c@{}}$0 < q_1 < 1$;\\ $0 < q_2 < 1$\end{tabular}  \\ \hline
$e\cos(\omega)$              & $-1 < e\cos(\omega) < 1$                 \\ \hline
$e\sin(\omega)$              & $-1 < e\sin(\omega) < 1$           \\ \hline
Poly Instrumental Variables  & Free (if present)                  \\ \hline
GP Instrumental Variables    & \begin{tabular}[c]{@{}c@{}}$-3 < \ln(L_x) < 0$ (if present);\\ $-3 < \ln(L_y) < 0$ (if present);\\$p(C) = {\rm Gam}(1, 100)$ (if present)\end{tabular}  \\ \hline
PLD Instrumental Variables   & \begin{tabular}[c]{@{}c@{}}$-3 < p_{i,1} < 3$ (if present);\\ $-500 < p_{i,2} < 500$ (if present)\end{tabular}  \\ \hline
BLISS Instrumental Variables & None                  \\
\end{tabular}
\end{table}

\begin{figure*}
    \centering
    \includegraphics[width=\linewidth]{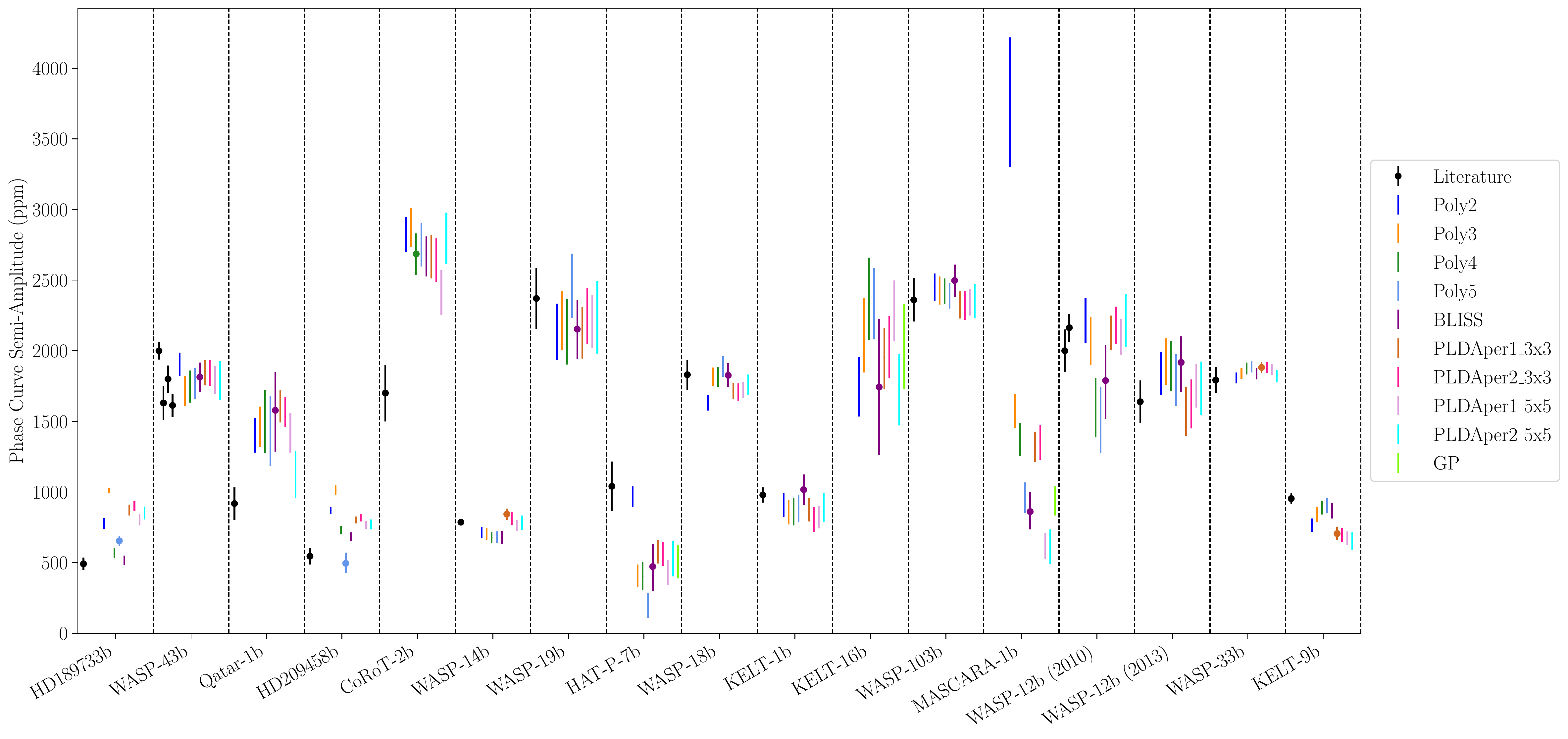}
    \caption{Phase curve semi-amplitudes for all detector models using FWM centroiding, and the previously published semi-amplitude for each phase curve. The first literature value for WASP-12b (2010) is from \citet{cowan2012}, and the second is from \citet{bell2019}. The literature values for WASP-43b are from \citet{stevenson2017}, \citet{mendonca2018a}, \citet{morello2019}, and \citet{may2020} from left to right.}
    \label{fig:amp}
\end{figure*}

\begin{figure*}
    \centering
    \includegraphics[width=\linewidth]{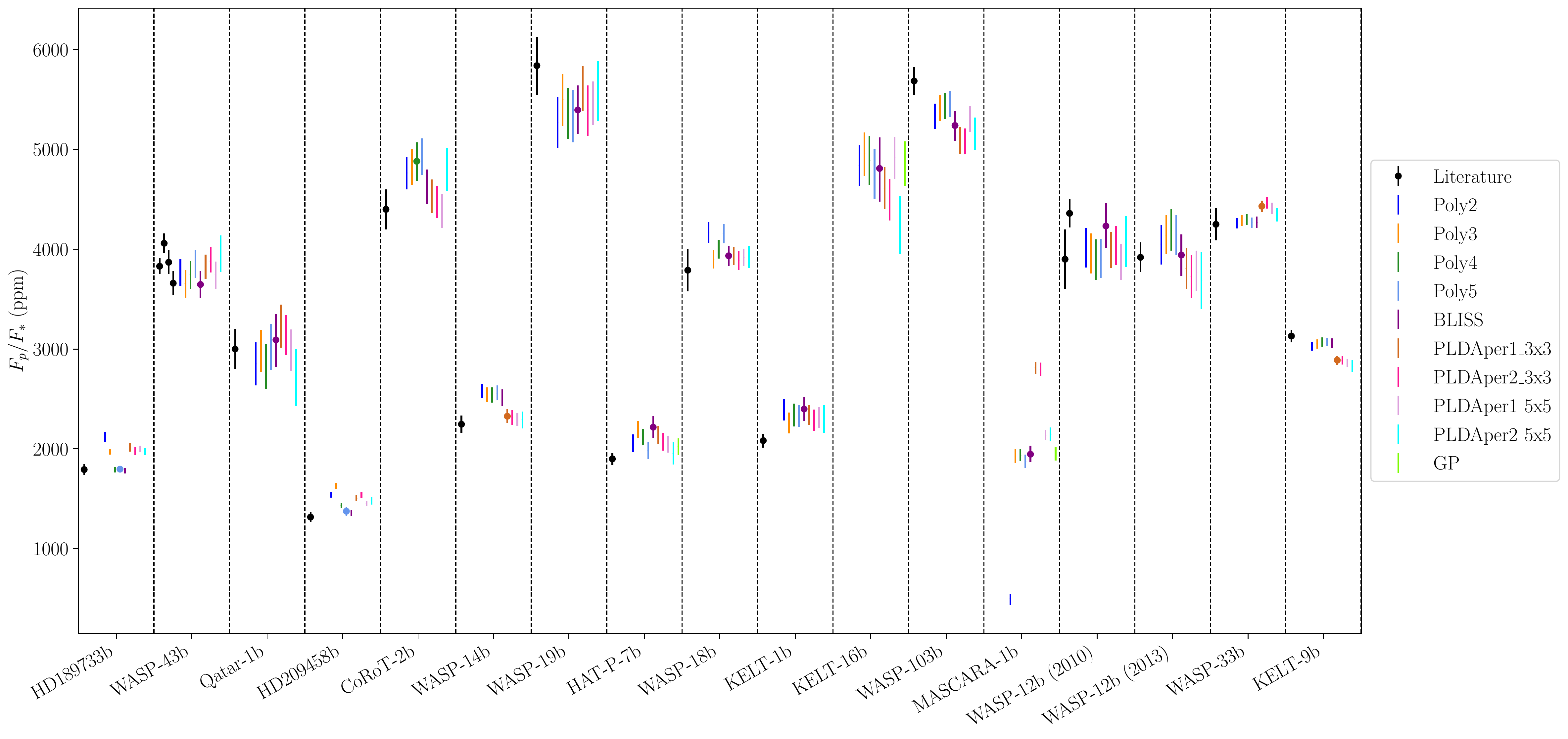}
    \caption{Dayside fluxes for all detector models using FWM centroiding, and the previously published dayside fluxes for each phase curve. The first literature value for WASP-12b (2010) is from \citet{cowan2012}, and the second is from \citet{bell2019}. The literature values for WASP-43b are from \citet{stevenson2017}, \citet{mendonca2018a}, \citet{morello2019}, and \citet{may2020} from left to right.}
    \label{fig:fday}
\end{figure*}

\begin{figure*}
    \centering
    \includegraphics[width=\linewidth]{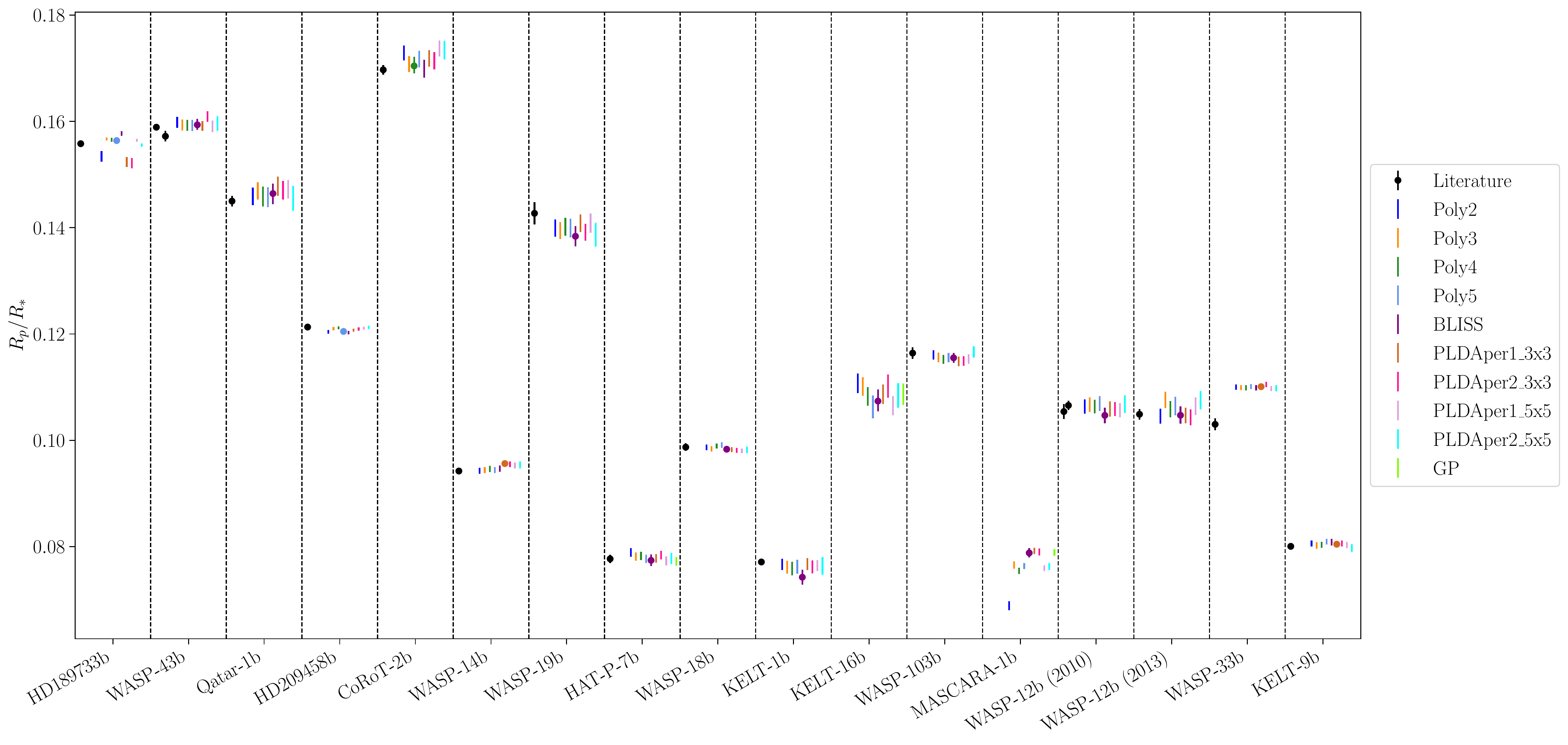}
    \caption{Radii for all detector models using FWM centroiding, and the previously published radii for each phase curve. The first literature value for WASP-12b (2010) is from \citet{cowan2012}, and the second is from \citet{bell2019}. The literature values for WASP-43b are from \citet{stevenson2017} and \citet{morello2019} from left to right.}
    \label{fig:rp}
\end{figure*}

\begin{figure*}
    \centering
    \includegraphics[width=\linewidth]{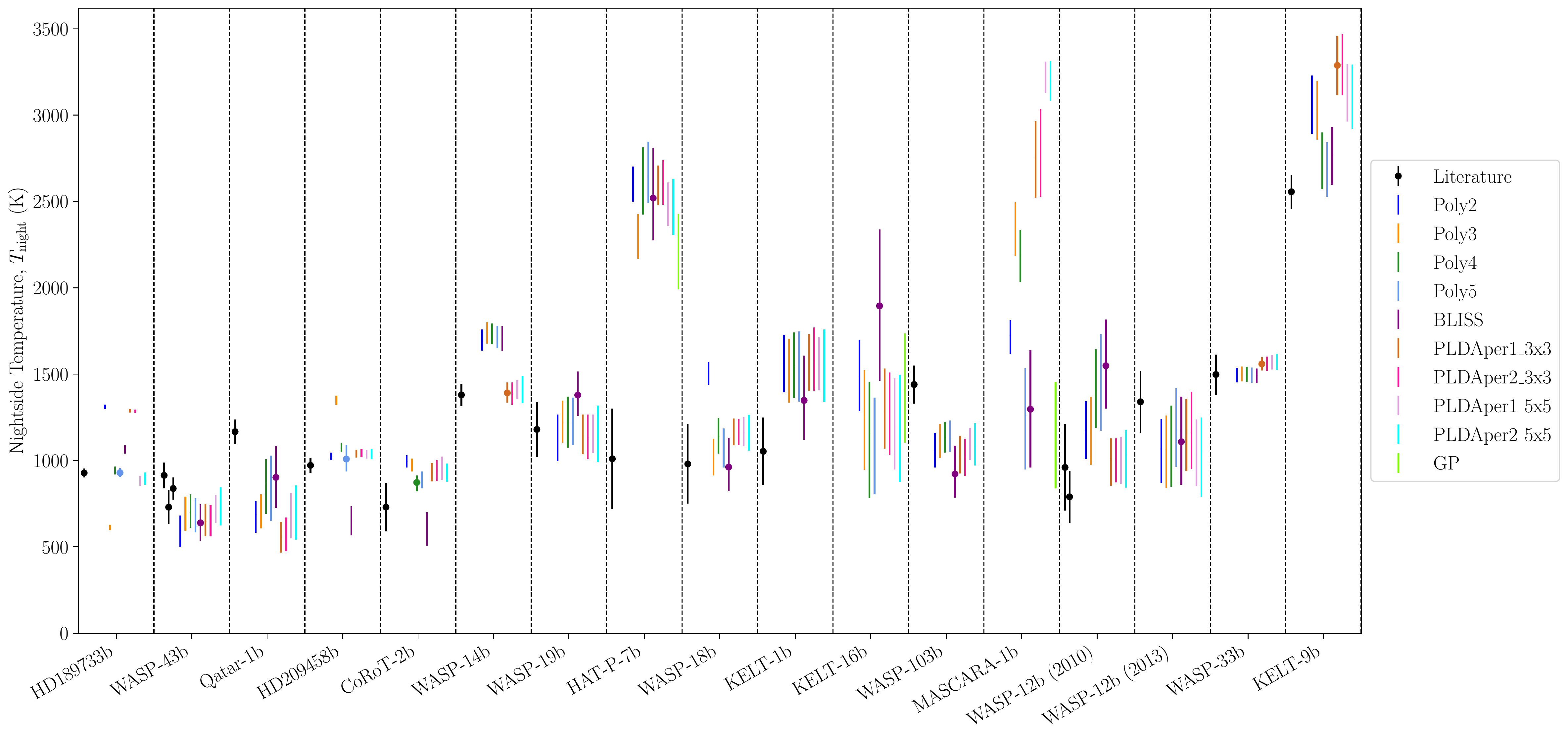}
    \caption{Nightside temperatures for all detector models using FWM centroiding, and the previously published nightside temperatures for each phase curve. The first literature value for WASP-12b (2010) is from \citet{cowan2012}, and the second is from \citet{bell2019}. The literature values for WASP-43b are from \citet{mendonca2018a}, \citet{morello2019}, and \citet{may2020} from left to right, while \citet{stevenson2017} found a 2$\sigma$ upper limit of 650~K.}
    \label{fig:tnight}
\end{figure*}

\clearpage
\section*{{\LARGE\textbf{Supplementary Information}}}

\begin{figure*}
    \centering
    \includegraphics[height=0.8\textheight]{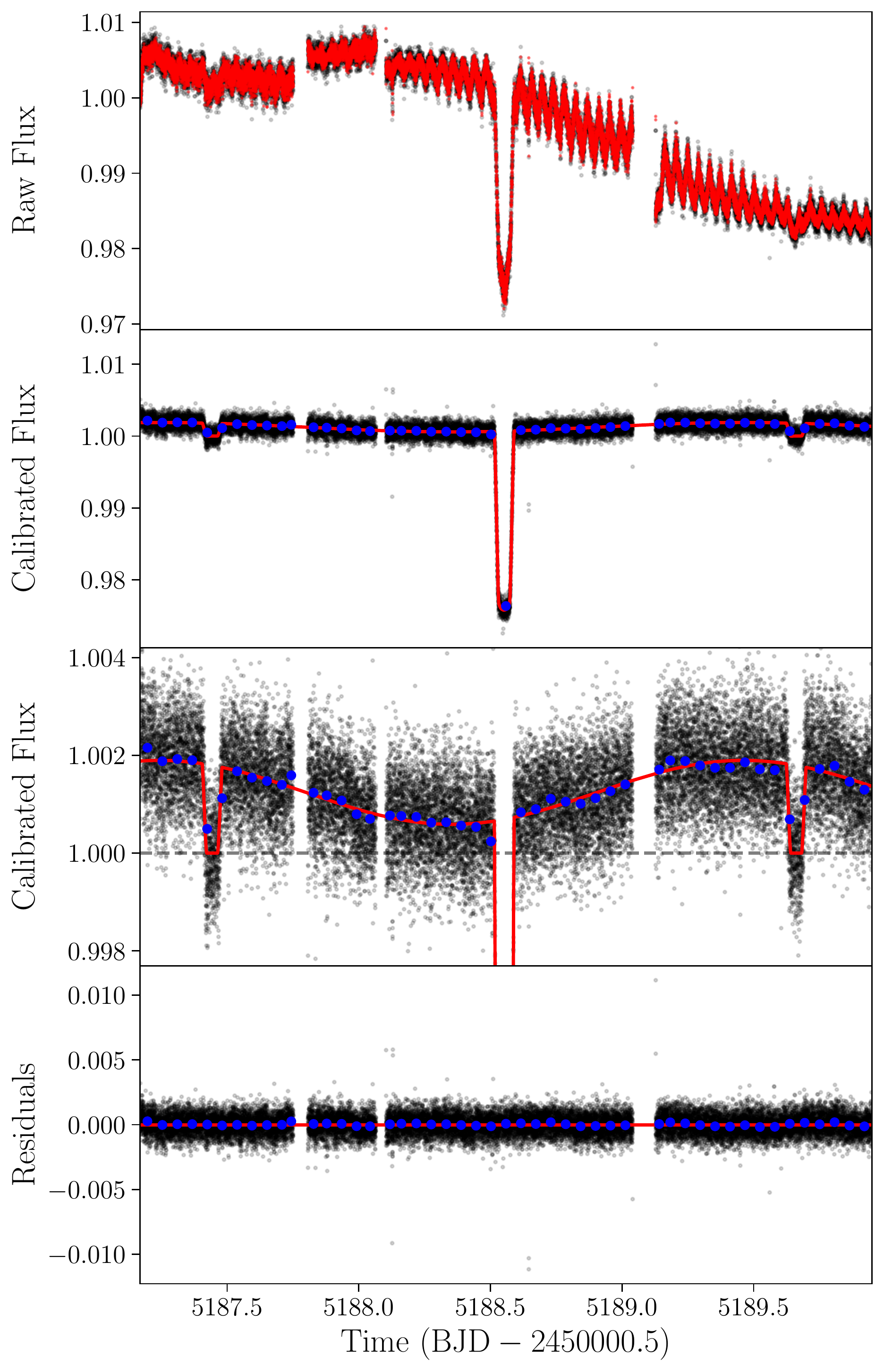}
    \caption{Preferred model fit (Poly5\textunderscore v1) for HD 189733b.}
\end{figure*}

\begin{figure*}
    \centering
    \includegraphics[height=0.8\textheight]{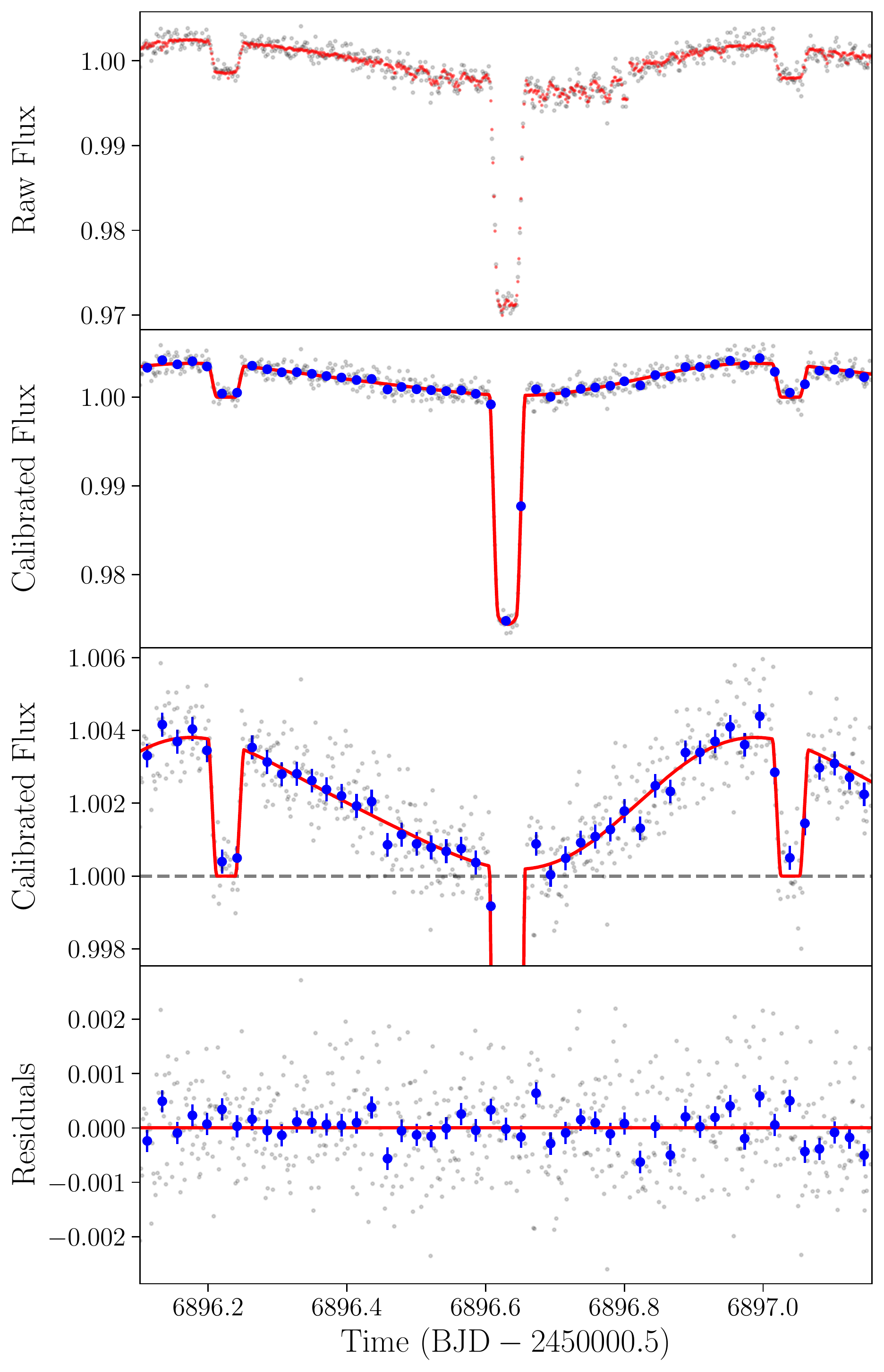}
    \caption{Preferred model fit (BLISS\textunderscore v2) for WASP-43b.}
\end{figure*}

\begin{figure*}
    \centering
    \includegraphics[height=0.8\textheight]{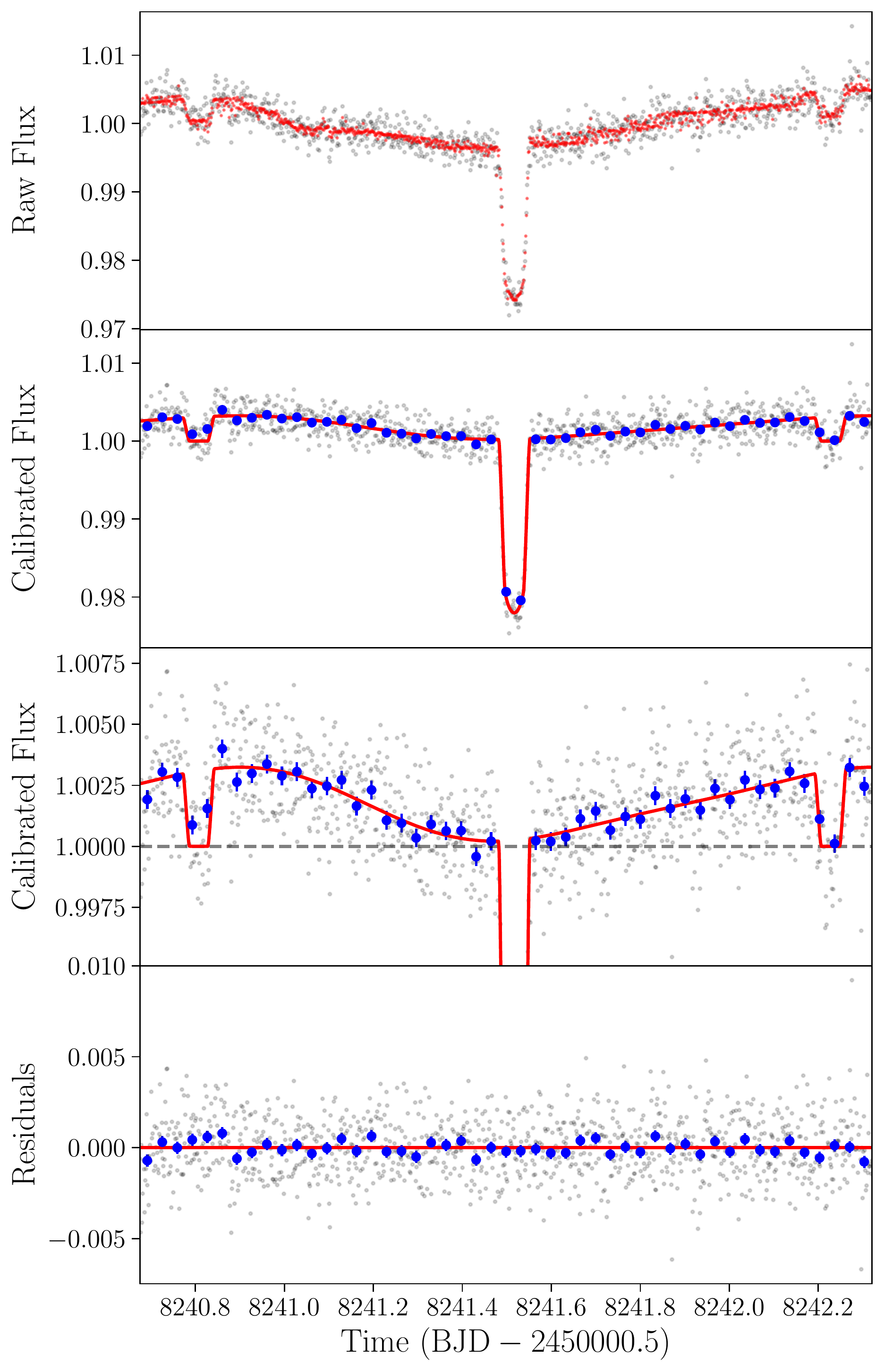}
    \caption{Preferred model fit (BLISS\textunderscore v2) for Qatar-1b.}
\end{figure*}

\begin{figure*}
    \centering
    \includegraphics[height=0.8\textheight]{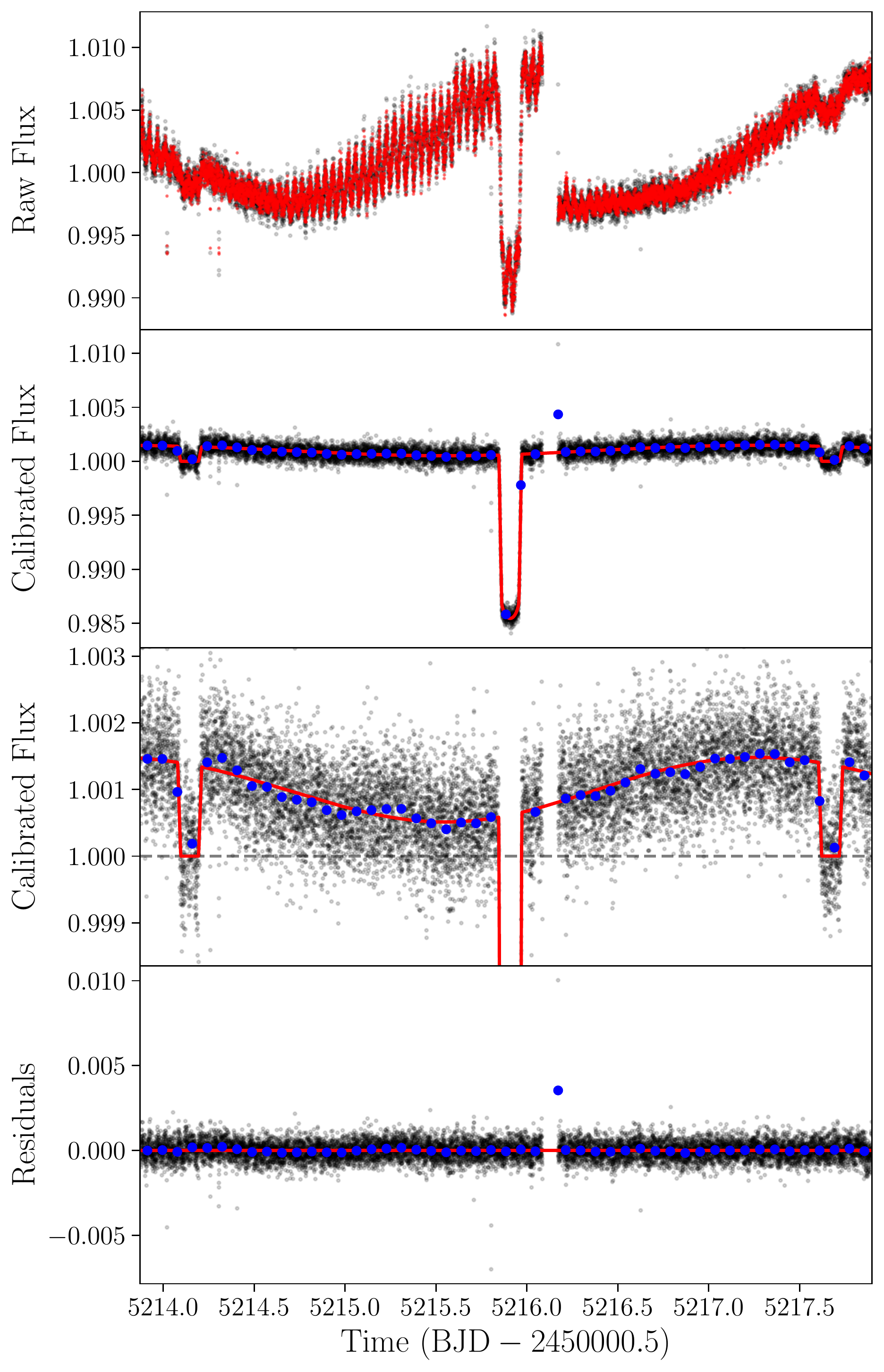}
    \caption{Preferred model fit (Poly5\textunderscore v1) for HD 209458b.}
\end{figure*}

\begin{figure*}
    \centering
    \includegraphics[height=0.8\textheight]{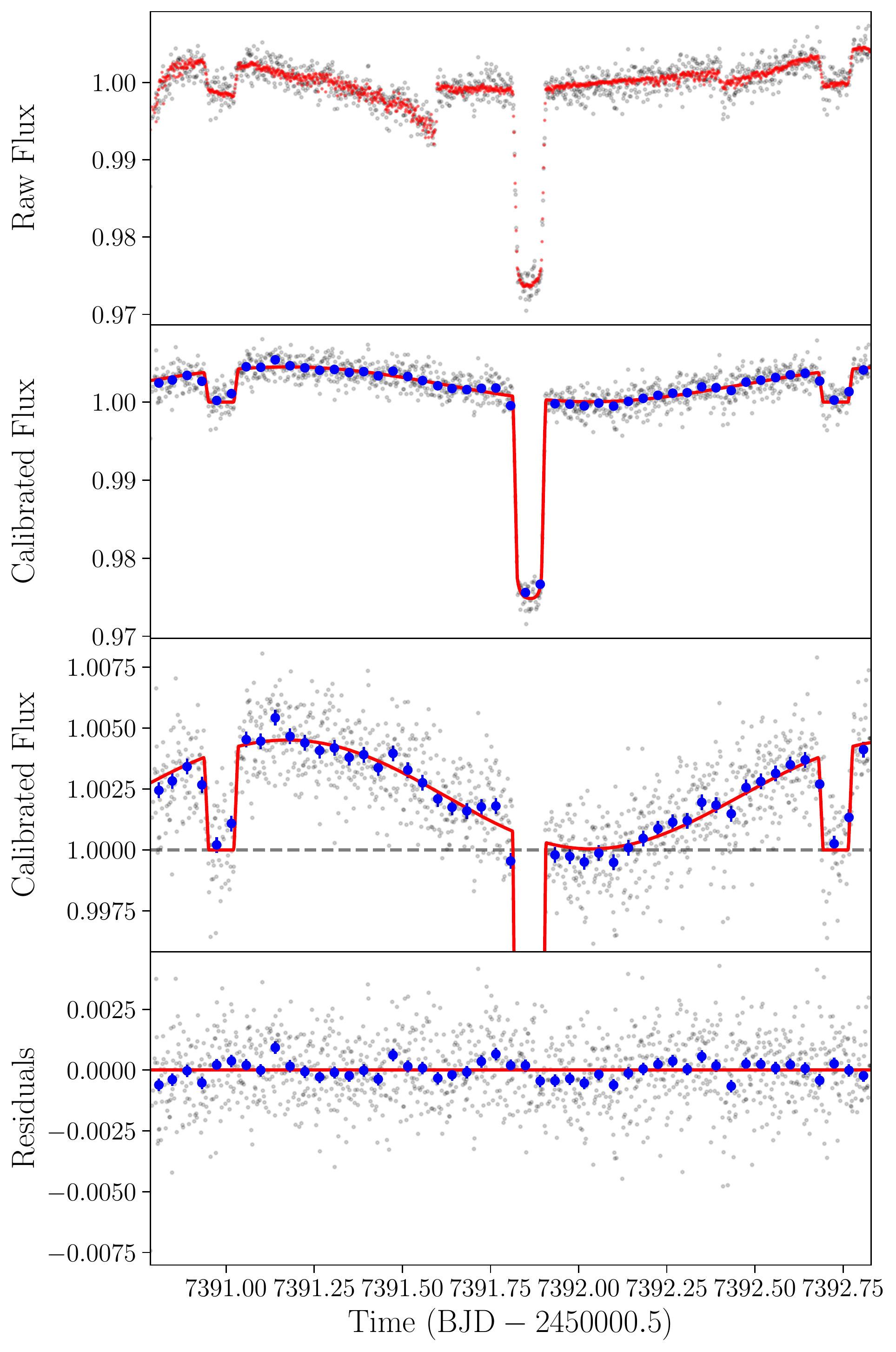}
    \caption{Preferred model fit (Poly4\textunderscore v1) for CoRoT-2b.}
\end{figure*}

\begin{figure*}
    \centering
    \includegraphics[height=0.8\textheight]{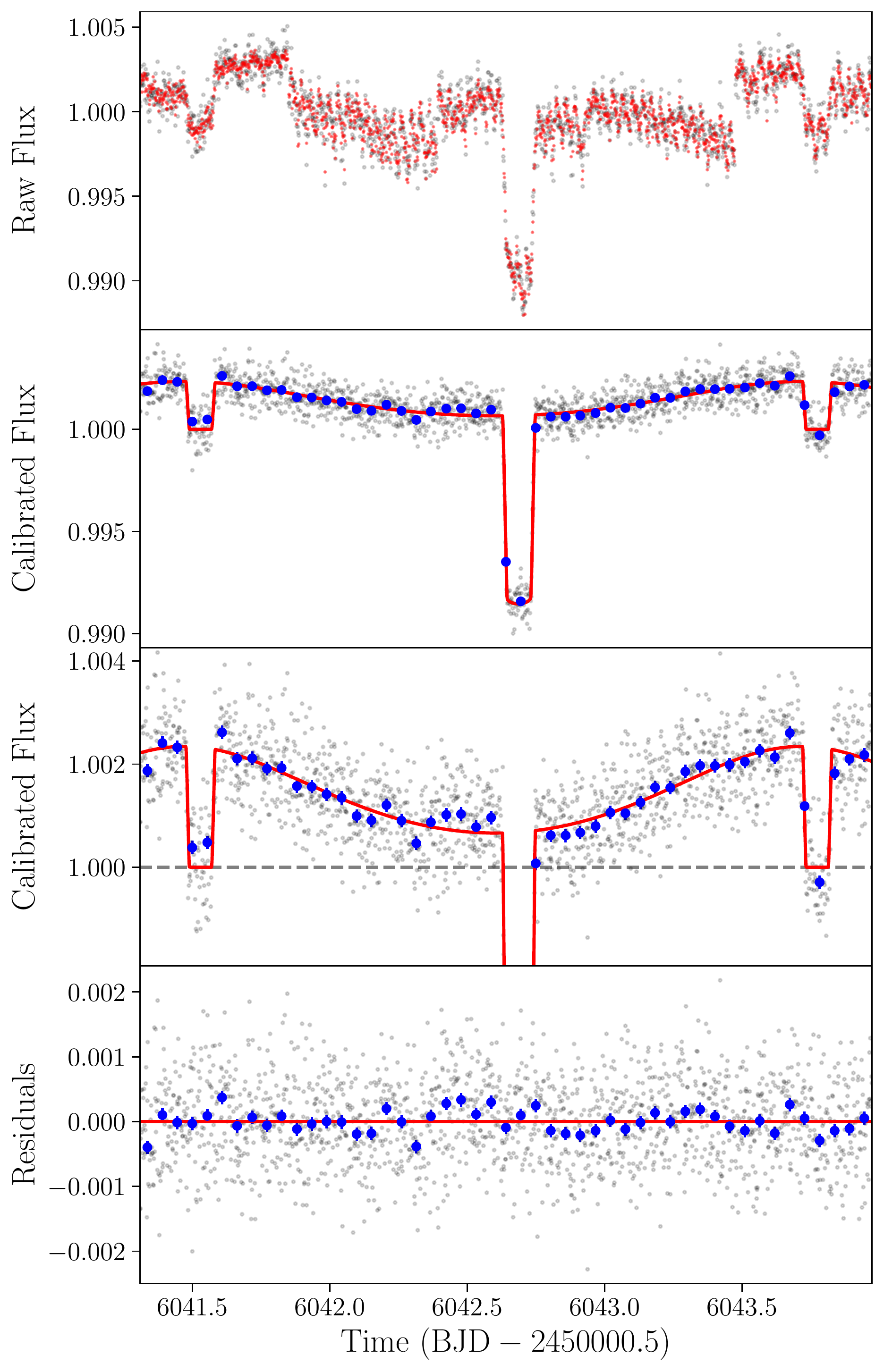}
    \caption{Preferred model fit (PLDAper1\textunderscore 3x3\textunderscore v1) for WASP-14b.}
\end{figure*}

\begin{figure*}
    \centering
    \includegraphics[height=0.8\textheight]{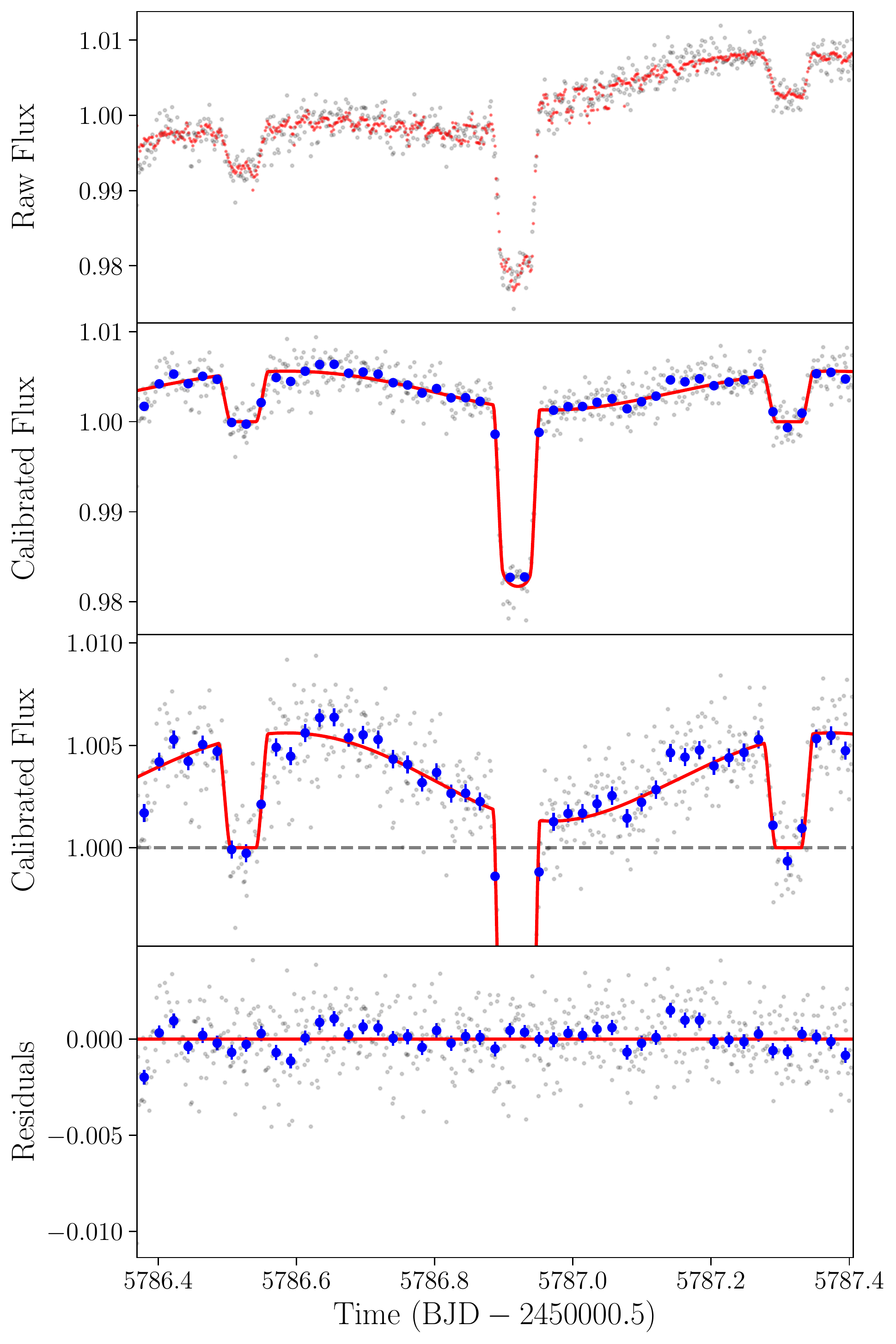}
    \caption{Preferred model fit (BLISS\textunderscore v1) for WASP-19b.}
\end{figure*}

\begin{figure*}
    \centering
    \includegraphics[height=0.8\textheight]{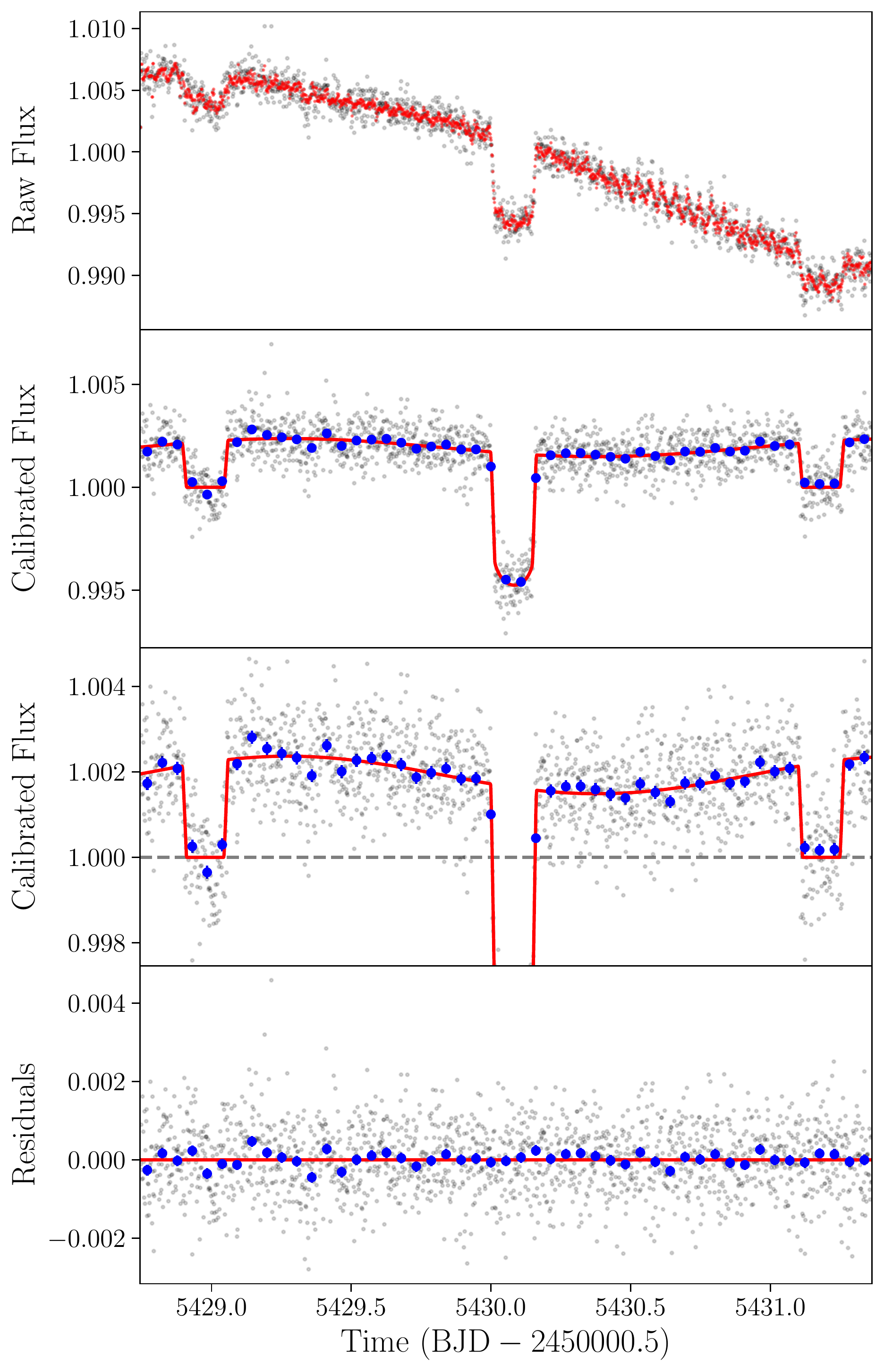}
    \caption{Preferred model fit (BLISS\textunderscore v1) for HAT-P-7b.}
\end{figure*}

\begin{figure*}
    \centering
    \includegraphics[height=0.8\textheight]{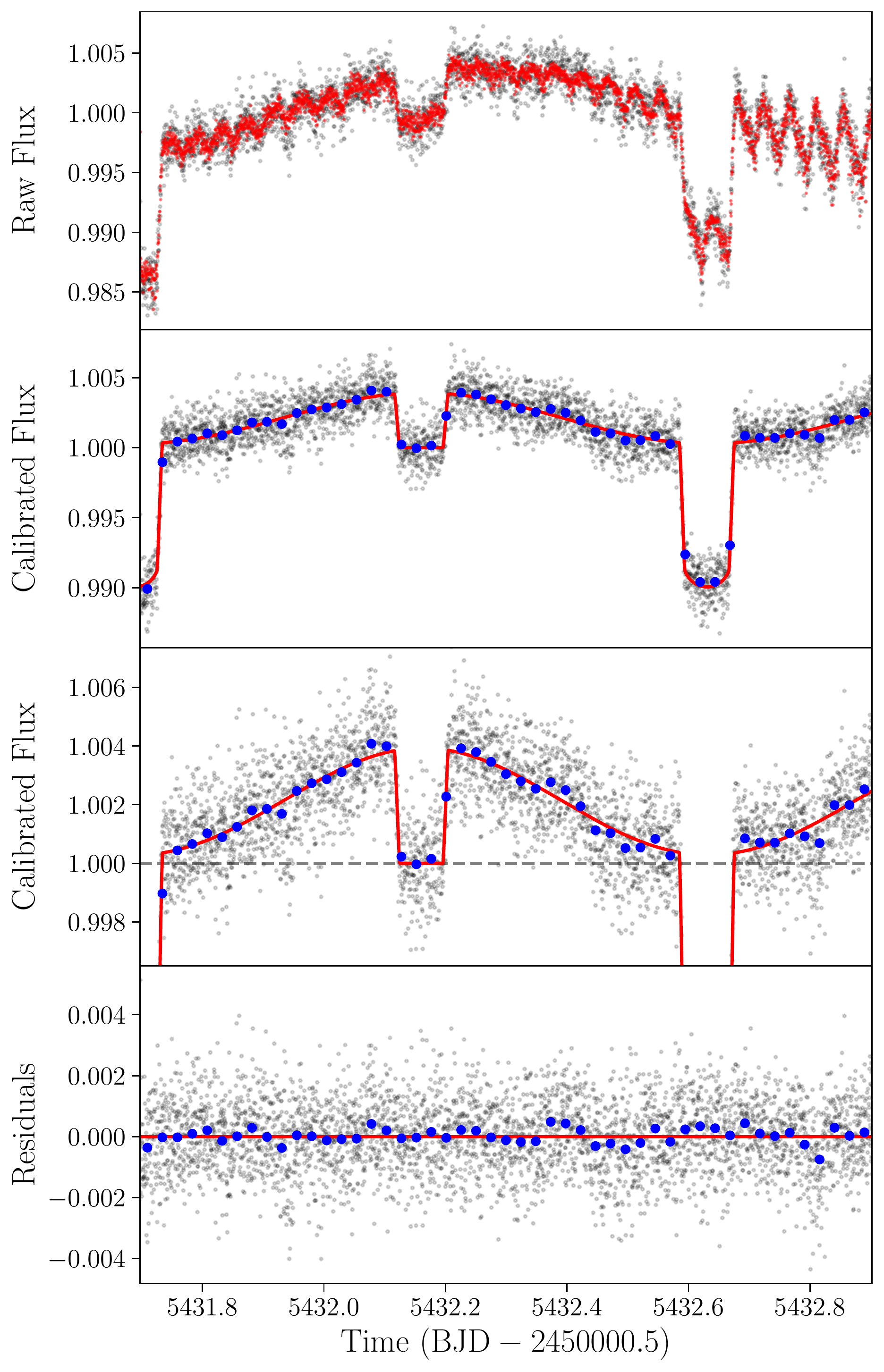}
    \caption{Preferred model fit (BLISS\textunderscore v1) for WASP-18b.}
\end{figure*}

\begin{figure*}
    \centering
    \includegraphics[height=0.8\textheight]{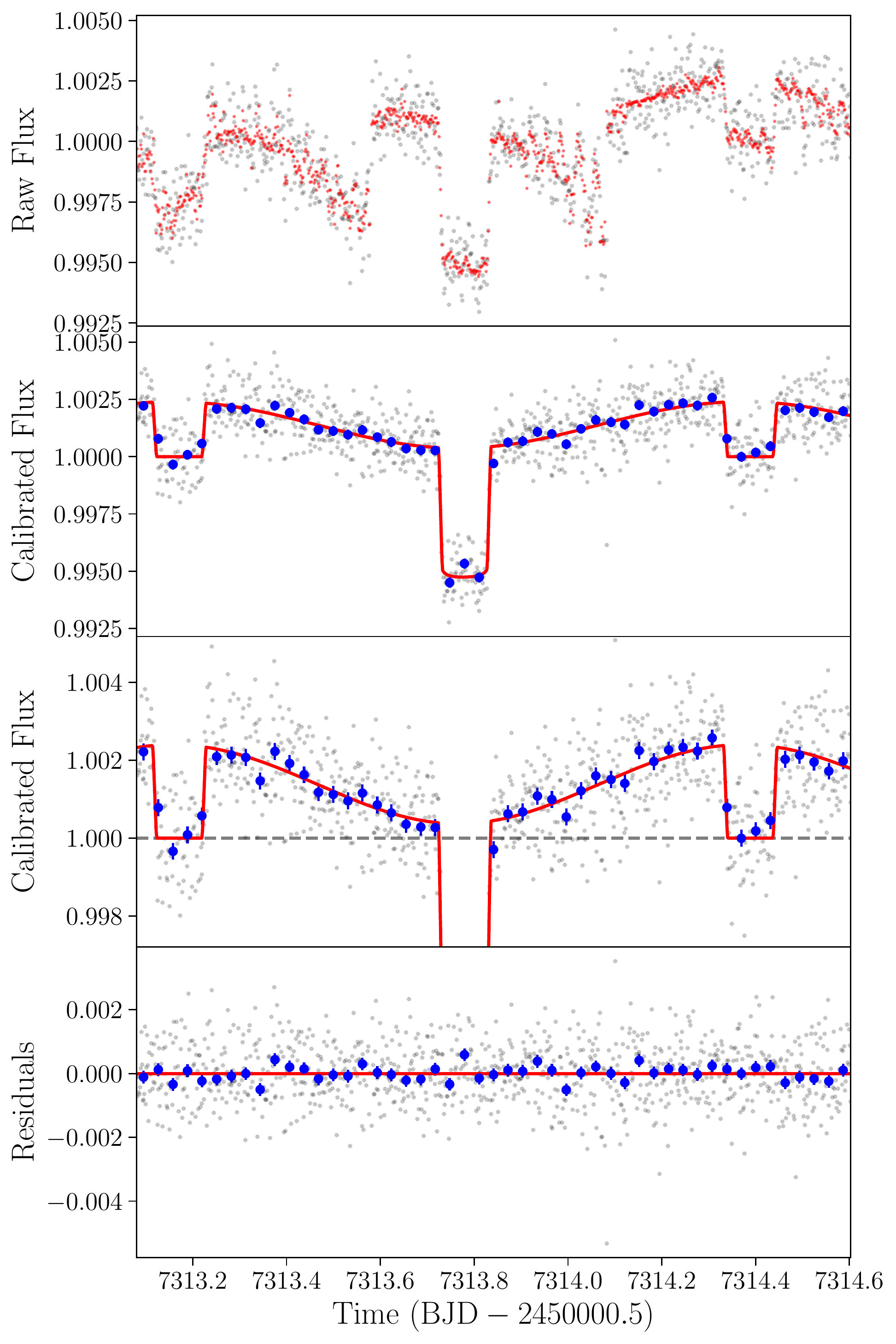}
    \caption{Preferred model fit (BLISS\textunderscore v1 with PSF centroiding) for KELT-1b.}
\end{figure*}

\begin{figure*}
    \centering
    \includegraphics[height=0.8\textheight]{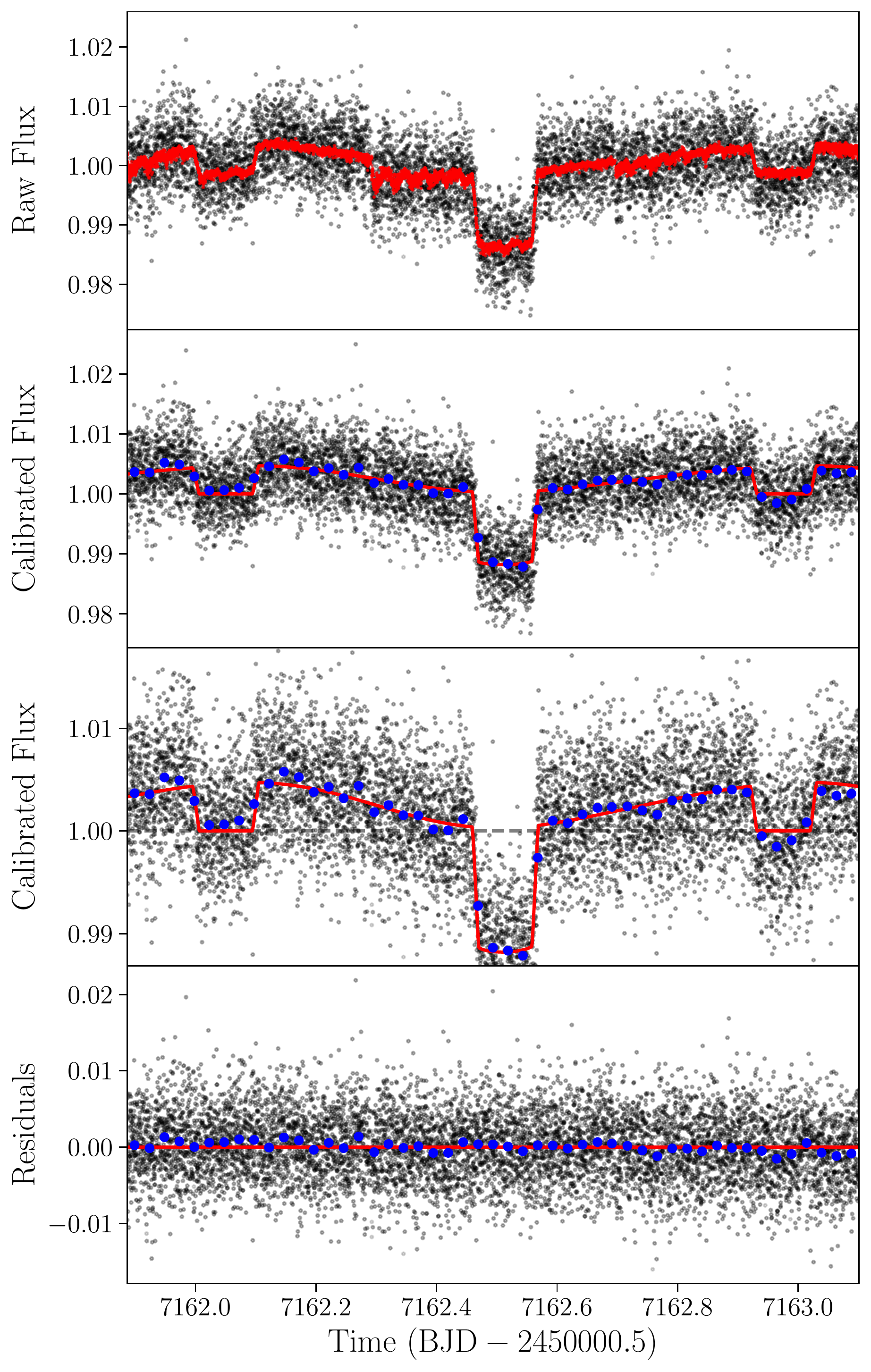}
    \caption{Preferred model fit (BLISS\textunderscore v2) for the unbinned WASP-103b photometry.}
\end{figure*}

\begin{figure*}
    \centering
    \includegraphics[height=0.8\textheight]{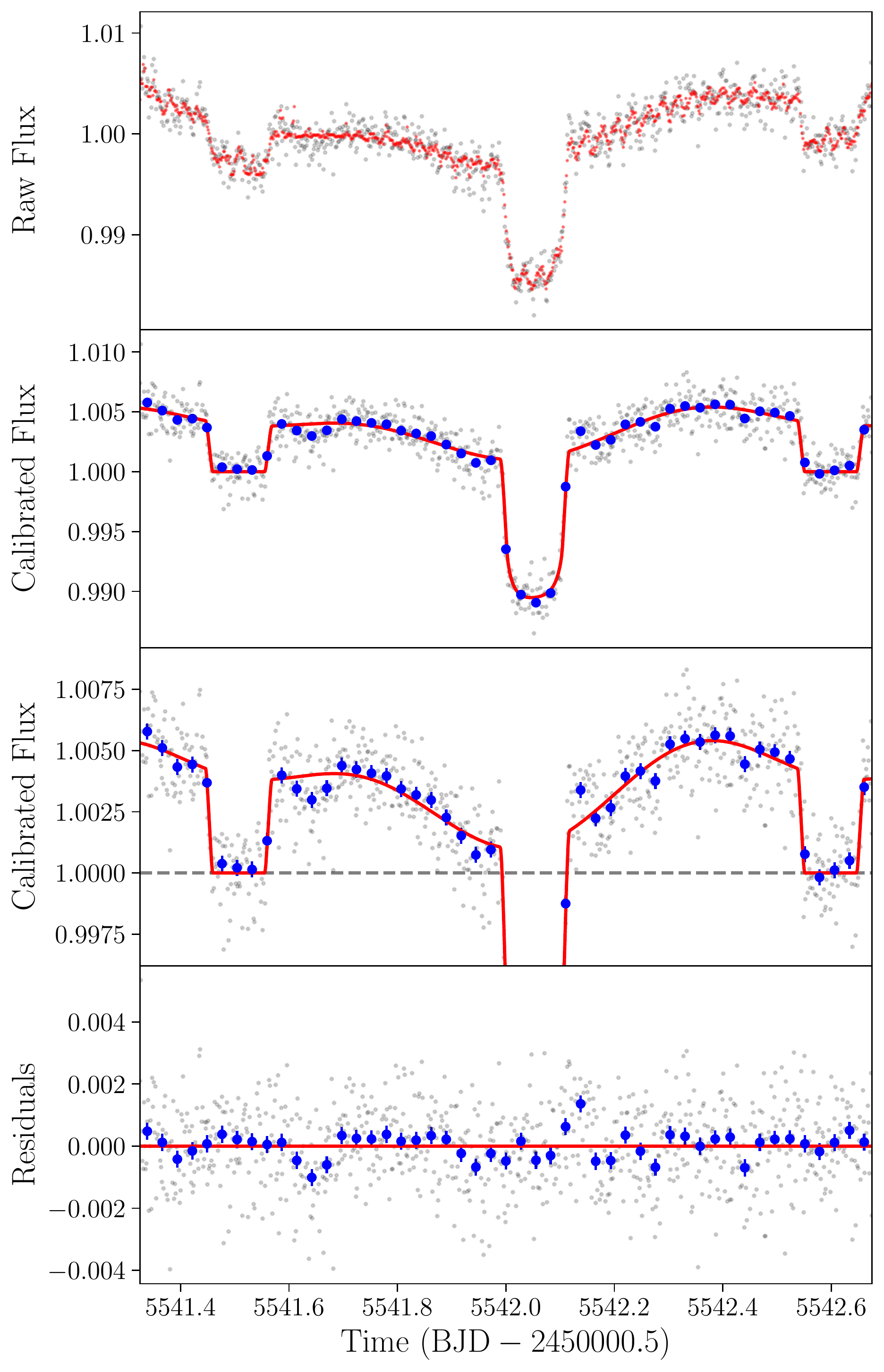}
    \caption{Preferred model fit (BLISS\textunderscore v2) for WASP-12b (2010).}
\end{figure*}

\begin{figure*}
    \centering
    \includegraphics[height=0.8\textheight]{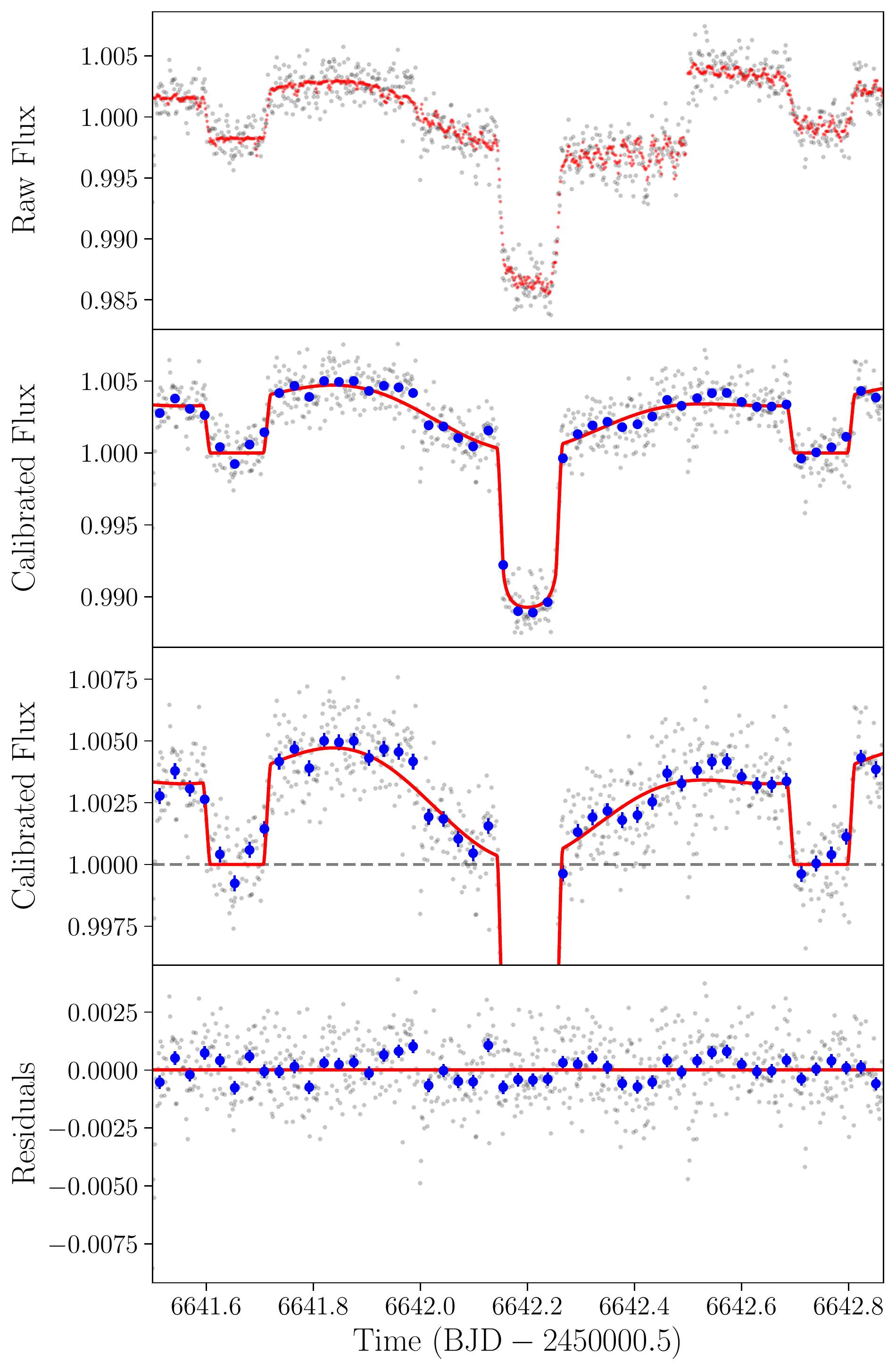}
    \caption{Preferred model fit (BLISS\textunderscore v2) for WASP-12b (2013).}
\end{figure*}

\begin{figure*}
    \centering
    \includegraphics[height=0.8\textheight]{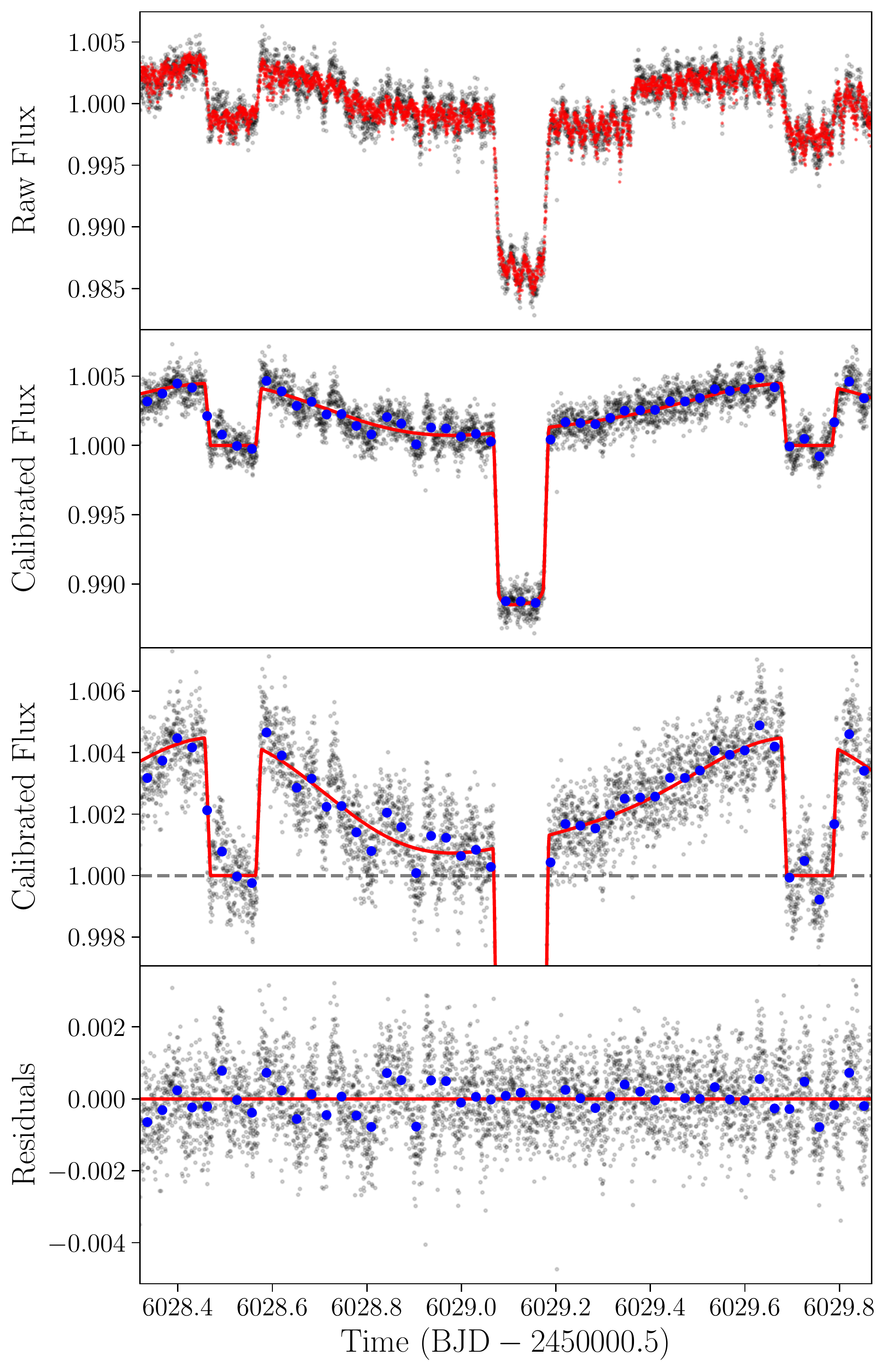}
    \caption{Preferred model fit (PLDAper1\textunderscore 3x3\textunderscore v2) for WASP-33b. The high frequency residual noise is caused by the unmodelled variability of the host star.}
\end{figure*}

\begin{figure*}
    \centering
    \includegraphics[height=0.8\textheight]{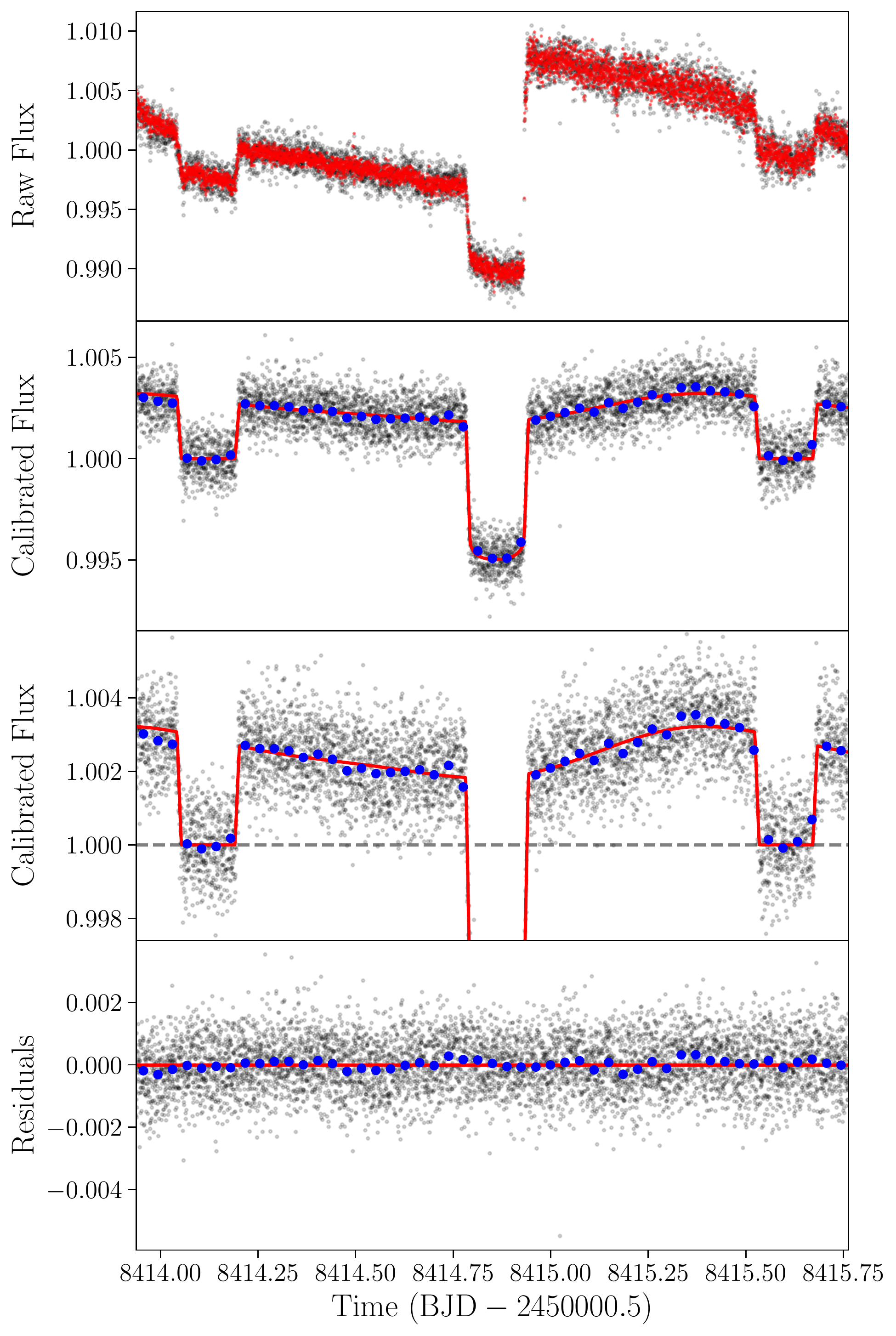}
    \caption{Preferred model fit (PLDAper1\textunderscore 3x3\textunderscore v2) for KELT-9b.}
\end{figure*}

% Don't change these lines
\bsp	% typesetting comment
\label{lastpage}
\end{document}